\documentclass[11pt]{article}
\usepackage{amsmath,amsthm,latexsym,amssymb,amsfonts,epsfig, psfrag,color}
\usepackage[small,nohug,heads=vee]{diagrams}
\diagramstyle[labelstyle=\scriptstyle]

\makeatletter \@addtoreset{equation}{section} \makeatother

\newtheorem{Theorem}{Theorem}[section]
\newtheorem{Definition}{Definition}[section]

\def\be{\begin{equation}}
\def\ee{\end{equation}}
\def\ba{\begin{eqnarray}}
\def\ea{\end{eqnarray}}

\newcommand\nn{\nonumber}
\newcommand\q{\quad}
\def\Nl{{\mathchoice
{\setbox0=\hbox{$\displaystyle\rm N$}\hbox{\hbox to0pt
{\kern0.4\wd0\vrule height0.9\ht0\hss}\box0}}
{\setbox0=\hbox{$\textstyle\rm N$}\hbox{\hbox to0pt
{\kern0.4\wd0\vrule height0.9\ht0\hss}\box0}}
{\setbox0=\hbox{$\scriptstyle\rm N$}\hbox{\hbox to0pt
{\kern0.4\wd0\vrule height0.9\ht0\hss}\box0}}
{\setbox0=\hbox{$\scriptscriptstyle\rm N$}\hbox{\hbox to0pt
{\kern0.4\wd0\vrule height0.9\ht0\hss}\box0}}}}
\def\Zl{{\mathchoice
{\setbox0=\hbox{$\displaystyle\rm Z$}\hbox{\hbox to0pt
{\kern0.4\wd0\vrule height0.9\ht0\hss}\box0}}
{\setbox0=\hbox{$\textstyle\rm Z$}\hbox{\hbox to0pt
{\kern0.4\wd0\vrule height0.9\ht0\hss}\box0}}
{\setbox0=\hbox{$\scriptstyle\rm Z$}\hbox{\hbox to0pt
{\kern0.4\wd0\vrule height0.9\ht0\hss}\box0}}
{\setbox0=\hbox{$\scriptscriptstyle\rm Z$}\hbox{\hbox to0pt
{\kern0.4\wd0\vrule height0.9\ht0\hss}\box0}}}}
\def\Ql{{\mathchoice
{\setbox0=\hbox{$\displaystyle\rm Q$}\hbox{\hbox to0pt
{\kern0.4\wd0\vrule height0.9\ht0\hss}\box0}}
{\setbox0=\hbox{$\textstyle\rm Q$}\hbox{\hbox to0pt
{\kern0.4\wd0\vrule height0.9\ht0\hss}\box0}}
{\setbox0=\hbox{$\scriptstyle\rm Q$}\hbox{\hbox to0pt
{\kern0.4\wd0\vrule height0.9\ht0\hss}\box0}}
{\setbox0=\hbox{$\scriptscriptstyle\rm Q$}\hbox{\hbox to0pt
{\kern0.4\wd0\vrule height0.9\ht0\hss}\box0}}}}
\def\Rl{{\mathchoice
{\setbox0=\hbox{$\displaystyle\rm R$}\hbox{\hbox to0pt
{\kern0.4\wd0\vrule height0.9\ht0\hss}\box0}}
{\setbox0=\hbox{$\textstyle\rm R$}\hbox{\hbox to0pt
{\kern0.4\wd0\vrule height0.9\ht0\hss}\box0}}
{\setbox0=\hbox{$\scriptstyle\rm R$}\hbox{\hbox to0pt
{\kern0.4\wd0\vrule height0.9\ht0\hss}\box0}}
{\setbox0=\hbox{$\scriptscriptstyle\rm R$}\hbox{\hbox to0pt
{\kern0.4\wd0\vrule height0.9\ht0\hss}\box0}}}}
\def\Cl{{\mathchoice
{\setbox0=\hbox{$\displaystyle\rm C$}\hbox{\hbox to0pt
{\kern0.4\wd0\vrule height0.9\ht0\hss}\box0}}
{\setbox0=\hbox{$\textstyle\rm C$}\hbox{\hbox to0pt
{\kern0.4\wd0\vrule height0.9\ht0\hss}\box0}}
{\setbox0=\hbox{$\scriptstyle\rm C$}\hbox{\hbox to0pt
{\kern0.4\wd0\vrule height0.9\ht0\hss}\box0}}
{\setbox0=\hbox{$\scriptscriptstyle\rm C$}\hbox{\hbox to0pt
{\kern0.4\wd0\vrule height0.9\ht0\hss}\box0}}}}
\def\Hl{{\mathchoice
{\setbox0=\hbox{$\displaystyle\rm H$}\hbox{\hbox to0pt
{\kern0.4\wd0\vrule height0.9\ht0\hss}\box0}}
{\setbox0=\hbox{$\textstyle\rm H$}\hbox{\hbox to0pt
{\kern0.4\wd0\vrule height0.9\ht0\hss}\box0}}
{\setbox0=\hbox{$\scriptstyle\rm H$}\hbox{\hbox to0pt
{\kern0.4\wd0\vrule height0.9\ht0\hss}\box0}}
{\setbox0=\hbox{$\scriptscriptstyle\rm H$}\hbox{\hbox to0pt
{\kern0.4\wd0\vrule height0.9\ht0\hss}\box0}}}}
\def\Ol{{\mathchoice
{\setbox0=\hbox{$\displaystyle\rm O$}\hbox{\hbox to0pt
{\kern0.4\wd0\vrule height0.9\ht0\hss}\box0}}
{\setbox0=\hbox{$\textstyle\rm O$}\hbox{\hbox to0pt
{\kern0.4\wd0\vrule height0.9\ht0\hss}\box0}}
{\setbox0=\hbox{$\scriptstyle\rm O$}\hbox{\hbox to0pt
{\kern0.4\wd0\vrule height0.9\ht0\hss}\box0}}
{\setbox0=\hbox{$\scriptscriptstyle\rm O$}\hbox{\hbox to0pt
{\kern0.4\wd0\vrule height0.9\ht0\hss}\box0}}}}
\newcommand{\cc}{\mathcal C}

\newcommand{\ch}{\mathcal H}

\newcommand{\cl}{\mathcal L}

\newcommand{\cp}{\mathcal P}
\newcommand{\cq}{\mathcal Q}

\newcommand{\ft}{\mathfrak{t}}

\newcommand{\eps}{\epsilon}

\def\nn{\nonumber}

\newcommand{\eqa}{\begin{eqnarray}}
\newcommand{\neqa}{\end{eqnarray}}

\newcommand{\p}{\partial}

\usepackage{bbm}

\def\Z{{\mathbbm Z}}

\let\eps=\epsilon

\def\q{{\quad}}

\definecolor{bianca}{rgb}{0,0.,0.8}

\begin{document}

{\renewcommand{\thefootnote}{\fnsymbol{footnote}}

\title{Canonical simplicial gravity}
\author{Bianca Dittrich\footnote{e-mail address: {\tt dittrich@aei.mpg.de}}  $^{1}$ and Philipp A H\"ohn\footnote{e-mail address: {\tt phoehn@perimeterinstitute.ca}}  $^{2}$\\
\small  $^1$ MPI for Gravitational Physics, Albert Einstein Institute,\\
 \small Am M\"uhlenberg 1, D-14476 Potsdam, Germany\\
\small   $^2$ Institute for Theoretical Physics, Universiteit Utrecht,\\
 \small  Leuvenlaan 4, NL-3584 CE Utrecht, The Netherlands 
}

\date{\small AEI--2011--055, ITP--UU--11/31, SPIN--11/24}

}


\setcounter{footnote}{0}
\maketitle

\vspace{-.9cm}

\begin{abstract}

A general canonical formalism for discrete systems is developed which can handle varying phase space dimensions and constraints. The central ingredient is Hamilton's principal function which generates canonical time evolution and ensures that the canonical formalism reproduces the dynamics of the covariant formulation following directly from the action. We apply this formalism to simplicial gravity and (Euclidean) Regge calculus, in particular. A discrete forward/backward evolution is realized by gluing/removing single simplices step by step to/from a bulk triangulation and amounts to Pachner moves in the triangulated hypersurfaces. As a result, the hypersurfaces evolve in a discrete `multi-fingered' time through the full Regge solution. Pachner moves are an elementary and ergodic class of homeomorphisms and generically change the number of variables, but can be implemented as canonical transformations on naturally extended phase spaces. Some moves introduce {\it a priori} free data which, however, may become fixed {\it a posteriori} by constraints arising in subsequent moves. The end result is a general and fully consistent formulation of canonical Regge calculus, thereby removing a longstanding obstacle in connecting covariant simplicial gravity models to canonical frameworks. The present scheme is, therefore, interesting in view of many approaches to quantum gravity, but may also prove useful for numerical implementations.

\end{abstract}

\section{Introduction}

Over the past decades, discrete gravity theories and models have been employed as useful tools in the study of numerous aspects of both classical and quantum gravity. The discrete structures usually provide, on the one hand, formulations which are amenable to numerical investigations \cite{baumshap,regge} and, on the other hand, a regulator which is particularly convenient for the construction and definition of (UV-finite) quantum gravity models \cite{rocekwilliams,lollreview,edt,cdt,barrett,spinfoam,newspinfoam}. An auxiliary discretization of a continuum theory can, however, introduce a number of discretization artifacts as well as break continuum symmetries, in particular, the continuum diffeomorphism symmetry of general relativity is generically broken in discrete gravity \cite{rocekwilliams,lollreview,bd1,dithoe1,loll, hamberwilliamsgauge,morse}. 

The vast majority of research in discrete gravity has been performed in the covariant setting, i.e.\ by means of formulations directly following from an action. Specifically, in the quantum theory one has thereby benefited from the many numerical methods devised for lattice gauge theories which are based on Euclidean path-integral techniques \cite{lollreview,edt,cdt}. In contrast to this, efforts to construct canonical formulations of discrete gravity have been few and far between; first attempts \cite{friedmann} were based on a discretized space, but a continuous time. However, consistency requires that the canonical theory reproduces exactly the dynamics and (possibly broken) symmetries of the covariant formulation, i.e.\ solutions of the canonical theory must replicate those following directly from the action. In particular, a canonical framework consistent with the discrete action ought to yield a discrete time evolution. A first consistent canonical formulation of simplicial gravity\footnote{By `simplicial gravity' we mean a gravitational theory based on simplices and triangulations, as opposed to the more general term `discrete gravity' which may encompass also other discretization schemes.} has only appeared recently \cite{bd1,dithoe1} and was based on ideas from the `consistent discretizations program'  \cite{gambini} and an evolution scheme presented in \cite{commi}. However, this canonical evolution scheme was only applicable to a special class of triangulations with both fixed topology and connectivity of the `spatial' triangulated hypersurfaces and no consistent canonical formulation of simplicial gravity has been constructed in full generality thus far.

One of the major obstacles in the quest of a general canonical formulation of simplicial gravity has been the issue of foliating a $D$-dimensional covariant solution by $(D-1)$-dimensional triangulated hypersurfaces which are generically comprised of different numbers of subsimplices which, in turn, carry the variables of the theory. The dimension of configuration or phase spaces associated to such discrete slices must therefore vary and a canonical evolution scheme thereby calls for mappings between phase spaces of different dimension. 

Why should one attempt to construct a general formulation of canonical simplicial gravity in the first place? The general answer is that classically this provides a better notion of discrete dynamics and time evolution, and, furthermore, in quantum gravity one is interested in, e.g.\ transition amplitudes between different three-geometries which generally requires to be able to identify the corresponding quantum states as elements of a Hilbert space of three-geometries. But there are further practical reasons why it appears useful to be able to treat a changing/evolving lattice with varying numbers of physical and gauge degrees of freedom in a canonical framework. For instance: (i) There are plenty of physical situations whose description may be facilitated via an adaptation of the discretization in time. For (discrete) gravity the prime example is an expanding or contracting universe, while a more extreme example is the `no boundary' proposal \cite{hawking}. (ii) Such a framework may foster the scope of numerical implementations. (iii) It may be advantageous for quantization and help in comparing or even connecting various approaches to quantum gravity. Most of the latter can be roughly divided into two categories, namely canonical continuum quantizations (e.g.\ Loop Quantum Gravity \cite{thiemannal,rovellibuch}) and covariant path integral formulations, regularized by a discrete lattice structure (e.g.\ Causal Dynamical Triangulations \cite{cdt}, Spin Foams \cite{spinfoam,newspinfoam,rovellibuch}, Quantum Regge Calculus \cite{rocekwilliams}...). A link between such approaches remains elusive thus far.\footnote{With the exception of the Loop-Quantum-Gravity- and Spin-Foam-formulations of the topological 3D-theory which were shown to be equivalent \cite{perez}.} It seems that a consistent canonical framework for simplicial gravity is a prerequisite for at least comparing these two categories, especially in view of Loop Quantum Gravity, on the one hand, which involves changing `spatial' graphs, and Spin Foams, on the other hand, which are based on triangulations.

Motivated by these potential benefits, it is the goal of the present article to overcome the technical obstacles first at the classical level by devising a canonical (discrete) evolution scheme for triangulations which
\begin{enumerate}
\item is general and applicable to arbitrary triangulations (of fixed `spatial' topology, but varying `spatial' hypersurfaces) satisfying the equations of motion following from an action, 
\item is local (i.e., for practical reasons, one may restrict oneself to a finite number of equations), and
\item may be interpreted entirely from the perspective of the $(D-1)$-dimensional hypersurfaces.
\end{enumerate}
The underlying basic idea is to build up a $D$-dimensional triangulation (satisfying the equations of motion) step by step by gluing at every discrete evolution step one single $D$-simplex onto the $(D-1)$-dimensional hypersurface of the previous piece of triangulation. Backward evolution is analogously obtained by removing single $D$-simplices from the bulk triangulation. In such a way --- in analogy to the evolution of hypersurfaces in canonical general relativity --- a `spatial' triangulated hypersurface evolves in a discrete `multi-fingered' or `bubble' time through the full D-dimensional solution. As we shall see, the gluing/removal moves which we permit can be interpreted within the hypersurfaces as Pachner moves, which are an elementary and ergodic class of topology preserving moves, mapping between any triangulations of the same topology by finite sequences. This local evolution scheme has to be constructed consistently in such a way that the equations of motion are satisfied at every step and requires the discrete action to be additive. The latter is fulfilled for a particular discretization of general relativity, namely length Regge Calculus, for which we will develop this evolution scheme here in 3D and 4D. However, the scheme is more general and, through a suitable change of variables, applicable to other simplicial gravity theories (with additive action) as well. 

In order to incorporate these ideas in a canonical framework, we must first of all extend the theory of discrete dynamics from the regular \cite{marsdenwest} (i.e., for systems without constraints) to the irregular case where constraints necessarily occur (see also \cite{gambini}), before applying this to simplicial gravity. The central idea of our canonical framework is to resort to Hamilton's principal function as a generating function for a canonical time evolution; in our irregular case, this time evolution map will only be defined on and between constraint surfaces. Additionally, equations of motion transform naturally into certain canonical constraints which permit to suitably extend phase spaces, associated to hypersurfaces in such a way as to implement the Pachner moves as canonical transformations. Because of the {\it a priori} (i.e.\ prior to extension) varying phase space dimensions, the space of initial data corresponding to a given hypersurface cannot, in general, correspond to the space of solutions (modulo gauge) and we generically have to expect a high degree of non--uniqueness. Indeed, as we shall discuss in more detail in the main body of this paper, not every Pachner move will invoke equations of motion and rather allow for data that can be {\it a priori} freely chosen at that step, but may become fixed {\it a posteriori} by additional constraints arising in subsequent moves. Nevertheless, the end result is a general and fully consistent formulation of canonical Regge Calculus.

This new canonical framework can also be viewed as a method to generate solutions and is, furthermore, susceptible to quantization. In particular, it admits the advantage that, by using the action as a generating function, the eventual canonical quantum theory should directly correspond to the path-integral formulation based on the same discrete action.

The rest of the article is organized as follows: section \ref{candisc} provides the basic ideas to deal with systems with changing phase space dimensions and summarizes all necessary facts for generating functions of relevant canonical transformations. Subsequently, in section \ref{Pachner} we outline the elementary evolution scheme for triangulations which amounts to Pachner moves in the hypersurfaces and review essential background knowledge of Regge Calculus. Section \ref{3D} discusses in detail the Regge dynamics and implementation of Pachner moves in 3D, elucidates how to start from a zero-dimensional phase space and exhibits in an example the reproduction of the so--called tent moves in 3D by sequences of the Pachner moves, while section \ref{4D} investigates the same points in 4D. In order to enhance readability of the article, we decided to move more technical aspects to the end of the manuscript where, after an introduction to the Lagrangian and canonical formalism for regular discrete systems in section \ref{appreg}, we develop in section \ref{sing} a general formalism for singular discrete systems, which, in particular, is also applicable to systems with varying phase space dimension. We discuss the relevant symplectic structure and specify in which sense symplectic forms are preserved, before applying this formalism to Pachner moves in section \ref{apppach}. Section \ref{intcon} offers a brief interpretation of the constraints arising in the new framework and, finally, section \ref{conc} contains the conclusions and an outlook to further work. 

The application of the present framework to the 4D linearized theory will appear in a companion paper \cite{dhta}.

\section{Canonical discrete dynamics}\label{candisc}

The central idea \cite{bd1,gambini,marsdenwest, bianca,baez,rovnew} for generating a canonical discrete dynamics is to employ Hamilton's principal function $\tilde S$ as a generating function for a canonical transformation, 
which implements time evolution.

Hamilton's principal function $\tilde S$ is the action $S$ evaluated on solutions. That is, we assume that the action defines a well defined boundary problem, with boundary data given by some configuration variables $x_{ini}$ and $x_{fin}$ associated with initial and final boundaries. (We will assume that we can split the boundary in two parts.) Hamilton's principal function $\tilde S$ is thus a function of these boundary data $x_{ini}$ and $x_{fin}$. Since it arises by integrating out (i.e.\ solving for) all the bulk variables, we might also use the term `effective action', instead of Hamilton's principal function.

Furthermore, we assume that the action is additive in an appropriate sense. More precisely, the action associated to a region which is comprised of two regions should be the sum of the actions associated to each of the two regions
\ba
S(\{x\}_{A \cup B})&=& S(\{x\}_{A})+S(\{x\}_{B})\q,
\ea
where $\{x\}_C$ denotes all the dynamical variables associated to the region $C$.
As a consequence of the fact that actions normally arise as space-time integrals over Lagrangians, additivity can usually be obtained by taking into account boundary terms.

If the regions are such that the boundary variables of $A,B$ are $\{x_{ini}, x_{inter}\}$ and $\{x_{inter},x_{fin}\}$, respectively, we find the
 following convolution property of Hamilton's principal function
\ba\label{b1}
\tilde S(x_{ini}, x_{fin})= \text{extr}_{x_{inter}}\left[   \tilde S(x_{ini}, x_{inter})+   \tilde S (x_{inter}, x_{fin})  \right] \q .
\ea
Here `extr' indicates that we look for the value of $x_{inter}$ that solves the variational problem of the functional on square brackets in (\ref{b1}), i.e.\ a solution to the dynamical problem defined by the action
\ba\label{b2}
S(x_{ini},x_{inter},x_{fin}):= \tilde S(x_{ini}, x_{inter})+   \tilde S (x_{inter}, x_{fin})  \q .
\ea
The property (\ref{b1}) can be proven by splitting the variational problem over all variables associated to the region $A\cup B$ (but keeping $x_{ini},x_{fin}$ fixed) into three parts: one involving only the variables inside $A$, the other one only involving the variables inside $B$ and, finally, varying with respect to the boundary data $x_{inter}$.

The action on the left hand side of (\ref{b2}) can, in fact, be understood as a discrete action of a problem with two time steps. Usually, we do not have Hamilton's principal function available (which requires the solution of the continuum problem), however, one can approximate it for small time intervals with some discrete (one time step) action, which is a function of initial and final configuration data. Enumerating the time steps with a discrete label $k$, we will denote such a choice by $S_k:=S(x_{k-1},x_{k})$.

In order to define a Legendre transform and time evolution from such a discrete action, we can appeal to the standard formalism for Hamilton's principal function. Hamilton's principal function is also a generating function of the first kind (i.e.\ depends on the old and new configuration coordinates) and thereby determines the canonical 
 time evolution:
\ba\label{b3}
{}^{-}p^{k-1}:= -\frac{\partial S_{k}}{ \partial x_{k-1}}  \q  , \q\q  {}^{+}p^{k}  := \frac{\partial S_k}{ \partial x_{k}}   \q .
\ea
(Recall that $S_k:=S(x_{k-1},x_{k})$.) 
We shall refer to the momenta ${}^{-}p$ as {\it pre--momenta} and to the momenta ${}^{+}p$ as {\it post--momenta} because a discrete formulation (\ref{b3}) allows to define two momenta at every time step:
\ba\label{b4}
{}^{-}p^{k}:= -\frac{\partial S_{k+1}}{ \partial x_{k}}  \q  , \q\q  {}^{+}p^{k}  := \frac{\partial S_{k}}{ \partial x_{k}}   \q .
\ea
Note that the requirement ${}^+p^k={}^-p^k$, which we term `momentum matching' implements the equations of motion (in the sense of (\ref{b1})) for the variables $x_k$. Or, conversely, the equations of motion implement a momentum matching of forward and backward momenta such that there are unique momenta for the variables at step $k$. Henceforth, we will often omit the superindices $+$ and $-$ at the momenta, implicitly assuming that momentum matching holds.

Accordingly, we can define two Legendre transformations from $\cq_{k-1}\times\cq_k$ to the phase spaces $\cp_{k-1}$ and $\cp_k$ \cite{marsdenwest}, where $\cq_k$ is the configuration space at step $k$,
\ba\label{b5}
&&\mathbb{F}^+S_k:  (x_{k-1},x_k) \mapsto (x_k,{}^{+}p^k)\,=\,\left(x_k,  \frac{\partial S_k}{ \partial x_{k}}\right) \nn\\
&&\mathbb{F}^-S_k:  (x_{k-1},x_k) \mapsto (x_{k-1},{}^{-}p^{k-1})\,=\,\left(x_{k-1},  -\frac{\partial S_k}{ \partial x_{k-1}}\right) \q .
\ea
These Legendre transformations are, in general, not isomorphisms because in the discrete we will allow for the general situation $\dim\cq_{k-1}\neq\dim\cq_{k}$. Constraints in phase space arise if the image of the Legendre transform is not given by the entire phase space, but by a submanifold, which is called the constraint manifold. Notice that here we have two Legendre transformations, consequently, we will encounter two different kinds of constraints, which we will call {\it pre--} and {\it post--constraints}. Pre--constraints result from $\mathbb{F}^-S_k$ and constitute the conditions that the time evolution step from $(k-1)$ to $k$ can take place. (In other words, that momentum matching can be applied at time step $(k-1)$.) Post--constraints, on the other hand, arise from $\mathbb{F}^+S_k$ and are relations on phase space that are automatically satisfied by the momenta ${}^+p^k$ (for all initial values) after having performed an evolution step from $(k-1)$ to $k$. By momentum matching these post--constraints will provide conditions for the momenta ${}^-p^k$, that is, the momenta for the next evolution step. The symplectic structure in the presence of these constraints will be discussed in detail in section \ref{app}.

The formalism outlined here can be easily applied to a discrete time evolution; at least to the case where space-time can be foliated into disjoint hypersurfaces, which, moreover, carry the same number of variables. However, for a triangulation this need not be the case. Furthermore, we are striving for a local notion of time evolution, i.e.\ one in which only a small region of a hypersurface is evolved, or pushed forward in time. Hypersurfaces will, consequently, in general overlap. The goal to devise a general discrete time evolution scheme therefore gives us the following problems to tackle:
\begin{itemize}
\item[(a)]  A time step may involve bulk variables. This will not severely complicate the situation, since the Hamilton--Jacobi formalism, as described here, automatically takes care of this.
\item[(b)] Different time steps, i.e.\ hypersurfaces defining instants of time, may involve coinciding subsets of variables. In other words, two hypersurface with different time labels, may partially overlap. This is an example of the `multi--fingered' time evolution encountered in diffeomorphism invariant theories. This issue can also be handled easily; here the additivity of the action plays an essential role.
\item[(c)]  The number of variables may actually differ from time step to time step. As a result, the phase space dimensions change in general and we have to reconsider `canonical evolution'.
\end{itemize}

Let us discuss these issues one by one.

\begin{figure}[hbt!]
\begin{center}
\psfrag{n}{ $n$}
\psfrag{n+1}{$n+1$}
\psfrag{sn0}{$\Sigma_{n-1}$}
\psfrag{sn}{ $\Sigma_n$}
\psfrag{sn1}{ $\Sigma_{n+1}$}
\psfrag{Sn}{$S_n$}
\psfrag{Sn1}{$S_{n+1}$}
\includegraphics[scale=.4]{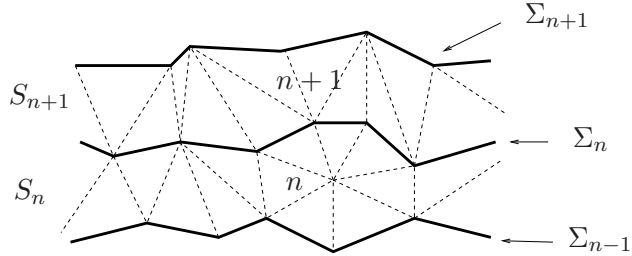}
\caption{\small Schematic illustration of a fat slicing of a triangulation. The triangulation can be built up step by step by single simplices where we count such elementary steps by $k \in \Z$. The elementary steps can be grouped up into fat slices which we now count by $n \in \Z$.}\label{fatslices}
\end{center}
\end{figure}

~\\
(a): Assume we can split the configuration variables $x_k$ at step $k$ into two sets: `true boundary variables' $x_k^t$ and `internal (bulk) variables' $x_k^i$. Such a situation can be easily constructed by starting from some elementary evolution scheme with time steps labelled by $k \in \Z$ and then defining an effective evolution scheme where we consider only every $n$--th time step (or consider time steps of different basic lengths) and henceforth label these rearranged steps by a different label $n\in \Z$. That is, for these steps we then have $x_n^t$ and $x_n^i$. For example, figure \ref{fatslices} depicts a triangulation which has been built up by simplices which can be grouped up into fat slices bounded by non-intersecting hypersurfaces $\Sigma_n$. Individual simplices are enumerated by $k\in \Z$, while fat slices are enumerated by the new label $n\in \Z$. Now assume this to be a Regge triangulation where the lengths of edges are the configuration variables and assign lengths of edges inside fat slice $n$ and of edges contained in $\Sigma_n$ to step $n$. Thus, at step $n$ we have the `bulk lengths' $l^i_n$ of edges which reside inside the fat slice $n$, as well as the `true boundary lengths' $l^e_n$ of edges $e\subset\Sigma_n$. Consequently, the action associated to fat slice $n+1$ does not depend on $l^i_n$ and we have the dependence $S_{n+1}(l^e_n,l^i_{n+1},l^{e'}_{n+1})$, etc.

\noindent

In such a situation, we can generally require that the action $S_{n}(x_{n-1},x_{n})$ does {\it not} depend on the internal variables $x_{n-1}^i$, as these variables only appear between times $(n-2)$ and $(n-1)$. The time evolution equations are then
\ba\label{b6}
{}^{-}p^{n-1}_t&:=& -\frac{\partial S_{n}}{ \partial x^t_{n-1}}  \q  , \q\q \q\q {}^{+}p^{n}_t  := \frac{\partial S_n}{ \partial x^t_{n}}   \nn\\
{}^{-}p^{n-1}_i&:= &-\frac{\partial S_{n}}{ \partial x^i_{n-1}} \,=\,0 \q  , \q\q  {}^{+}p_i^{n}  := \frac{\partial S_n}{ \partial x^i_{n}} \q .
\ea
Hence, we have ${}^-p_i^{n-1}=0$ and, by momentum matching, also  ${}^+p_i^{n-1}=0$ (which may lead to further `secondary' constraints). If the same kind of splitting into internal and boundary variables occurs for all time steps $n$, we will obtain ${}^-p_i^{n}=0$ for all $n$. The equations of motion for the internal variables are then implemented via momentum matching
\be
0={}^-p^n_i={}^+p^n_i=\frac{\partial S_n}{ \partial x^i_{n}} \q .
\ee
 This situation will always arise if some class of variables only appears in the action $S_n$ for only one label $n$.

On the other hand, here we see the first occurrence of a pre--constraint; the constraint hypersurface is defined by $p_i\equiv 0$. This is a general feature: constraints appear as equations of motion which involve only canonical data from one time step. By redefining the time label of variables (here by summarizing several time steps enumerated by $k$ into one labeled by $n$), one can transform equations of motion into constraints.

The internal variables can be integrated out and we can define an effective action $\tilde S_n$ which only depends on the `true boundary' variables $x^t_{n-1}$ and $x^t_{n}$. One can easily show that this leads to an equivalent time evolution for the remaining variables.

~\\
(b):
Assume we have the same variables $x^b$ appearing in different time steps,
 for instance $x^b_{k+1} \equiv x^b_{k}$ for some set of variables $x^b$ (we return here to a counting of time steps by $k$). This generally happens if we implement a local time evolution, i.e.\ if the hypersurface in every time step only changes in a specific region.

First let us consider the case that $x^b$ is not dynamically involved in the time evolution at all, that is, we want to implement the evolution equations
 \ba\label{c1}
 x^b_{k+1}\,=\,x^b_k \q,  \q\q p_b^{k+1}\,=\,p_b^k  \q .
 \ea
It is not possible to implement these equation by using the action as a generating function of the first kind, as the fact that the $x^b$ are not dynamically involved means that neither $x^b_k$ nor $x^b_{k+1}$ appear in the action $S_{k+1}$. However, for these variables we can use the identity transformation generated either by a generating function of the second (depending on old configuration and new momentum variables) or the third kind (depending on new configuration and old momentum variables):
 \ba\label{c2}
 &&G_2(x^k_b,p^{k+1}_b)\,=\,-x^b_k p_b^{k+1} \q,\q\q p_b^k\,=\,-\frac{\partial G_2}{\partial x_b^k}\,=\,p_b^{k+1}\q ,\q\q
 x_b^{k+1}=\,-\frac{\partial G_2}{\partial p_b^{k+1}}\,=\,x_b^k   \nn\\
 &&G_3(x^{k+1}_b,p^{k}_b)\,=\,\;\, x^b_{k+1} p_b^{k} \q,\q\q p_b^{k+1}\,=\,\;\,\frac{\partial G_3}{\partial x^b_{k+1}}\,=\,p_b^{k}\q ,\q\q
 x_b^{k}\,=\,\,\;\frac{\partial G_3}{\partial p^b_{k}}\,=\,x^b_{k+1}\, . \q\q
 \ea

Another case which will appear is that some configuration variables do not evolve $x^e_{k}=x^e_{k+1}$, while the associated momenta, however, change as either $x^e_k$ or $x^e_{k+1}$ appear in the action $S_{k+1}$ (but not both for the same index $e$). That is, in accordance with (\ref{b5}), we want to implement either
 \ba\label{c3}
  p_e^{k}\,=\,p_e^{k+1}-\frac{\partial S_{k+1}(x_k)}{\partial x_{k}^e} \q \q \text{or}\q\q
 p_e^{k+1}\,=\,p_e^k+\frac{\partial S_{k+1}(x_{k+1})}{\partial x_{k+1}^e}\q ,
 \ea
which we call `momentum updating'. This can still be implemented via a generating function of the second or the third kind by simply adding either $G_2$ or $G_3$, respectively, to the action $S_{k+1}$. As $S_{k+1}$ either only depends on the old configuration variables $x_k^e$ or only on the new configuration variables $x_{k+1}^e$, the type of the generating function, in fact, does not change. One can check that these generating functions lead to the evolution equations we wished for.

\begin{figure}[htbp!]\begin{center}             
   \psfrag{S}{ $\Sigma_k$}
   \psfrag{Sf}{$S_{k-}$}
   \psfrag{Sp}{\textcolor{white}{$S_{k+}$}}
   \includegraphics[scale=.3]{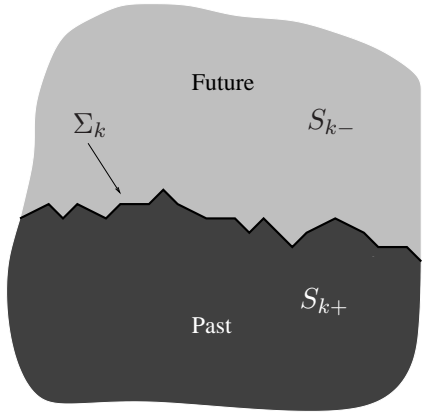} 
\caption{\small Hypersurface $\Sigma_k$ separating `past' and `future' region at step $k$.}\label{hyp}
\end{center}
\end{figure}

Note that with these definitions, all momenta at any given time step can also be defined by
\ba\label{b8}
{}^{-}p^{k}:= -\frac{\partial S_{k-}(x^-_k)}{ \partial x_{k}}  \q  , \q\q  {}^{+}p^{k}  :=\frac{\partial S_{k+}(x^+_k)}{ \partial x_{k}}   \q ,
\ea
where $S_{k-}(x^-_k)$ denotes the action associated to the region which lies to the future of the hypersurface with label $k$ and $S_{k+}(x^+_k)$ the action associated to the region in the past of this hypersurface (see figure \ref{hyp}). Here $x^-_k$ denotes all variables associated to $\Sigma_k$ and the future region, while $x^+_k$ denotes all variables associated to $\Sigma_k$ and the past of it. Moreover, $x_k$ can be {\it any} configuration variable.\footnote{In (\ref{b8}), if the same variable $x$ appears with multiple time labels $x_k=x_{k+1},\ldots$, it is understood that the action is expressed only as a function of $x=x_k$.} 

This will also include the case where, say, variable $x$ appears in the past of time $k$, but not in the future. The equation of motion is then automatically implemented by ${}^{-}p^k=0$ and momentum matching. In other words, in general we can actually add variables which are not `true dynamical variables', i.e.\ which are not associated to the hypersurface $\Sigma_k$ but rather to the future or past of it. Then one of the associated momenta ${}^{+}p^k$ or ${}^-p^k$ will be zero, thus enforcing the equations of motion by momentum matching ${}^+p^k={}^-p^k$.

This  allows us to deal with problem 

~\\
(c), namely the case where the number of configuration variables and, therefore, the dimensions of the phase spaces associated to different hypersurfaces $\Sigma_{k-1}$ and $\Sigma_{k}$ differ. We can formally extend the phase spaces by including any variables into the phase space at time $(k-1)$ which only appear at time $k$ but do not have a natural associated variable at time $(k-1)$, and vice versa.  Then we can have phase spaces of equal dimensions at the two time steps  and we can implement time evolution by a canonical transformation.

Say, we have a new variable $x^{n}_{k+1}$ which appears at time step $(k+1)$ but not at time step $k$ (for a similar situation in 3D Regge Calculus see figure \ref{3dpach}) and the action $S_{k+1}$ only depends on $x^{n}_{k+1}$ but not on $x^{n}_k$. Accordingly, we extend the phase space at time $k$ by the pair $(x_k^n,p^k_n)$ and may use the action as a generating function of the first kind, resulting in the evolution equations
\ba\label{c4}
{}^-p^{k}_{n}\,=\, -\frac{\partial S_{k+1}}{\partial x^{n}_{k}}\,=\,0 \q ,\q\q  {}^+p^{k+1}_{n}\,=\, \frac{\partial S_{k+1}}{\partial x^{n}_{k+1}} \q .
\ea

The variable $x^n_k$ remains undetermined, for it does not appear at all in the action $S_{k+1}$. We may fix its value to $x^n_k=x^n_{k+1}$, such that it appears as an initial datum for time step $k$, determining the data at time step $k+1$. Furthermore, in (\ref{c4}), the pre--constraint ${}^-p^k_n=0$ shows up as a constraint on the variable $p^k_n$, which appears at time $k$ for the `first time'. Hence, by momentum matching it does not place any restrictions on the dynamical variables at step $k$. (If one also extended the phase spaces at previous time steps by this variable, the momentum $p_n$ would also be vanishing.) Additionally, we will also encounter a post--constraint: Assuming that $N$ further variable pairs $(x_k^i,p^k_i)$ are involved, the time evolution will map from a $(2N+2)$--dimensional phase space to another $(2N+2)$--dimensional phase space (using that $x^n_k=x^n_{k+1}$). However, this map is not defined on the full phase space, because the pre--constraint $p^k_n=0$ has to hold. As a consequence, the image of the map can be maximally $(2N+1)$--dimensional, implying the occurrence of also a post--constraint. Indeed, in the examples in the sequel it will turn out that the second equation in (\ref{c4}) only involves variables from time step $(k+1)$ and, hence, constitutes a constraint.

Conversely, the case of an old variable $x^o_k$ that does not have an equivalent at time $(k+1)$ can be treated analogously, that is, we extend the phase space at time $(k+1)$ by the pair $(x_{k+1}^o,p^{k+1}_o)$. This time we obtain a post--constraint ${}^+p^{k+1}_n=0$.
 Likewise, we will also find a pre--constraint, which in the case of the Pachner moves will always be given by
\ba\label{c4a}
{}^-p_o^{k}&=&-\frac{\partial S_{k+1}}{\partial x^o_k}  \q ,
\ea
i.e.\ this equation will only involve variables from time step $k$.
Via momentum matching, ${}^-p_o^{k}={}^+p_o^{k}$, this constraint effectuates the equation of motion $\partial S/\partial x^o=0$, as $x^o$ appears in $S_{k+1}$ for the `last time'.

In the following section, we will implement Pachner moves as time evolution maps in simplicial gravity and, with the help of these examples, explain the concepts encountered here in greater depth.

\section{Pachner moves as time evolution maps}\label{Pachner}

In the previous section, we presented a general method to derive a canonical time evolution scheme from a given discrete action. The advantage of this method is that the canonical evolution equations will exactly reproduce the equations of motion of the (discrete) covariant formalism. This is, in general, different from the procedure of first deriving a continuum canonical formalism from the continuum action and subsequently discretizing this continuum canonical formalism \cite{friedmann}. In the case of general relativity, where diffeomorphism symmetry plays a fundamental role, the latter method usually leads to (classically) anomalous constraint algebras \cite{lollreview, loll,friedmann}.

A canonical evolution scheme for Regge calculus which reproduces the covariant equations of motion has been presented in \cite{bd1,dithoe1,bianca}. However, this scheme was only applicable to a special class of triangulations, namely those which can be evolved from a hypersurface by so--called tent moves. These tent moves \cite{commi} are special local evolution moves that do not change the topology and the triangulation (that is the connectivity of the triangulation) of the `equal time' hypersurfaces. Hence, in this case one does not encounter phase spaces of different dimensions.

In the present work, we will advance and devise a general canonical framework applicable to arbitrary triangulations (of fixed `spatial' topology) by implementing an elementary and ergodic class of local evolution moves, the so--called Pachner moves \cite{pachner}. Pachner moves are elementary in that they involve only a fixed number of simplices during each move and are, furthermore, topology-preserving and ergodic, i.e.\ can map between any triangulations of the same topology by finite sequences. In particular, the tent moves (see sections \ref{sectent1} and \ref{tent2}), which can involve an arbitrary number of simplices, can, in fact, be decomposed into Pachner moves.

\begin{center}\begin{figure}[htbp!]
\psfrag{sk}{$\Sigma_k$}
\psfrag{sk1}{$\Sigma_{k+1}$}
\psfrag{t}{$t$}
$\begin{array}{cccc}
\text{3D perspective:}&\hspace{.5cm}\includegraphics[scale=.2]{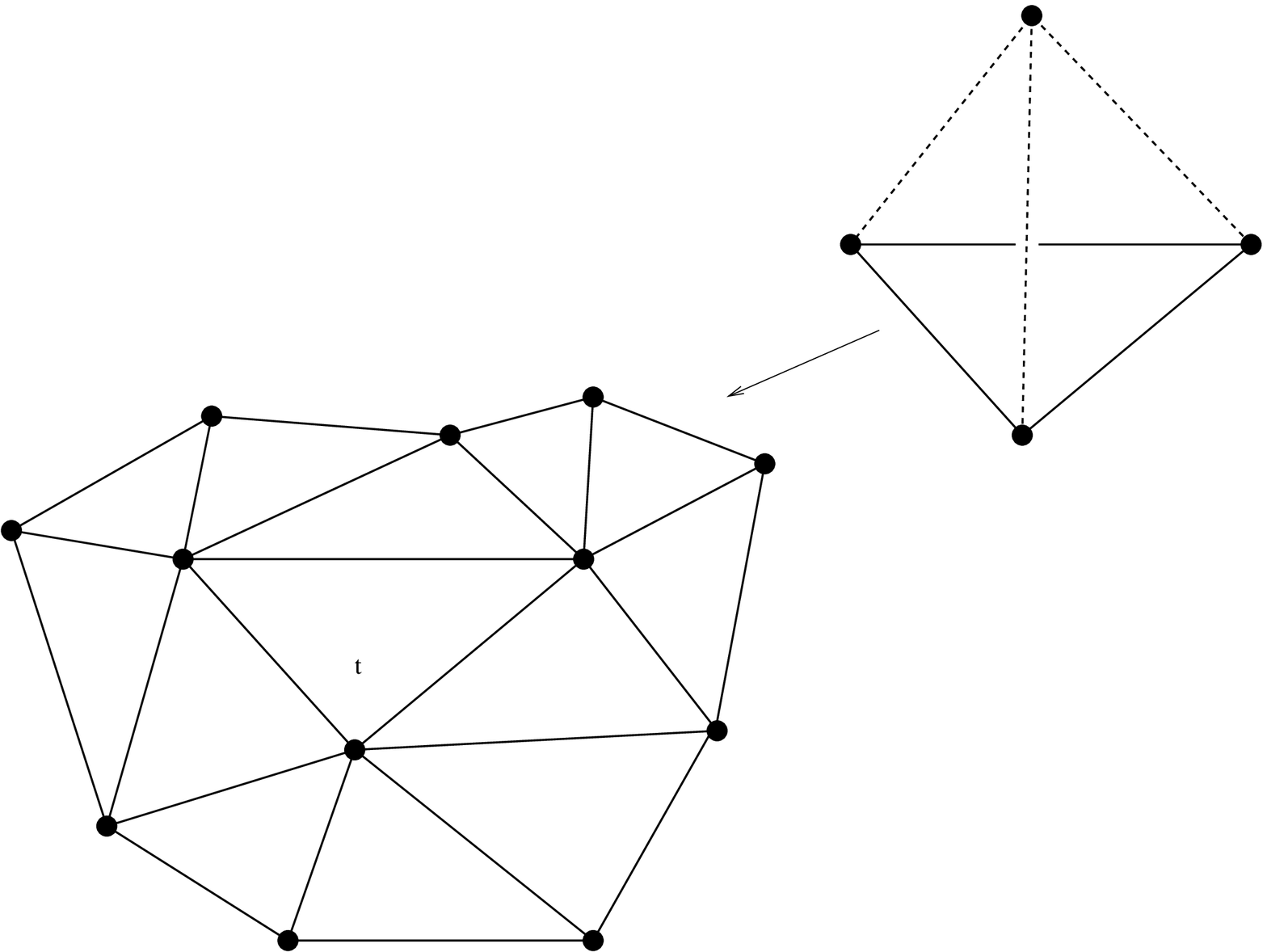} &\longrightarrow& \hspace{-.7cm}\includegraphics[scale=.2]{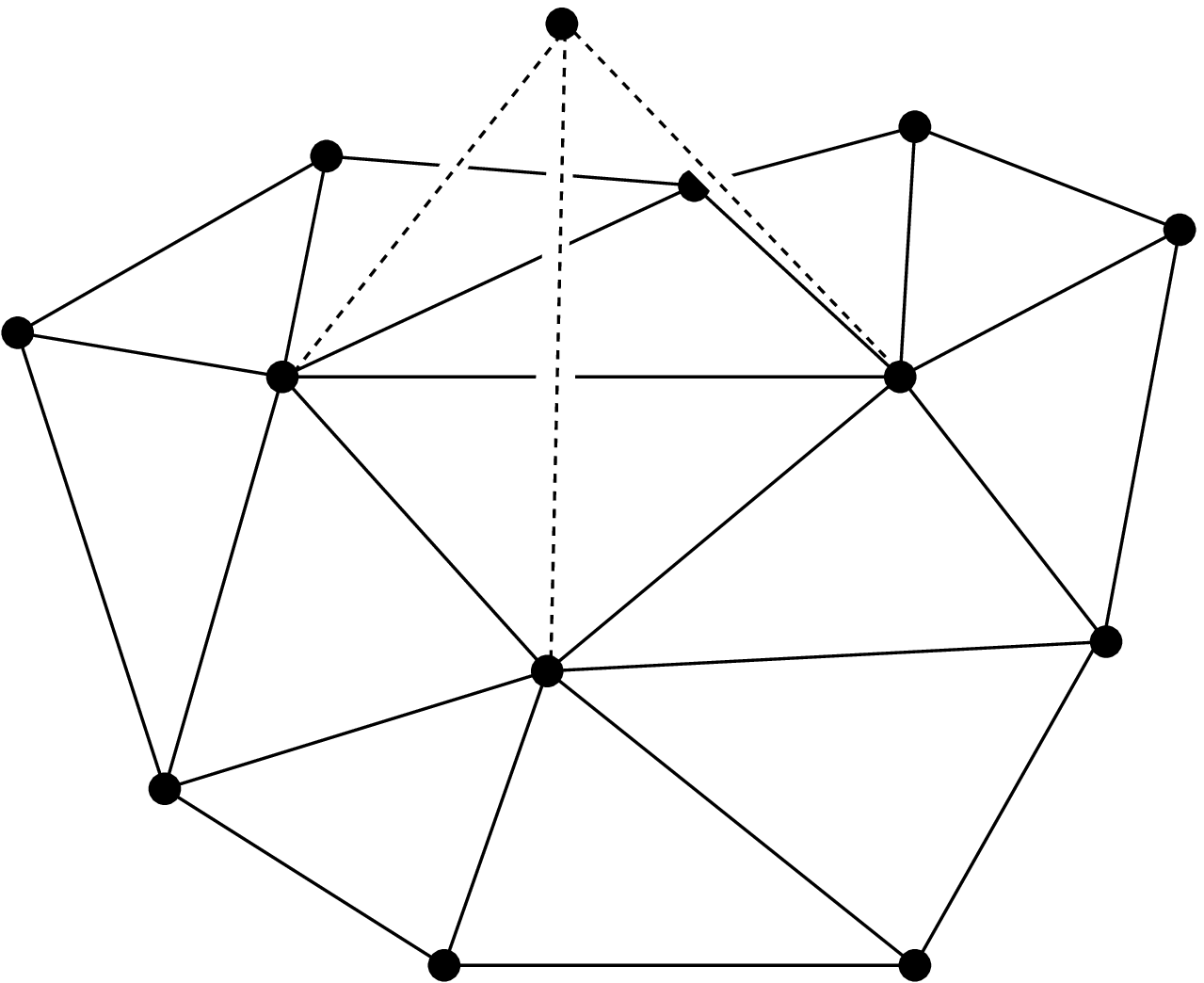} 
\vspace{1cm}\\ 
\text{2D perspective:}&\hspace{-.5cm}\includegraphics[scale=.2]{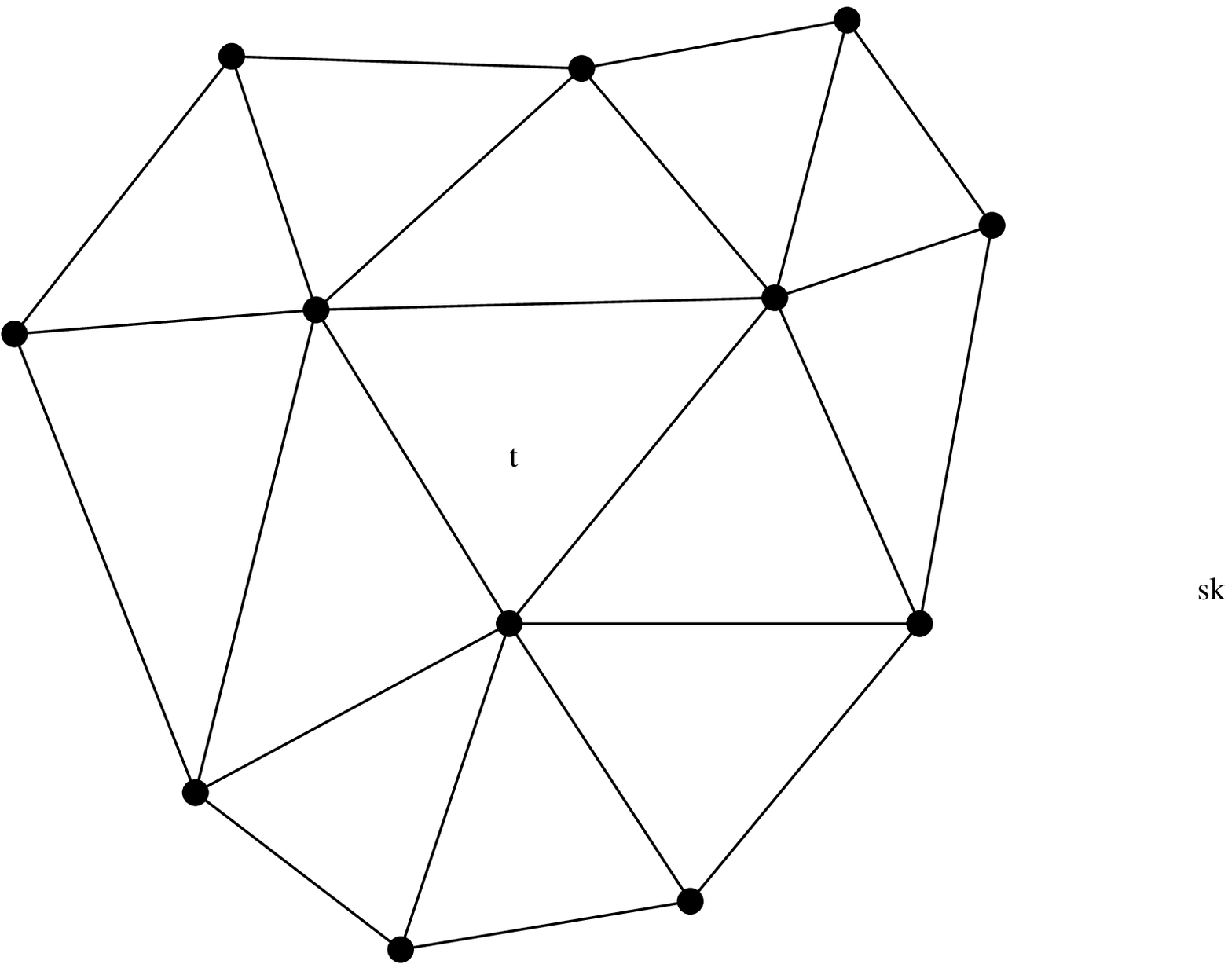} 
&\longrightarrow&
\includegraphics[scale=.2]{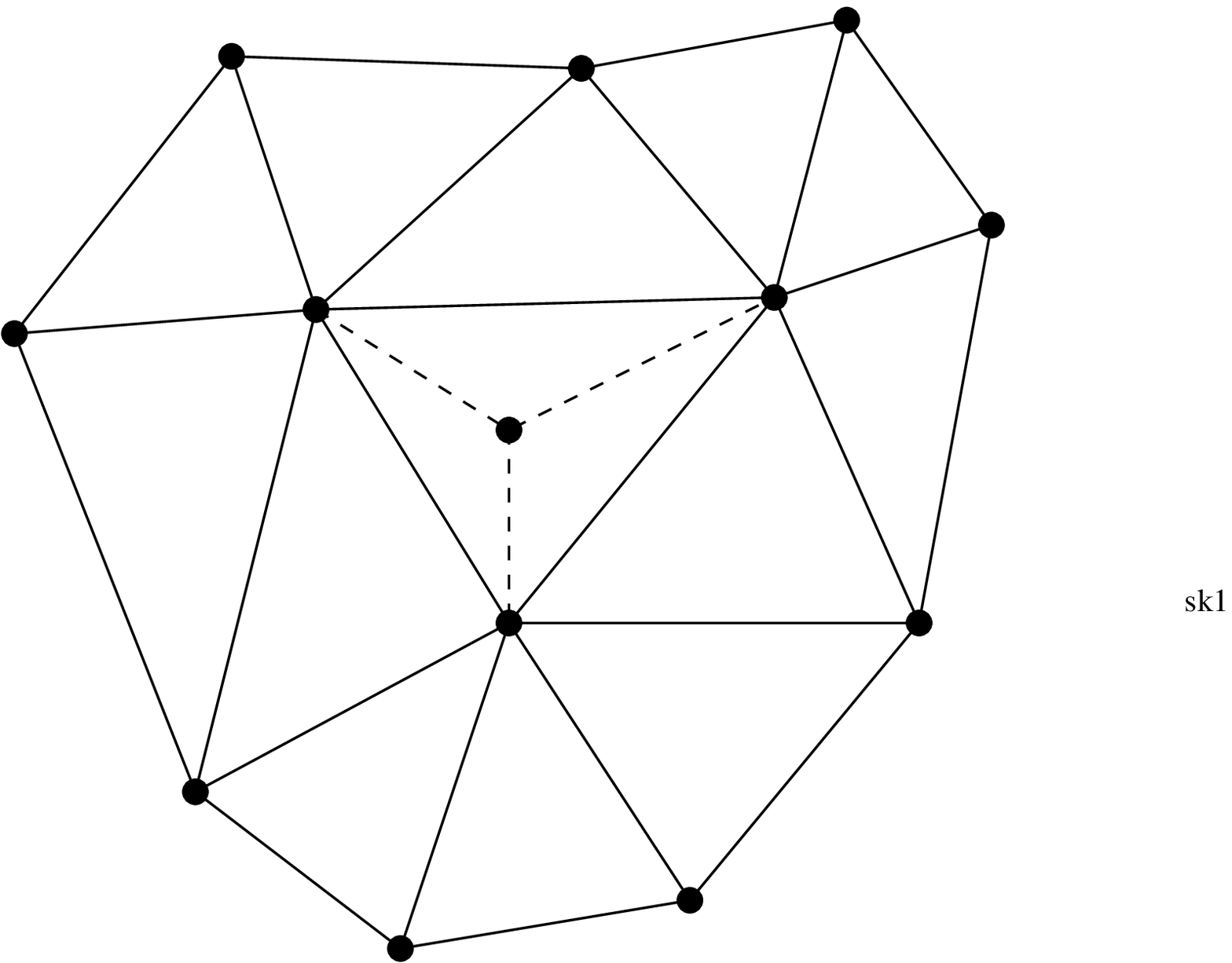} \\
&& \text{1--3 Pachner move}&
\end{array}$
\caption{\small 3D example: gluing a single tetrahedron onto a single triangle in the 2D boundary hypersurface of a 3D bulk triangulation. From the perspective of the 2D hypersurface this gluing move appears as a subdivision of the triangle $t$. That is, the move appears as a 1--3 Pachner move in the hypersurface.
\newline This is a specific example of the situation described earlier where the numbers of variables associated to the two hypersurfaces before and after the new move differ due to the three new edges.}\label{3dpach}
\end{figure}
\end{center}

\begin{figure}[htbp!]
\begin{center}
 \includegraphics[scale=.35]{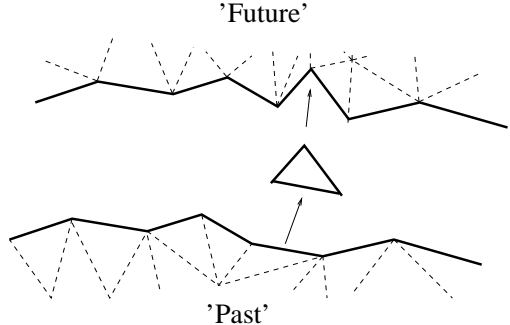}
 \end{center}
 \caption{\small Removal moves for the `past triangulation' are equivalent to gluing moves for the `future triangulation' and vice versa.}\label{glueqrem}
 \end{figure}

More precisely, the evolution moves to be implemented below can be interpreted entirely within the `spatial' hypersurfaces (as local changes) and appear as Pachner moves therein: If we take the $(D-1)$--dimensional hypersurface $\Sigma$ as the boundary of a $D$--dimensional bulk triangulation, the Pachner moves in $\Sigma$ arise by gluing a single $D$--simplex to, or removing one $D$--simplex from the bulk triangulation \cite{pachner}. Notice that here we only allow for gluing processes of top-dimensional simplices which identify faces of one dimension less of a given simplex with faces of equal dimension in the hypersurface (and analogously for removal procedures). 
For instance, figure \ref{3dpach} depicts the 3D example of the 1--3 Pachner move in the 2D hypersurface. All other Pachner moves in $(D-1)$ dimensions can be similarly produced by gluings or removals of single $D$-simplices in the $D$--dimensional bulk triangulation.
This elementary procedure provides a more compelling connection between the covariant and canonical picture: the discrete evolution of the hypersurface can be reinterpreted as building up the bulk triangulation, and, hence, the (discrete) space-time, step by step by simplices. In analogy to the situation in canonical general relativity, the triangulated hypersurface evolves in a discrete `multi-fingered' (or `bubble') time through the full (discrete) space-time solution. In this light, every gluing move at a given evolution step $k$ which adds a simplex to the `past triangulation' can also be viewed as a removal move which subtracts a simplex from the `future triangulation', and vice versa (see figure \ref{glueqrem} for a schematic illustration). Formally, this is guaranteed by momentum matching.
 
Since Pachner moves have inverses and, thus, in order to establish a genuine implementation of Pachner moves into the mathematical formalism, we have to ensure that every phase space map ${}^{\text{Pachner}}f$ corresponding to the action of a Pachner move on a hypersurface has an inverse which corresponds to the action of its inverse move (of course, via the gluing/removal correspondence applied to the same simplex), i.e.\ ${}^{\text{Pachner}}f\circ{}^{\text{Pachner}^{-1}}f=\text{id}$. This, indeed, will be the case.

As Pachner moves do not change the topology of a triangulation, a $(D-1)$--dimensional hypersurface with topology $\mathfrak T$  evolved by Pachner moves will lead to a $D$--dimensional triangulation of topology $[0,1]\times \mathfrak T$. Hence, just as usual canonical evolution schemes, also the present scheme assumes or implements a non--changing topology of the `equal time' triangulations. This might be advantageous, for instance, for quantization, if one wants to suppress topology changes of spatial hypersurfaces (interpreted as the production of baby universes). Specifically, the approach of Causal Dynamical Triangulations (CDT) \cite{cdt} shows that this can lead to a much more regular large scale limit than in the case of (Euclidean) Dynamical Triangulations \cite{edt,dt}, where arbitrary topology changes are allowed for.\footnote{The evolution moves implemented here yield triangulations akin to the triangulations employed in CDT. The splitting off of baby universes, aka spatial topology changes, on the other hand, would require the implementation of moves distinct from the Pachner moves. For instance, in D dimensions one could allow for a move which identifies all $D+1$ $(D-1)$--faces of a $D$--simplex with $D+1$ $(D-1)$--simplices in a $(D-1)$--dimensional hypersurface. Furthermore, by only allowing top-dimensional simplices to be glued onto faces of one dimension less, we are disallowing singular moves which, e.g., only identify a single vertex of the new simplex with a single vertex of the hypersurface.}

On the other hand, a Pachner move evolution scheme allows for a change of the number of variables `in time'. There are numerous physical situations, where it might be interesting to adapt the discretization (density) in time, for instance, for an expanding or contracting universe. In fact, we will show that, akin to the `no--boundary' proposal \cite{hawking}, we can start with an `empty triangulation', obtain the boundary of a $D$--simplex in the next time step and evolve this to a bigger and bigger triangulation. 

Furthermore, Pachner moves implement the idea of a `fluctuating lattice' in a controlled manner, that is, the idea that the lattice, or discretization, is not fixed in time (or space--time) but is either summed over or is determined by dynamical considerations. Indeed, we will see that even the classical dynamics may prefer or suppress certain Pachner moves and, hence, determine the evolution of the connectivity of the triangulation itself.

\subsection{Regge calculus}\label{secregcal}

In this article, we will develop this evolution scheme for a particular discretization of general relativity, namely Regge calculus \cite{regge,reggewilliams}. However, the basic ideas are general and straightforwardly adaptable to other discretization schemes fulfilling the basic prerequisite of additivity of the action; whenever we glue or remove an elementary building block, i.e.\ a simplex, we have to add or subtract the corresponding piece of action. 

To begin with, we will review some basic notions of Regge calculus based on length variables, before studying the details of the individual Pachner moves in sections \ref{3D} and \ref{4D} below. For formulations using other geometric variables see, for instance, \cite{dittrichspeziale, Bahr:2009qd}. Standard Regge calculus is based on a fixed triangulation where the basic variables are simply the length variables associated to the edges of this triangulation. These length variables (assuming generalized triangle inequalities are satisfied, which we will always do) specify a piecewise linear geometry for the triangulation in question. The equations of motion are determined by a variational principle from an action, the Regge action, which can be viewed as a discretization of the Einstein--Hilbert action.

The (Euclidean) Regge action (without a cosmological constant term) for
a $D$--dimensional triangulation $T$ with boundary $\partial T$ and interior
$T^\circ:=T\backslash \partial T$ is given \cite{regge,hartle} by
\begin{eqnarray}\label{regge1}
S\;=\;\sum_{h\subset
T^\circ}V_h\,\epsilon_h\;+\;\sum_{h\subset\partial T}V_h\,\psi_h\q,
\end{eqnarray}
 where $h$ stand for the  $(D-2)$--dimensional subsimplices
(sometimes called ``hinges'') of $T$, that is edges $e$ in $3$D and triangles $t$ in 4D.  
$V_h$ 
denotes the volume of the hinge $h$. The deficit angles
$\epsilon_h$ and exterior angles $\psi_h$ are given by
\begin{eqnarray}\label{regge2}
\epsilon_h\;&=&\;2\pi\;-\;\sum_{\sigma\subset
h}\theta_h^\sigma\qquad\text{for }h\subset T^\circ\q,\\[5pt]\nonumber
\psi_h\;&=&\; k_h \,\pi\;-\;\sum_{\sigma\subset
h}\theta_h^\sigma\qquad\text{for }h\subset\partial T,
\end{eqnarray}

\noindent and in both cases the sum ranges over all $D$--dimensional simplices $\sigma$ which contain $h$,
 and $\theta_h^\sigma$ is the
interior dihedral angle in the simplex $\sigma$ between the two
$(D-1)$--dimensional subsimplices that meet at $h\subset\sigma$. The number $k_h$ is fixed according to how many pieces $l$ of triangulations are glued along the hinge in question:
Because of the boundary term in (\ref{regge1}), the action is {\it additive} if we glue two pieces of triangulations together and choose $k_h=1$. In the canonical evolution scheme we will glue top--dimensional simplices to bulk triangulations. That is, the smallest triangulations we consider is just one tetrahedron $\tau$ in $3D$ or one 4--simplex $\sigma$ in $4D$, of which the actions are just boundary terms,
\ba\label{regge3}
S_\tau&=&   \sum_{e \subset \tau} l^e \left(k_e \pi - \theta^\tau_e \right) \nn\\
S_\sigma&=&   \sum_{t \subset \sigma} A_t \left(k_t \pi - \theta^\sigma_t \right)\q,
\ea
where $l^e$ denotes the length of the edge $e$ and $A_t$ the area of the triangle $t$. Again, $k_e$ and $k_t$ are to be determined by the gluing process. In the canonical evolution scheme presented below, we will either have $k_e,k_t=1$ or $k_e,k_t=0$.

The equations of motion are obtained by varying the action
(\ref{regge1}) with respect to the
lengths $l^e$ of the edges in $T^\circ$ (while keeping the edge lengths in the boundary $\partial T$ fixed). The boundary term in
(\ref{regge1}) ensures that,
for all edges $e\in T^\circ$, the Regge equations read
\begin{eqnarray}\label{regge4}
\sum_{h\supset e}\frac{\partial V_h}{\partial l^e}\,\epsilon_h=\;0  \q .
\end{eqnarray}
Note that in 3D the equations of motions enforce $\epsilon_e=0$, that is, vanishing deficit angles and therefore vanishing curvature.

Finally, the variation of the deficit angles appearing in the action vanishes because of the Schl\"afli identities
\ba\label{schlaefli}
\sum_{e \subset \tau}  l^e \frac{\partial \theta_e^\tau}{\partial l^{e'}} &=& 0 \q ,\nn\\
\sum_{t \subset \sigma}  A_t \frac{\partial \theta_t^\sigma}{\partial l^{e'}} &=& 0   \q .
\ea

\section{Pachner moves for 3D Regge calculus}\label{3D}

In the next subsections we will spell out the details of the dynamics of Pachner moves for 3D Regge gravity (without a cosmological constant). Before delving into the details, we will make a few remarks on the dynamics of 3D gravity. As mentioned in the previous section, the equations of motion for gravity require that all deficit angles vanish and thereby that the piecewise flat geometry is flat. In the canonical description we describe the dynamics of the 2D hypersurface, that is, the changes in the intrinsic and extrinsic geometry of the hypersurface evolving through the 3D space-time. As the latter is flat, we will essentially describe different embeddings of a 2D hypersurface into flat 3D space.

This allows us to determine Hamilton's principal function: since the deficit angles vanish for solutions, only the boundary term remains
\ba
\tilde S(l^e) = \sum_{e \subset \partial T} l^e \psi_e(l^e)   \q .
\ea
Here the $l^e$ are the lengths of the edges in the boundary of the triangulation. The momenta are given by the derivatives of Hamilton's principal function with respect to the edge lengths. Because of the Schl\"afli identity, the terms with derivatives of the exterior angles vanish and we obtain (the different signs in (\ref{b3}) can be taken into account by taking the appropriate orientation of the exterior angles)
\ba
p_e=\psi_e \q .
\ea
Thus, the momenta conjugate to the length variables are given by the exterior angles. The same result was obtained in \cite{waelbroeckzapata} starting from a first order (connection) formulation of discrete gravity.

As a consequence of the fact that the Pachner move dynamics allows to add degrees of freedom during the evolution, we can start from a small triangulated hypersurface and evolve to a much bigger one. In particular, we can start from an empty triangulation at time $k=0$. This can be seen as just a technical method, however, might also be useful (in 4D) to explore the `no--boundary proposal' \cite{hawking}.

In order to produce a triangulation with a hypersurface of, e.g., spherical topology, we can start with an evolution move from the empty triangulation to the boundary of a tetrahedron.\footnote{We could equally well begin by producing initial hypersurfaces of more complicated topology.} A tetrahedron has six boundary edges $l^n_1,\,n=1,\ldots,6$, hence, the phase space at time $k=1$ is 12--dimensional. Accordingly, we can extend the phase space at time $k=0$ by the six pairs $(l^n_0,p_n^0)$.

Using the one--tetrahedron action as generating function of the first kind
\ba\label{thu1}
 G_{0-1}(l^n_0,l^n_1)&=&S_\tau(l^n_1) \;=\;  \sum_{n \subset \tau} l^n_{1} \left( \pi - \theta^\tau_n (l^n_{1}) \right) \q ,
\ea
the equations of motion amount to
\ba\label{thu2}
&&p^{0}_n\,=\,0 \q,\q\q\q p^{1}_n\,=\, \frac{\partial S_\tau}{\partial l^n_{1}}\,=\, \pi-\theta_n(l^n_{1})  \q .
\ea
That is, the $l^n_1$ remain undetermined. However, we can set $l^n_0=l^n_1$ with the  understanding that the $l^n_1$ are initial data which appear only at time step $k=1$. The momenta at $k=0$ are constrained to vanish. But also at time $k=1$ we obtain a post--constraint because all the momenta are given as functions of the length variables at time $k=1$. Indeed, this equation just expresses the fact that the momenta are exterior angles, which for a flat tetrahedron are determined by the intrinsic geometry of the boundary surface, i.e.\ the length variables. Hence, we have a totally constrained system, as all the momenta are determined as functions of the length variables.

This hypersurface can be evolved into a more complicated hypersurface (with spherical topology) by means of the Pachner moves described below. Momentum updating will then take care of the geometrical updating of the exterior angles, i.e.\ the exterior angles change according to the dihedral angles that are added/subtracted by gluing/removing tetrahedra onto/from the hypersurface.

The dynamics of 3D gravity is special because the fact that the system is totally constrained does not even change for more complicated triangulations (of spherical topology); through constraints all momenta at time $k$ will be determined by the length variables of step $k$. These constraints express the fact that we are always dealing with a 2D hypersurface embedded into 3D flat space and that, furthermore, the momenta are the exterior angles. Accordingly, if we considered a parallel transport of a 3D vector along a small loop around a vertex $v$ of the hypersurface, we should obtain an identity transformation. This parallel transport can be expressed \cite{tHooft, kadarloll} as a sequence of rotations
\ba\label{thu3}
P_v= R(\alpha_{e_1e_2}) R(\psi_{e_2}) R(\alpha_{e_2 e_3}) R(\psi_{e_3}) \cdots R(\psi_{e_1}) \; \stackrel{!}{=} \text{Id}\q,
\ea
where $e_1,e_2,\ldots$ denotes some cyclic ordering of the edges around the vertex $v$, $R(\alpha_{e_i e_{i+1}})$ denotes the rotation in the plane spanned by the two edges $e_i, e_{i+1}$ with the angle $\alpha_{e_i e_{i+1}}$ and $R(\psi_{e_i})$ denotes the rotation around the edge $e_i$ by an angle $\psi_{e_i}$. Note that the exterior angles equal the momenta and that the interior angles $\alpha_{ee'}$ can be expressed as functions of the length variables (in the 2D star of the vertex).  Hence, the condition that the $3D$ parallel transport matrix should be the identity gives us three (as the matrix is in $SO(3)$) constraints on the phase space data for every vertex in the hypersurface. For a triangulation of spherical topology we have $3\sharp v=\sharp e+6$ for the number of vertices $\sharp v$ and the number of edges $\sharp e$. Therefore, we have at least as many constraints as configuration (or momentum variables). In fact, there are six more constraints than edges, because there exist six relations between the constraints (as can be checked explicitly for the example of the tetrahedron). These six relations correpsond to the three global rotations and three global translations which change the embedding of the 2D triangulation in 3D flat space, but do not change any of the geometrical data, i.e.\ neither lengths or exterior angles.

The constraints (\ref{thu3}) will be preserved by the Pachner moves, as these Pachner moves will implement the equations of motions, i.e.\ flatness of the 3D triangulation. Furthermore, momentum updating will ensure that the momenta are always given by the exterior angles of the 2D hypersurface. Consequently, the canonical data at every time step will describe a 2D triangulation embedded into flat 3D space for which the relations (\ref{thu3}) hold. Moreover, as we shall see shortly, the 1--3 Pachner move generates one vertex, and in conjunction with this vertex also three (post--) constraints of the form (\ref{thu2}), which are just a rewriting of the form (\ref{thu3}) for three--valent vertices.\footnote{The difference between the two forms of constraints is that (\ref{thu2}) is linear in the momenta, whereas (\ref{thu3}) involves $\cos p_e$. Indeed, the constraints should rather involve the square (or the $\cos$) of the momenta as in the continuum. This, however, can also be taken into account in (\ref{thu2}): Constraints quadratic in the momenta indicate that the constraint hypersurface has two pieces corresponding to the two roots of the quadratic equations. The two pieces correspond to the possibility of allowing for both orientations of the tetrahedron (i.e.\ allowing both signs for the exterior angles). We can replace (\ref{thu2}) by $\cos p_n=\cos \psi_n$ to take care of this fact.} The 2--2 Pachner move and the 3--1 Pachner move will not generate vertices and therefore also no additional constraints. The dynamics prescribed by these moves will, however, preserve the constraints of the form (\ref{thu3}) for the reasons just given.
~\\

As a mnemonic, for the description of the Pachner move dynamics in both 3D and 4D we will use the following edge indices in order to label and appropriately distinguish the various length and momentum variables:
\begin{itemize}
\item $e$ labels {\it edges} contained in the $D$--simplex of the Pachner move which occur in both $\Sigma_k$ and $\Sigma_{k+1}$,
\item $n$ labels {\it new} edges introduced by a Pachner move which occur in $\Sigma_{k+1}$ but not $\Sigma_k$,
\item $o$ labels {\it old} edges removed by a Pachner move which occur in $\Sigma_k$ but not $\Sigma_{k+1}$,
\item $b$ labels edges contained in both $\Sigma_k$ and $\Sigma_{k+1}$ which are not involved in the Pachner move and from that perspective may be considered as {\it boundary} edges.
\end{itemize}

\subsection{The 1--3 Pachner move}\label{13sec}

Consider a 3D triangulation with a boundary $\Sigma_k$ that we will consider as a $2D$ hypersurface at time $k$. Glue to this boundary a tetrahedron $\tau$ such that one of the triangles is identified with a triangle $t$ in the 2D hypersurface $\Sigma_k$. We obtain a new boundary $\Sigma_{k+1}$. From the perspective of the hypersurface this gluing can be interpreted as a 1--3 Pachner move, i.e.\ the triangle $t$ is replaced by three triangles, that share one vertex $v$ in the middle and have the same 1D boundary, consisting of three edges as the original triangle $t$ (see figure \ref{3dpach}).

\begin{figure}[htbp!]
\begin{center}
\psfrag{sk}{$\Sigma_k$}
$\begin{array}{cc}
\includegraphics[scale=.3]{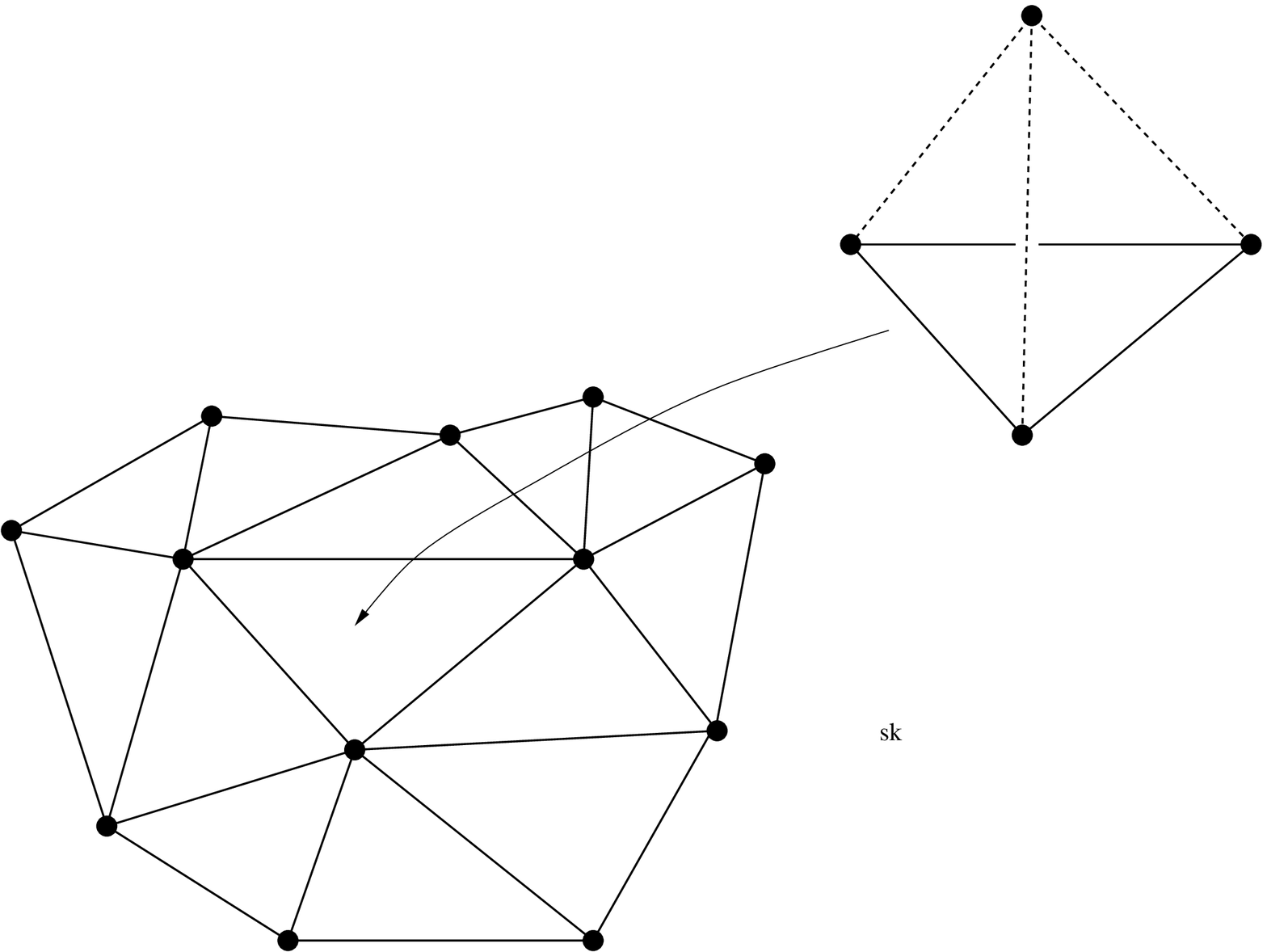}& \includegraphics[scale=.3]{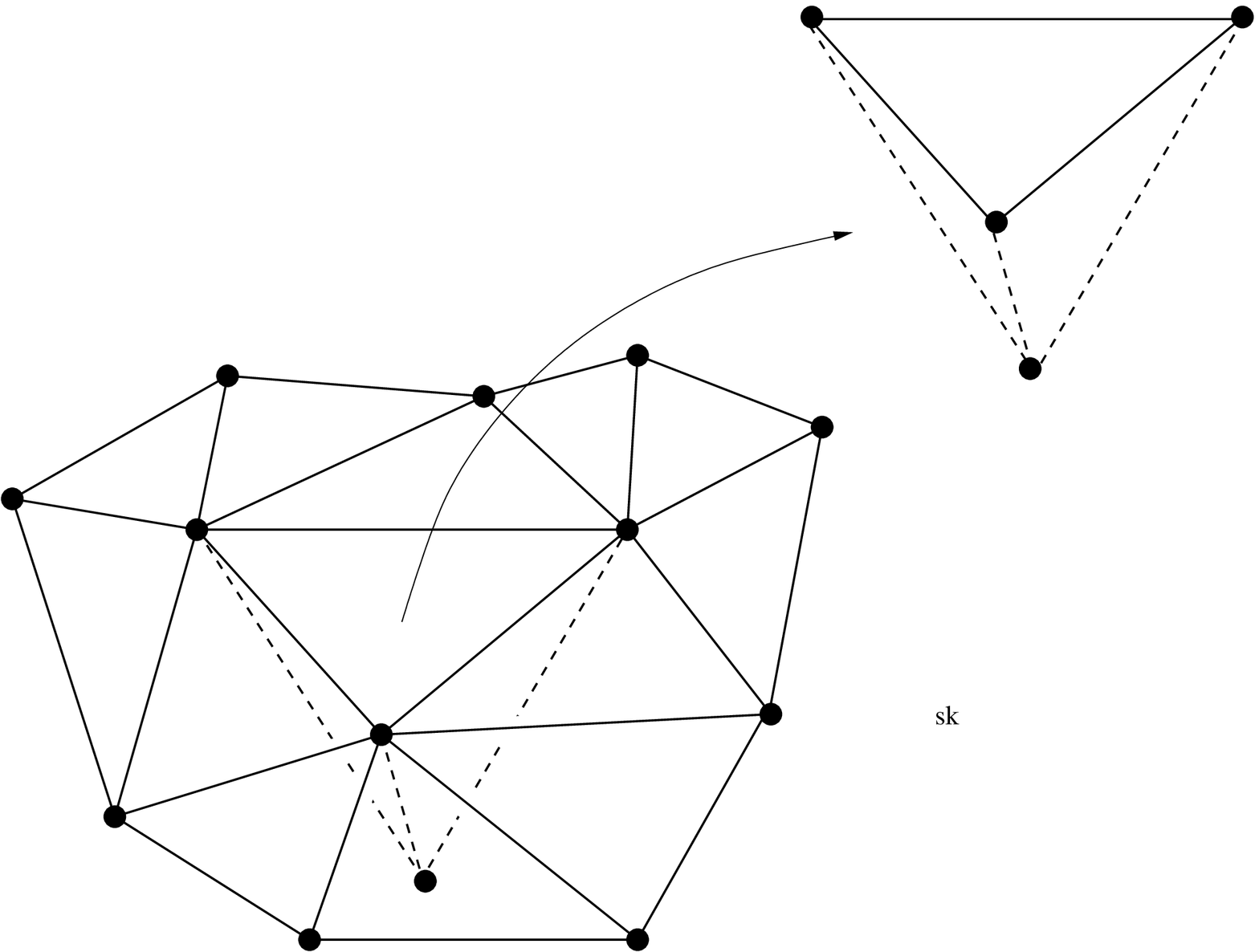} \\
a)&b)
      \end{array}$
      \end{center}\caption{\small a) The 1--3 gluing Pachner move, b) The 1--3 removal Pachner move.\newline The dashed edges are the three new edges.}\label{13glurem}
\end{figure}

Note that the tetrahedron can be glued with two different orientations to the hypersurface. This can be interpreted as gluing the tetrahedron either on the upper side (with a future pointing tip) or on the bottom side (with a past pointing tip) to the hypersurface, or, alternatively, as gluing or removing a tetrahedron to or from the bulk triangulation, respectively (see figure \ref{13glurem}).
The Regge actions for the different orientations of the tetrahedron just differ by a global sign (which can be understood to arise from the oriented exterior angles). This agrees with the interpretation of gluing or removing a tetrahedron with, say, positive orientation: if we remove a tetrahedron from the bulk triangulation, we would have to subtract the action for the positively oriented tetrahedron, the alternative is to add the action (glue the tetrahedron) with the negative orientation.

Every Pachner move in 2D and 3D  can be interpreted as gluing a simplex in one of the two different orientations to the 3D and 4D bulk, respectively. These two possibilities can, in general, also be seen as gluing and removing a simplex with, say, positive orientation, respectively. We will henceforth assume that the two possibilities are encoded in the orientations and therefore in the signs of the exterior angles in the action (\ref{regge3}). Thereby we can summarize these two cases into just one, which we will mostly refer to `as gluing a simplex' to the bulk triangulation or hypersurface.

\begin{figure}[htbp!]
\psfrag{1}{ 1--3}
\psfrag{2}{ 3--1}
\psfrag{e}{\tiny $e_n$}
\psfrag{v}{$v$}
\begin{center}\includegraphics[scale=.5]{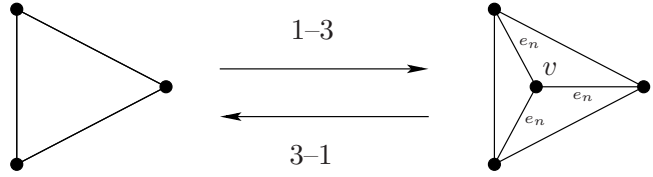} 
\end{center}
\caption{\small The 1--3 Pachner move and its inverse, the 3--1 Pachner move.}\label{13m} \end{figure}

In the 1--3 Pachner move one triangle $t$ is replaced by the same triangle subdivided into three new triangles (see figure \ref{13m}). 
Therefore, for this Pachner move we will have edges of three different types: edges $e_b$ not participating in the dynamics, three edges $e$ in the boundary of the triangle $t$, for which $l^e_k=l^e_{k+1}$ and the three new edges $e_n$ with lengths $l^n_{k+1}$ which only appear at time step $(k+1)$, but not at $k$. We will regard the action
\ba
S_\tau &=&   \sum_{e \subset \tau} l^e_{k+1} \left( - \theta^\tau_e(l^e_{k+1},l^n_{k+1})  \right) +    \sum_{n \subset \tau} l^n_{k+1} \left( \pi - \theta^\tau_n (l^e_{k+1},l^n_{k+1}) \right)
\ea
associated to the tetrahedron glued to the hypersurface as a function of the three new edge lengths $l^n_{k+1}$ and the three edge lengths $l^e_{k+1}$. Note that we fixed the factors $k_e$ appearing in (\ref{regge3}) to $k_n=1$ for the new edges $l^n$ and to $k_e=0$ for the `boundary edges' $l^e$. This will also work for the other moves, we will choose $k_e=0$ for all boundary edges and $k_n=1$ or $k_o=1$
for edges which either appear or `disappear' during the evolution move. In this way the $\pi$ factors add up correctly to $2\pi$ after a sufficient number of evolution moves.

According to the discussion below (\ref{c3}), we add the generating function $G_3(l^e_{k+1},p^k_e)$ for the edges of type $e$ and $b$ to the action. Hence, the generating function is
\ba\label{c5}
G_{1-3}(l^b_{k+1},p_b^k;\,l^e_{k+1},p_e^{k};\,l^n_k,l^n_{k+1}) &=& \sum_b l^b_{k+1}p_b^k + \sum_e l^e_{k+1}p_e^k  \, + \, S_\tau(l^e_{k+1},l^n_{k+1})  \q .
\ea
The evolution equations are then
\ba\label{c6}
&&l^b_k\,=\,l^b_{k+1}  \q ,\q\q p^{k+1}_b\,=\,p^k_b \q , \\
&&l^e_k\,=\,l^e_{k+1}\q,\q\q p^{k+1}_e\,=\,p^k_e+\frac{\partial S_\tau}{\partial l^e_{k+1}}\,=\, p^k_e-\theta_e(l^e_{k+1},l^n_{k+1}) \q ,\\
&&p^{k}_n\,=\,0 \q,\q\q\q p^{k+1}_n\,=\, \frac{\partial S_\tau}{\partial l^n_{k+1}}\,=\, \pi-\theta_n(l^e_{k+1},l^n_{k+1})  \q . \label{c6c}
\ea

The momenta $p_e$ are just updated to agree with the exterior angles of the evolved spatial hypersurface. The momenta $p_n^k$ at time step $k$ have to vanish, since the $l^n_k$ are not dynamical variables at this time step. As was mentioned in the discussion at the end of section \ref{candisc}, we have to expect a post--constraint for every new edge variable $l^n$. Indeed, equations (\ref{c6c}) are constraints, requiring that the new momenta $p_n^{k+1}$ are again given by the exterior angles of the new hypersurface (coinciding with the three exterior angles of the added tetrahedron) which, however, can be expressed as functions of the length variables $l^e_{k+1},l^n_{k+1}$ only.

Notice that for this evolution step not only the $l^n_k$ remain undetermined, but also the $l^n_{k+1}$ can be chosen arbitrarily (the generalized triangle inequalities have to be satisfied though).  We will set $l^n_k=l^n_{k+1}$, so that the $l^n_k$ can be interpreted as initial data that determine the data at time $(k+1)$. Later we shall see that also for some types of Pachner moves (in 4D) the edge lengths of  new edges  can be chosen arbitrarily.  This freedom can be understood as a choice of initial data, which becomes only relevant at the time step, at which the new edges appear. On the other hand, we will also see (in 4D) that pre--constraints, appearing in consecutive evolution moves, might fix these edge lengths {\it a posteriori}.


\subsection{The 2--2 Pachner move}\label{22sec}

\begin{figure}[htbp!]
\psfrag{t1}{$t_1$}
\psfrag{t2}{$t_2$}
\psfrag{s}{$\Sigma_k$}
\begin{center}\includegraphics[scale=.4]{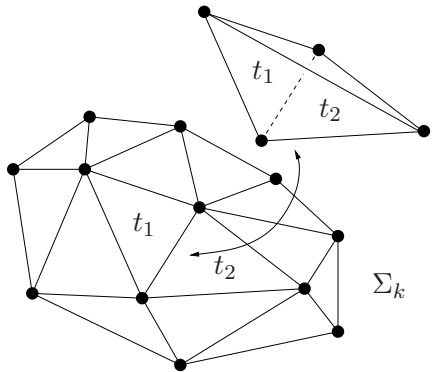} 
\end{center}
\caption{\small The gluing/removal 2--2 Pachner move involves two triangles $t_1$ and $t_2$ in hypersurface $\Sigma_k$.}\label{f22m} \end{figure}

\begin{figure}[htbp!]
\psfrag{1}{ 2--2}
\psfrag{2}{ 2--2}
\psfrag{o}{$o$}
\psfrag{n}{$n$}
\begin{center}\includegraphics[scale=.5]{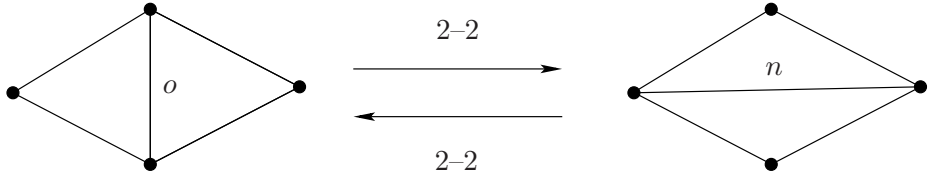} 
\end{center}
\caption{\small The 2--2 Pachner move is its own inverse.}\label{22m} \end{figure}

Consider the situation in which a tetrahedron $\tau$ is glued to $\Sigma_k$ in such a way that two of its triangles, $t_1$ and $t_2$, and thus five of its edges are identified with two neighbouring triangles and their five edges in $\Sigma_k$. This gluing move (with positive orientation) is only possible if the extrinsic dihedral angle $\psi^k_o$ at the edge $e=o$ along which the two triangles in $\Sigma_k$ are identified is negative. If it is positive, it is only possible to remove the corresponding tetrahedron $\tau$ from the triangulation, or, alternatively, to add a tetrahedron with negative orientation, see figure \ref{f22m}. 
From the perspective of the hypersurface, these elementary moves appear as 2--2 Pachner moves. That is, two triangles sharing an edge $e=o$ are replaced by two triangles sharing a new edge $e=n$ (see figure \ref{22m}).
Note that while this move removes one edge $o$ from the hypersurface and, instead, introduces the new edge $n$, it does not introduce a new vertex. Again, the two pairs of triangles have the same boundary of four edges $e$ and four vertices. The edges $o,n$ are transversal, that is $o$ and $n$ are connecting the two opposite pairs of vertices. Additionally, the 2--2 Pachner move is its own inverse, which from the 3D perspective is taken into account via the gluing/removing convention (and the different global signs in the action for the cases with different orientations) for the Pachner moves.


There are four kinds of edge lengths: apart from $l^b_{k/k+1},l^e_{k/k+1}$ and one length $l^n_{k+1}$, we also have one length $l^o_{k}$ which appears only at time step $k$ but not at $(k+1)$.  Thus, we will extend the phase spaces at time $k$ and $(k+1)$ by the pairs $(l^n_k,p^k_n)$ and $(l^o_{k+1},p_o^{k+1})$, respectively.

Accordingly, we will choose a generating function of the first kind in the variables $l^n,l^o$ and of the third kind 
in the variables $l^b,l^e$:
\ba\label{c7}
G_{2-2}(l^b_{k+1},p_b^k;\,l^e_{k+1},p_e^{k};\,l^o_k,l^o_{k+1};\,l^n_k,l^n_{k+1})
&=&\!\!
\sum_b l^b_{k+1}p_b^k + \sum_e l^e_{k+1}p_e^k  \, + \, S_\tau(l^e_{k+1},l^o_k,l^n_{k+1}) \, , \q\q
\ea
where
\ba
S_\tau(l^e_{k+1},l^o_k,l^n_{k+1})&=&
  \sum_{e \subset \tau} l^e_{k+1} \left( - \theta^\tau_e(l^e_{k+1},l^o_k,l^n_{k+1})  \right) +   l^n_{k+1} \left( \pi - \theta^\tau_n(l^e_{k+1},l^o_k,l^n_{k+1}) \right)  + \nn\\&&\q\q\q \q\q\q\q\q \q\q\q\q\q\q l^o_{k} \left( \pi - \theta^\tau_o(l^e_{k+1},l^o_k,l^n_{k+1}) \right) \, .
\ea
The evolution equations are given by
\ba\label{c8}
l^b_k&=&l^b_{k+1}  \q ,\q\q p^{k+1}_b\,=\,p^k_b \q , \\
l^e_k&=&l^e_{k+1}\q,\q\q p^{k+1}_e\,=\,p^k_e+\frac{\partial S_\tau}{\partial l^e_{k+1}}\,=\, p^k_e-\theta_e(l^e_{k+1},l^n_{k+1}) \q ,\\
p^{k}_n&=&0 \q,\q\q\q p^{k+1}_n\,\,=\, \frac{\partial S_\tau}{\partial l^n_{k+1}}\,\q\q\;=\, \pi-\theta_n(l^e_{k+1},l^o_k,l^n_{k+1}) \q , \label{c8c}\\
p^{k+1}_o &=& 0 \q,\q\q\q p^{k}_o\q\,\,=\, -\frac{\partial S_\tau}{\partial l^n_{k+1}}\,\q\;\;=\,- \pi+\theta_n(l^e_{k+1},l^o_k,l^n_{k+1}) \, .\q \label{c8d}
\ea

As discussed generally, we have $p_o^{k+1}=0$ and $p_n^k=0$.  In contrast to the 1--3 move, where the new edge lengths $l^n_k,l^n_{k+1}$ were undetermined at both time step, here the edge length $l^n_{k+1}$ is determined by equation (\ref{c8d}) as a function of the initial data, which also involve $l^o_k$ at time $k$. (Again, we can define $l^n_k=l^n_{k+1}$ but this time the interpretation of $l^n_k$ as additional new data does not apply, rather $l^n_k$ is in this case constrained, that is determined by the other initial data at time $k$. The same applies to $l^o_{k+1}$ if we go backwards in time.) Note that---via momentum matching---(\ref{c8d}) implements the Einstein equations $\epsilon_o=0$ for the edge $o$ which becomes a bulk edge in the course of the 2--2 (gluing) move. Equation (\ref{c8d}) demands that ${}^-p^k_o$ is given by the (oriented) exterior angle of the tetrahedron that is glued to the hypersurface. Momentum matching, on the other hand, imposes ${}^-p^k_o={}^+p^k_o$, where ${}^+p^k_o$ is determined by the previous moves and given by the (oriented) exterior angle of the hypersurface determined by the bulk, and thus that these two exterior angles add up to a vanishing deficit angle around the edge $o$. This condition will, in general, fix the edge length $l^n_{k+1}$ as the exterior (or dihedral) angle at the opposite edge $o$ of the tetrahedron depends on this length.

The difference to the 1--3 move is that here we have both a `new edge' and an `old edge'. Hence, the argument in section \ref{candisc}, according to which we have to expect a pre--constraint (because of the `old edge') and a post--constraint (because of the `new edge'), does not apply, as this argument relied on determining the maximal dimension of the image of the time evolution map. Here the counting changes since we have both a new and an old variable and due to the conditions $p_n^k=0$ and $p^{k+1}_o=0$, we do not necessarily expect a further constraint based on this argument (where we ignore that $l^o_{k+1}$ remains undetermined).
{\it A priori}, with the exception of $p^{k+1}_o=0$, all momenta at time $(k+1)$ involve edge lengths from time $k$ and time $(k+1)$, in particular, (\ref{c8c}) for $p^n_{k+1}$.\footnote{Since $l^o$ remains undetermined at time $(k+1)$ and $l^n$ at time $k$, one might wonder whether it is possible to summarize the two variables $l^o,l^n$ into just one variable (with two different time labels). Indeed, it is possible \cite{draft} to define a canonical time evolution in this way; the method presented here, however, is more similar to the 1--3 and 3--1 move and, moreover, to the Pachner moves in 4D. Nevertheless, this remark shows that canonical time evolution maps can be redefined by just relabeling or identifying variables with each other. We will see a similar situation in the tent move evolution in sections \ref{sectent1} and \ref{tent2} below, where several Pachner move steps are grouped together and labeled as a single tent move time step.} 

The fact that there are no new post--constraints arising in the 2--2 move should, however, not be confused with the feature that the system under consideration is totally constrained (that is, for spherical topology of $\Sigma$ all momenta are constrained as functions of the edge lengths). Rather, the 2--2 move preserves the (flatness) constraints, which encode that the hypersurface in question bounds a flat bulk triangulation, by implementing the flatness condition for the edge that becomes a bulk edge.

\subsection{The 3--1 Pachner move}\label{31sec}

\begin{figure}[htbp!]
\psfrag{1}{ 1--3}
\psfrag{2}{ 3--1}
\psfrag{e}{\tiny$e_o$}
\psfrag{v}{$v$}
\begin{center}\includegraphics[scale=.5]{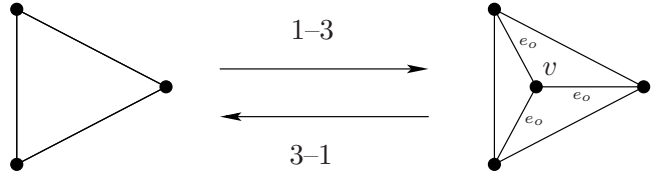} 
\end{center}
\caption{\small The 1--3 and 3--1 Pachner moves.}\label{31m} \end{figure}

The 3--1 move is the inverse of the 1--3 move: Consider, therefore, a three-valent vertex $v$ in $\Sigma_k$ whose adjacent three edges $e_o$ are equipped with extrinsic dihedral angles $\psi_o$ which are all negative, i.e.\ the vertex is pointing into the hypersurface and represents the tip of a tetrahedron which is upside down. In this situation we can glue a tetrahedron $\tau$ to this surface by identifying the three triangles sharing vertex $v$ in $\tau$ with the corresponding ones in $\Sigma_k$. Consider now the opposite situation, where all $\psi_o$ at the three-valent vertex are positive. In this case the vertex represents the tip of a tetrahedron $\tau$ which is sticking out of the hypersurface and we may remove this tetrahedron (or equivalently glue a tetrahedron with opposite orientation). These situations are depicted in figure \ref{13glurem} if the orientation of the arrows is reversed. From the perspective of the hypersurface, these elementary moves appear as a 3--1 Pachner move, see figure \ref{31m}.
Prior to this move, all six edges involved in the move are edges in $\Sigma_k$. During the move the vertex $v$, as well as the $e_o$ are removed (as these become internal to the bulk triangulation).

By virtue of the fact that there are no new edges introduced during the move, the equations of motion play rather the role of constraints in this case which have to be satisfied in order for this move to be allowed. Since up to $\Sigma_k$ we have always solved the equations of motion during the elementary evolution steps, the entire triangulation up to $\Sigma_k$ will be embedded in a flat 3D manifold. The three equations of motion of the edges which become internal, therefore, do not add any new constraints and are thus trivially satisfied.

The 3--1 move is the inverse to the 1--3 move, hence, we have three kinds of edge lengths: apart from $l^b_{k/k+1},l^e_{k/k+1}$ there are three edge lengths of type $l^o_{k}$ which appear only at time step $k$ but not at $(k+1)$. We therefore extend the phase space associated to $\Sigma_{k+1}$ by three pairs $(l^o_{k+1},p^{k+1}_o)$.

We choose the action to be a function of the edge lengths at time $k$ and, accordingly, use for edges of type $e$ and $b$ a generating function of second type. According to the general description, we have
\ba\label{c9}
G_{3-1}(l^b_k,p_b^{k+1}; l^e_k,p_e^{k+1}; l^o_k,l^o_{k+1})&=&
-\sum_b l^b_{k}p_b^{k+1} - \sum_e l^e_{k}p_e^{k+1}  \, + \, S_\tau(l^e_{k},l^o_k)
\ea
with
\ba
 S_\tau(l^e_{k},l^o_k)&=&
   \sum_{e \subset \tau} l^e_{k} \left( - \theta^\tau_e(l^e_{k},l^o_{k})  \right) +    \sum_{o \subset \tau} l^o_{k} \left( \pi - \theta^\tau_o (l^e_{k},l^o_{k}) \right) \q .
\ea
The evolution equations are then given by
\ba\label{c10}
&&l^b_{k+1}\,=\,l^b_{k}  \q ,\q\q p^{k}_b\,=\,p^{k+1}_b \q , \\
&&l^e_{k+1}\,=\,l^e_{k}\q,\q\q p^{k}_e\,=\,p^{k+1}_e-\frac{\partial S_\tau}{\partial l^e_{k}}\,=\, p^{k+1}_e+\theta_e(l^e_{k},l^o_{k}) \q ,\label{c9b}\\
&&p^{k+1}_o\,=\,0 \q,\q\q p^{k}_o\,=\,- \frac{\partial S_\tau}{\partial l^o_{k}}\,=\,- \pi+\theta_o(l^e_{k},l^o_{k})  \q .\label{c9c}
\ea

The second equation in (\ref{c9b}) defines the updated momenta $p^{k+1}_e$ as a function of the phase space variables at time $k$, 
while the second equation of (\ref{c9c}) is a pre--constraint, this time on the phase space variables at time $k$. That is, to perform this 3--1 move, this condition on the phase space variables has to be satisfied. Again, as for the 2--2 move these pre--constraints implement the equations of motions, namely flatness for the edges which become bulk edges during the move. In general, there are two possibilities: either i) the pre--constraints are automatically satisfied, if we consider a hypersurface which has been evolved in the manner described here by Pachner moves, or ii) the constraints are only satisfied for specific initial data, including the kind of additional initial data which arise, for instance, by the 1--3 Pachner move. For the 3D case, only the first possibility takes place as argued above. In 4D, on the other hand, we will generally encounter the second possibility, namely data which are {\it a priori} free to choose in a certain Pachner move may become fixed by pre--constraints arising in later moves.

\subsection{Example: 3D tent moves}\label{sectent1}

The Pachner moves generically change the connectivity and the number of edges in the triangulation. There are, however, combinations of Pachner moves, e.g.\ the so--called tent moves  \cite{commi,dithoe1,bd1}, which do {\it not} change the connectivity of the triangulations and therefore induce a canonical dynamics in the standard interpretation, that is, between phase spaces of equal dimensions.

\begin{figure}[hbt!]
\begin{center}
    \psfrag{vn}{$v_{n}$}
    \psfrag{vn+1}{$v_{n+1}$}
    \psfrag{1}{\small$1$}
    \psfrag{2}{\small$2$}
    \psfrag{3}{\small$3$}
    \psfrag{4}{\small$4$}
    \includegraphics[scale=0.3]{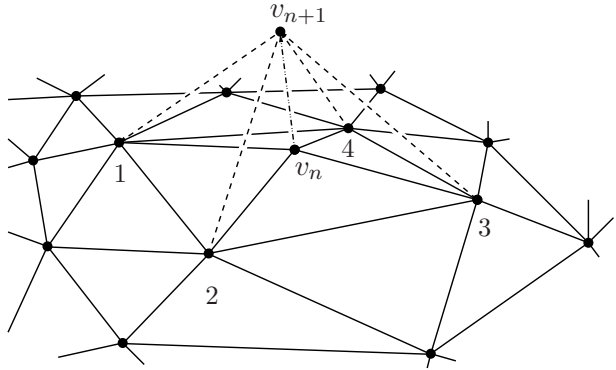}
    \end{center}
    \caption{\label{3dtm1}{\small A 3D tent move applied to a four-valent vertex $v_n$ in a 2d Cauchy hypersurface. This four-valent tent move can be reproduced through a gluing sequence of four tetrahedra which correspond to a sequence of one 1--3, two 2--2 and one 3--1 Pachner moves in the hypersurface. }}
\end{figure}

A tent move can be constructed by picking some vertex $v_n$\footnote{Note that henceforth, in order to avoid confusion, we will enumerate tent moves by $n\in\Z$, while the elementary Pachner moves into which the tent moves can be decomposed are counted by $k\in\Z$.}  in a $(D-1)$--dimensional triangulated hypersurface $\Sigma_n$ and subsequently defining a new vertex $v_{n+1}$ to the `future' of $v_n$ which must be connected by an edge to $v_n$ (the `tent pole'). Denote all other vertices in $\Sigma_n$ to which $v_n$ is connected by $1,\ldots,N$. Connect $v_{n+1}$ to each of these vertices $1,\ldots,N$ by $N$ edges. The new vertex $v_{n+1}$ then lies in the new hypersurface $\Sigma_{n+1}$ and is $N$-valent in $\Sigma_{n+1}$ as $v_n$ is in $\Sigma_n$ (the tent pole is now internal) and for every $(D-1)$-simplex $\sigma(v_nij\ldots)$, $i,j\in1,\ldots,N$, there is now a $(D-1)$-simplex $\sigma(v_{n+1}ij\ldots)$. That is, the triangulations of the two hypersurfaces are the same. Notice that each tent move only involves the $(D-1)$ star of the vertex $v_n$ in $\Sigma_n$. The situation for a four-valent 3D tent move is illustrated in figure \ref{3dtm1}.
\noindent

The new piece of $D$-dimensional triangulation between $\Sigma_n$ and $\Sigma_{n+1}$ with the $N+1$ new edges $e(v_{n+1}i)$ and $e(v_nv_{n+1})$ can be decomposed into a sequence of gluings of single $D$-simplices and the tent move evolution may therefore be described in terms of a sequence of Pachner moves in the hypersurface. In particular, in 3D simply pick an N-valent vertex in some $\Sigma_n$ and perform one 1--3 Pachner move, $(N-2)$ 2--2 Pachner moves and a final 3--1 Pachner move in order to generate $\Sigma_{n+1}$.

To define the Pachner move dynamics, we used the Regge action as (part of) a generating function, so that the resulting canonical dynamics leads to the same equations of motions as the ones obtained by varying the action. We can again employ the Regge action to define the canonical evolution corresponding directly to the tent moves \cite{bd1, dithoe1,bianca}. By construction, this dynamics will coincide with the one obtained by performing the sequence of Pachner moves (matching appropriately the edge labels). This highlights the remarks put forward earlier that canonical time evolution maps may be redefined by relabeling variables. For the tent move dynamics we can either keep the length of the tent pole as an internal variable in the sense of problem (a) in section \ref{candisc}, see equations (\ref{b6}), or, alternatively, integrate this length out and work with Hamilton's principal function as a generating function:

To this end, notice that the 3D triangulation---the `tent'---we are gluing to the hypersurface has two 2D boundaries given by the 2D stars of the vertices $v_n$ and $v_{n+1}$. These two 2D boundaries meet in a 1D boundary, given by the (cyclically ordered) vertices $1,\ldots, N$ and the edges connecting these vertices. The tent consists of $N$ tetrahedra sharing the tent pole as an edge, and therefore each having vertices $v_n,v_{n+1}$ and $i,i+1$. The tent pole is an internal edge, whereas the 2D boundary has edges $e(v_n,i),e(v_{n+1},i)$ and edges $e(i,i+1)$ with edge lengths $l^i_n,l^i_{n+1},\,i=1,\ldots N$ and $l^{i,i+1}_n= l^{i,i+1}_{n+1}$, respectively.  We can solve the equation of motion for the tent pole as a function of the boundary edges. This equation of motion will just require that the deficit angle around the tent pole is vanishing. Using this result in the action, we obtain Hamilton's principal function for the `tent'
\ba
\tilde S_{tent}(l^i_n,l^i_{n+1},l^{i,i+1}_{n+1})&=&\!\!\!-\sum_i l_{n+1}^{i,i+1}\theta_{i,i+1}^{\tau(v_n,v_{n+1},i,i+1)} + \sum_i l_n^i\left(\pi-  \theta_{v_ni}^{\tau(v_n,v_{n+1},i,i+1)} - \theta_{v_ni}^{\tau(v_n,v_{n+1},i-1,i)}\right)  +\nn\\
&&\q\q\q\q \sum_i l_{n+1}^i\left(\pi-  \theta_{v_{n+1}i}^{\tau(v_n,v_{n+1},i,i+1)} - \theta_{v_{n+1}i}^{\tau(v_n,v_{n+1},i-1,i)}\right)  \q .
\ea
Taking into account  the additional terms $l^{i,i+1}_{n+1} p^n_{i,i+1}$ in the generating function we obtain the following equations of motion
\ba
l^{i,i+1}_{n}&=&l^{i,i+1}_{n+1}  \q , \nn\\
 p^{n+1}_{i,i+1}&=&p^{n}_{i,i+1}- \theta_{i,i+1}^{\tau(v_n,v_{n+1},i,i+1)} \q , \label{fria0} \\
p^{n}_i &=&   -\pi+  \theta_{v_ni}^{\tau(v_n,v_{n+1},i,i+1)} + \theta_{v_ni}^{\tau(v_n,v_{n+1},i-1,i)}  \q ,    \label{fria1}  \\
 p^{n+1}_i &=&   \pi-  \theta_{v_{n+1}i}^{\tau(v_n,v_{n+1},i,i+1)} - \theta_{v_{n+1}i}^{\tau(v_n,v_{n+1},i-1,i)} \; ,\q\q\q\label{fria2}
\ea
where again the Schl\"afli identity was used. Now, {\it a priori} the dihedral angles on the right hand side of the equations (\ref{fria1},\ref{fria2}) depend on both sets of edge lengths $l^i_n$ and $l^{i+1}_n$. But, for instance, for the tent move at a three valent vertex the right hand side of (\ref{fria2}) does not depend on the lengths $l^i_n$ and we obtain three post--constraints. (The geometry around  the new vertex is the one around a vertex of a flat tetrahedron and the dihedral angles can be expressed as functions of the six edge lengths $l^i_{n+1}, l^{i,i+1}_{n+1}$.) Likewise, (\ref{fria1}) will give us three pre--constraints as the right hand side can be expressed as a function of $l^i_n,l^{i,i+1}_n$ only. For tent moves at higher valent vertices we can still conclude that we have three pre--constraints and three post--constraints of the form (\ref{thu3}). To this end one just has to use the same arguments as for (\ref{thu3}) applied to the triangulation of the tent, which itself is a flat triangulation.

Hence, we will have three pre-- and three post--constraints for the tent move. From this we can conclude that three of the edge lengths $l^i_{n+1}$ remain undetermined by the equations of motion (\ref{fria0}--\ref{fria2}). (Correspondingly, there is a three parameter set of initial data at step $n$ which can be evolved to the same data at time $(n+1)$, and vice versa.) These are the same three parameters that are left undetermined by the 1--3 move in section \ref{13sec} and correspond to lapse and shift degrees of freedom. The three post--constraints of the 1--3 move are of the form as encountered here and the present discussion shows that these constraints remain preserved through all the additional Pachner moves which make up the tent move. Also the pre--constraints which need be fulfilled so that the final 3--1 move can be performed, will automatically be satisfied, if the triangulation has been correctly evolved by Pachner moves.

Finally, we remark that the lapse and shift degrees of freedom which remain free in the tent move can be chosen to be infinitesimally small. This allows to recover a continuous time evolution (plus Hamiltonian and constraints generating this evolution, which coincide with (\ref{thu3})) from the discrete time evolution presented here.

\section{Pachner moves for 4D Regge calculus}\label{4D}

While the 3D Regge dynamics leads to flat geometries, the 4D Regge equations allow for curved solutions. This renders the 4D dynamics significantly more complicated than in the 3D case, in particular, the preservation of the constraints will generally not hold. Related to this is the fact that lapse and shift degrees of freedom will, in general, not remain free but will become fixed by pre--constraints (unless one considers initial data which lead to flat solutions).

The Regge action for a 4D triangulation (without cosmological constant) is given by (\ref{regge1},\ref{regge2}) for $h=t$,
\ba\label{y1}
S=\sum_{t\subset T^\circ} A_t \epsilon_t + \sum_{t\subset \partial T} A_t \psi_t  \q .
\ea
The equations of motion for an inner edge $e$ read
\ba\label{y2}
\sum_{t \supset e} \frac{\p A_t}{\p l^e} \,\,\epsilon_t &=&0  \q ,
\ea
which---depending on the boundary data---allow for flat solutions $\epsilon_t=0$ as well as solutions with curvature $\epsilon_t\neq 0$.

Despite the latter, we can show that Hamilton's principal function reduces to a boundary term (as is the case in the continuum). To this end, multiply each of the equations (\ref{y2}) with the length $l^e$ and sum over all (inner) edges. We will employ the Euler identity (see, e.g., \cite{Bahr:2009qc})
\ba\label{y3}
\sum_{e \subset t} l^e \frac{\p A_t}{\p l^e} &=& 2 A_t
\ea
for the area of a triangle. Using the equations of motion (\ref{y2}), we can write
\ba\label{y4}
0&=& \sum_{e \subset T^\circ } l^e \sum_{t\subset e} \frac{\p A_t}{\p l^e} \epsilon_t \nn\\
&=&  2 \sum_{t \subset T^\circ_t}  A_t \epsilon_t \,+\, \sum_{t \subset T^\p_t}\; \sum_{e \subset t \, \cap\, e\subset T^\circ}  l^e \frac{\p A_t}{\p l^e}\eps_t
\ea
where $T^\circ_t$ denotes the set of triangles which have all three edges contained in the bulk $T^\circ$ and $T^\p_t$ denotes the set of triangles in $T^\circ$ where at least one of the edges is in the boundary $\p T$. (Note that an inner triangle can have either one, two or all three edges in the boundary.) Employing equations (\ref{y3}) and (\ref{y4}) in the action (\ref{y1}), we discover that Hamilton's principal function
\ba\label{y5}
\tilde S&=& \frac{1}{2} \sum_{e\subset \p T} l^e \left[ \sum_{t \supset e\,\cap\,t\subset T^\p_t} \frac{\p A_t}{\p l^e}\epsilon_t\;+\; \sum_{t \supset e\,\cap\,t\subset \p T}  \frac{\p A_t}{\p l^e} \psi_t \right]  \q,
\ea
is given by a boundary term. The form of this last expression suggests that the momenta are given by the combination of exterior angles and deficit angles `near' the boundary in square brackets on the right hand side of (\ref{y5}),
\ba\label{y6}
p_e&=&\sum_{t \supset e\,\cap\,t\subset T^\circ} \frac{\p A_t}{\p l^e}\epsilon_t\;+\; \sum_{t \supset e\,\cap\,t\subset \p T}  \frac{\p A_t}{\p l^e} \psi_t
\ea
Indeed, momentum updating during the Pachner move evolution will confirm this.

As in 3D, we can produce a Regge triangulation with boundary topology given by the 3--sphere\footnote{Once more, we could equally well start by producing an initial hypersurface of more complicated topology.} by considering first an evolution move from the empty hypersurface to the boundary of a 4--simplex. The generating function would simply be given by the one--simplex action
\ba
G_{0-1}(l^n_0,l^n_1)&=& S_\sigma(l^n_1) \;=\;\sum_t A_t (\pi-\theta_t^\sigma)\q,
\ea
and gives rise to the following equations of motion
\ba
p^0_n=0 \q ,\q\q\q p^1_n =\frac{\p S_\sigma}{\p l^n_1} = \sum_{t\supset e} \frac{\p A_t}{\p l^n_1} \left( \pi- \theta_t^\sigma \right) \q .
\ea

Again, just as in the 3D case, we also obtain post--constraints here which determine all 10 momenta at time $k=1$ as a function of the edge lengths at time $k=1$. Hence, the boundary of a simplex is a totally constrained system. This property does not change under 1--4 Pachner moves (yielding so--called {\it stacked spheres}) which will introduce four post--constraints for every four new edge lengths added as configuration variables. This sector of the 4D dynamics behaves very similarly to 3D gravity and was also discussed in \cite{BiancaJames}. Furthermore, a derivation of the symplectic structure for this sector, starting from a first order formulation of canonical discrete gravity, may be found in \cite{BiancaJames}. The results obtained there are in agreement with (\ref{y6}). The reason why this sector is totally constrained is that the 1--4 Pachner moves neither lead to inner triangles nor inner edges. Hence, the hypersurface triangulations produced by means of this move allow for bulk triangulations without any inner triangles which could carry inner curvature. That is, we are considering again just flat triangulations.

This will, in general, change if in addition we consider 2--3, 3--2, and 4--1 Pachner moves. As the 1--4 Pachner move will not lead to any additional inner triangles or edges but adds four boundary edges, we can consider it as just adding lapse and shift degrees of freedom. Indeed, the four new edge lengths will remain free parameters, at least before doing any other moves. The 2--3 Pachner move, on the other hand, generates an inner triangle but no inner edge. We shall see that this move can be interpreted as adding a curvature degree of freedom (or physical degree of freedom as opposed to the lapse and shift gauge degrees of freedom). The 3--2 move, furthermore, produces an inner edge for which the Regge equation of motion---canonically in the form of a pre--constraint---must be satisfied. Consequently, this move can be considered as the `true evolution step' in the sense that it requires the solution of an equation of motion, involving curvature degrees of freedom. Finally, the 4--1 Pachner move generates six inner triangles and four inner edges and will feature four pre--constraints. For the same reasons as in the 3D case, the pre--constraints will be automatically satisfied if we consider initial data leading to a flat solution. For curved solutions, however, these pre--constraints will fix some free parameters introduced in the previous steps.

\subsection{The 1--4 Pachner move}

This move is the 4D analogue of the 1--3 Pachner move discussed in section \ref{13sec}. Glue a 4-simplex $\sigma$ onto a given 3D hypersurface $\Sigma_k$ in such a way that its `bottom tetrahedron' $\tau$ is identified with a tetrahedron in $\Sigma_k$. Similarly, consider the situation in which one simplex $\sigma$ that shares one tetrahedron $\tau$ with $\Sigma_k$ is removed from a given triangulation (equivalent to adding a tetrahedron with opposite orientation).

\begin{figure}[htbp!]
\psfrag{1}{1--4}
\psfrag{2}{4--1}
\begin{center}\includegraphics[scale=.5]{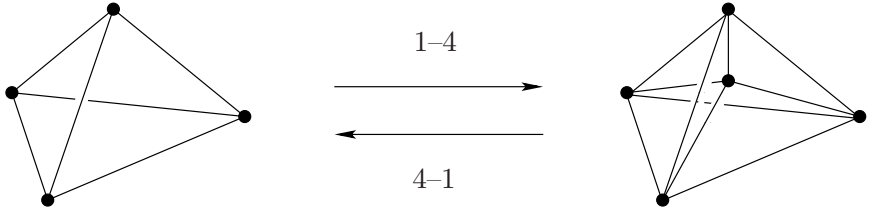} \end{center}
\caption{\small The 1--4 Pachner move and its inverse, the 4--1 Pachner move.}\label{14m}
 \end{figure}

This move acts as a 1--4 Pachner move on $\Sigma_k$. It introduces one new vertex and four new edges into the new hypersurface, but it does not  render any edges internal. The 1--4 Pachner moves replaces the tetrahedron $\tau$ with the subdivided tetrahedron (consisting of four smaller tetrahedra); the boundary of the tetrahedron $\tau$ does not change and we have the same four triangles $t^e$ and six edges $l^e$ before and after the move. Through the subdivision there will appear four new edges $e_n$ and six new triangles $t^n$ adjacent to these new edges, see figure \ref{14m}.

Since there are no edges which become internal, there are no equations of motion to be satisfied. As a consequence, the lengths of the four new edges, labeled by $n$, can be freely chosen. That is, one has a fourfold freedom in choosing the `tip' of $\sigma$, which can be parametrized by lapse $N$ and shift $N^\alpha$ variables. As opposed to the 3D case (and the continuum) this freedom will, in general, be restricted by the appearance of pre--constraints in later moves (see also the discussion in section \ref{tent2} about tent moves).

We proceed in the same way as for the 1--3 move and use the generating function
\ba\label{4c5}
G_{1-4}(l^b_{k+1},p_b^k;\,l^e_{k+1},p_e^{k};\,l^n_k,l^n_{k+1}) &=& \sum_b l^b_{k+1}p_b^k + \sum_e l^e_{k+1}p_e^k  \, + \, S_\sigma(l^e_{k+1},l^n_{k+1})  \q,
\ea
where we use the action of a single simplex
\ba
S_\sigma&=& \sum_{t^e \subset \sigma} A_{t^e}(l^e_{k+1},l^n_{k+1})\left( - \theta^\sigma_{t^e}(l^e_{k+1},l^n_{k+1})  \right) +    \sum_{t^n \subset \sigma} A_{t^n} (l^e_{k+1},l^n_{k+1})   \left( \pi - \theta^\sigma_{t^n} (l^e_{k+1},l^n_{k+1}) \right)  \; .\q\q
\ea
The evolution equations are then
\ba\label{4c6}
&&l^b_k\,=\,l^b_{k+1}  \q ,\q\q p^{k+1}_b\,=\,p^k_b \q , \\
&&l^e_k\,=\,l^e_{k+1}\q,\q\q p^{k+1}_e\,=\,p^k_e+\frac{\partial S_\sigma}{\partial l^e_{k+1}}\,=\, p^k_e -   \sum_{t^e\subset \sigma}
 \frac{\p A_{t^e}}{\p l^e_{k+1}} \theta_{t^e}^\sigma
 +\sum_{t^n\subset \sigma} \frac{\p A_{t^n}}{\p l^e_{k+1}} \left( \pi-\theta_{t^n}^\sigma\right)
  \,,\q\\
&&p^{k}_n\,=\,0 \q,\q\q\q p^{k+1}_n\,=\, \frac{\partial S_\sigma}{\partial l^n_{k+1}}\q\q\;\,=\, \q\q
\sum_{t^n\subset \sigma} \frac{\p A_{t^n}}{\p l^n_{k+1}} \left( \pi-\theta_{t^n}^\sigma\right)
. \label{4c6c}
\ea

As for the 1--3 move the $l^n_k$ and $l^n_{k+1}$ remain arbitrary, nevertheless, we can equire $l^n_k=l^n_{k+1}$. Again, we encounter post--constraints for every new edge, namely the second equation in (\ref{4c6c}). These post--constraints fix the momenta as a specific combination of the exterior angles. At a four--valent vertex, which can be seen as the tip of a four--simplex, this combination of exterior angles can be expressed as a function of the adjacent length variables (namely the lengths of the simplex $\sigma$). Such a post--constraint will always appear if we produce a boundary edge in the course of evolution at which no bulk triangles (as potential carriers of curvature) are hinging.

In the later discussion we shall see that these constraints are, in general, not preserved by the other moves. Additionally, the free data, that is the lenghts of the four new edges, may, in general, become fixed by pre--constraints of later moves. This is related to the fact, that the Regge discretization does not preserve the diffeomorphism symmetry of the continuum \cite{bd1,dithoe1}. The constraints and the free data discussed here, therefore, correspond to the Hamiltonian and diffeomorphism constraints and the gauge freedom (of lapse and shift) of the continuum. Through the breaking of diffeomorphism symmetry by discretization, the constraints encountered here will generally not be preserved by the other Pachner moves but, instead, turned into so--called pseudo--constraints \cite{bd1,dithoe1,gambini,bianca}. These are equations of motion for canonical data of two different times, in which these data are only weakly coupled to each other. (More precisely, the eigenvalues of the appropriate Hessian matrix associated to these equations of motion will turn out to be small compared to those of a Hessian associated to proper evolution equations.)

\subsection{The 2--3 Pachner move}

Next, let us discuss the 2--3 move which turns out to generate a dynamical (or physical) degree of freedom.

Consider the situation in which a 4--simplex $\sigma$ is glued to $\Sigma_k$ in such a way that two of its (adjacent) tetrahedra are identified with two (adjacent) tetrahedra in $\Sigma_k$. This gluing process is only possible when these two tetrahedra in $\Sigma_k$ are {\it not} part of the same 4--simplex in the underlying triangulation (for the five vertices of these two tetrahedra are already the five vertices of $\sigma$). In contrast to the 2--2 move in 3D, the extrinsic curvature angle around the triangle $t^o$ along which the two adjacent tetrahedra are identified need {\it a priori} (i.e.\ kinematically) not be negative.\footnote{Due to the flat embedding in 3D Regge Calculus (without a cosmological constant), the extrinsic angle $\psi^k_o$ around the edge $e=o$ in the configuration of the 2--2 Pachner move in section \ref{22sec} must be negative for a 2--2 gluing move to be possible such that eventually $\epsilon_o=0$ (for non-degenerate simplices, the dihedral angles are always positive). } On the other hand, we can remove such a simplex from the triangulation (or, equivalently, add a simplex with opposite orientation), if the extrinsic angle around $t^o$ is positive, the two adjacent tetrahedra reside in the same $\sigma$ and there is a piece of triangulation underneath $\Sigma_k$. From the perspective of the hypersurface, this evolution move amounts to a 2--3 Pachner move.
\begin{figure}[htbp!]
\psfrag{1}{2--3}
\psfrag{2}{3--2}
\psfrag{t}{$t^o$}
\psfrag{n}{$n$}
\begin{center}\includegraphics[scale=.5]{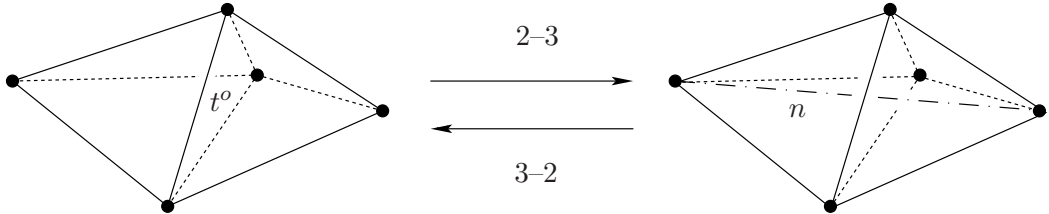} \end{center}
\caption{\small The 2--3 Pachner move and its inverse, the 3--2 Pachner move.}\label{23m}
 \end{figure}
 
Two tetrahedra sharing a triangle $t^o$ and having a boundary consisting of six triangles $t^e$ and nine edges $e$ are replaced by three tetrahedra sharing a new edge $n$ and sharing pairwise three new triangles $t^n$, see figure \ref{23m}. As for all the Pachner moves, the boundary of the three new tetrahedra is the same as for the two tetrahedra we started with.
Consequently, during this move no edge will become internal and, therefore, there will be no additional equations of motion which we could solve for the length of the new edge. Furthermore, the triangle $t^o$ becomes a bulk triangle in the gluing move. This is important, since after the move, the bulk triangle can carry a non--vanishing deficit angle, that is curvature. This deficit angle depends on the value of the new edge and as this can be freely chosen, we are generating a curvature degree of freedom.\footnote{For instance, it is possible to generate the complete 4D star of a  triangle, i.e.\ rendering it internal, by starting out with one 4-simplex and performing a sequence of 1--4 Pachner moves on it and a final 2--3 move. During this sequence, the triangle shared by all simplices has become internal, but there are only boundary edges. This star of the triangle, therefore, carries curvature without any internal edges. Hence, there are no Regge equations to be satisfied.} This is different from the four free edge lengths that arise in the 1--4 move which rather correspond to lapse and shift, and, therefore, gauge variables, in the continuum.

Because of the new edge at step $(k+1)$, we extend the phase space at step $k$ by the pair $(l^n_k,p_n^k)$. We use the generating function
\ba\label{c12}
G_{2-3}(l^b_{k+1},p_b^k;\,l^e_{k+1},p_e^{k};\,l^n_k,l^n_{k+1})
&=&
\sum_b l^b_{k+1}p_b^k + \sum_e l^e_{k+1}p_e^k  \, + \, S_\sigma(l^e_{k+1},l^n_{k+1}) \q\q,
\ea
where in this case the action of the single 4--simplex under consideration is
\ba
S_\sigma(l^e_{k+1},l^n_{k+1})& =&
 \sum_{t^e \subset \sigma} A_{t^e}(l^e_{k+1},l^n_{k+1})\left( - \theta^\sigma_{t^e}(l^e_{k+1},l^n_{k+1})  \right) + A_{t^o} (l^e_{k+1},l^n_{k+1})   \left( \pi - \theta^\sigma_{t^o} (l^e_{k+1},l^n_{k+1}) \right) +\nn\\
 &&\q\q\q\q\q\q\q     \sum_{t^n \subset \sigma} A_{t^n} (l^e_{k+1},l^n_{k+1})   \left( \pi - \theta^\sigma_{t^n} (l^e_{k+1},l^n_{k+1}) \right)  \q .
\ea
The evolution equations read
\ba\label{c13}
l^b_k&=&l^b_{k+1}  \q ,\q\q p^{k+1}_b\,=\,p^k_b \q , \\
l^e_k&=&l^e_{k+1}\q,\q\q p^{k+1}_e\,=\,p^k_e+\frac{\partial S_\sigma}{\partial l^e_{k+1}}\,=\, p^k_e
-\sum_{t^e\subset\sigma} \frac{\p A_{t^e}}{\p l^e_{k+1}} \theta^\sigma_{t^e} +\frac{\p A_{t^o}}{\p l^e_{k+1}}(\pi-\theta^\sigma_{t^o} )  \nn\\
&&\q\q\q\q\q\q\q\q\q\q\q\q\q\q\q\q\q\q\q\q\q
+\sum_{t^n\subset \sigma} \frac{\p A_{t^n}}{\p l^e_{k+1}} (\pi-\theta^\sigma_{t^n})
\; ,    \label{c13b}  \\
p^{k}_n&=&0 \q,\q\q\q p^{k+1}_n\,=\, \frac{\partial S_\sigma}{\partial l^n_{k+1}}\,\q\q\;=\, \sum_{t^n\subset \sigma} \frac{\p A_{t^n}}{\p l^n_{k+1}} (\pi-\theta^\sigma_{t^n})
\q .\label{c13c} 
\ea

The second equation in (\ref{c13b}) determines the momenta $p_e^{k+1}$ at time $k+1$ as a function of the ones at time $k$, as well as the new edge lengths $l^n_{k+1}$ and $l^e_{k+1}=l^e_k$. The new lengths $l^n_{k},l^n_{k+1}$ remain undetermined, but we may again choose $l^n_k=l^n_{k+1}$. Accordingly, we find a post--constraint at time $(k+1)$, namely the second equation in (\ref{c13c}). It determines the new momentum variable as a function of the edge lengths at step $(k+1)$. Geometrically this constraint arises as by construction there are no bulk, but only boundary triangles hinging at the new edge. Hence, we can compute the exterior angle from the boundary geometry alone, i.e.\ the length variables of the hypersurface $\Sigma_{k+1}$.

\subsection{The 3--2 Pachner move}

\begin{figure}[htbp!]
\psfrag{1}{2--3}
\psfrag{2}{3--2}
\psfrag{t}{$t^n$}
\psfrag{n}{$o$}
\begin{center}\includegraphics[scale=.5]{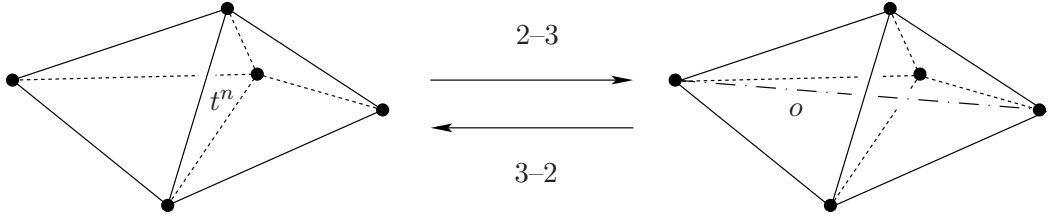} \end{center}
\caption{\small The 2--3 Pachner move and its inverse, the 3--2 Pachner move.}\label{32m}
 \end{figure}

The 3--2 Pachner move is the inverse of the 2--3 move: a 4-simplex $\sigma$ is glued onto $\Sigma_k$ in such a way that three of its (adjacent) tetrahedra, sharing an edge $o$, are identified with three adjacent tetrahedra in $\Sigma_k$. This gluing is possible only when these three tetrahedra reside in three distinct 4--simplices of the triangulation underlying $\Sigma_k$. As in the case of the 2--3 move and, by virtue of the possible absence of a flat embedding, the extrinsic angles at the three triangles $t^o$ which are each shared by two of the three tetrahedra in $\Sigma_k$ need {\it a priori} not be negative; the values of the resulting deficit angles around these triangles will eventually be determined by the dynamics. Likewise, one can remove a simplex $\sigma$ from the triangulation underlying $\Sigma_k$ (or, equivalently, glue a simplex with opposite orientation to $\Sigma_k$), if the three extrinsic angles are positive and the three tetrahedra are part of the same 4--simplex. These gluings/removals appear as 3--2 Pachner moves, that is three tetrahedra sharing one edge $o$ and pairwise three triangles $t^o$ are replaced with two tetrahedra sharing one triangle $t^n$, such that the boundary of the two complexes, consisting of six triangles $t^e$ and nine edges $e$ remains unchanged (see figure \ref{32m}).
The edge $o$ will become internal, in consequence we will have to satisfy the equation of motion corresponding to this edge. This equation of motion will be implemented via a pre--constraint and momentum matching. It may either not be possible at all to satisfy this pre--constraint, or only for specific choices of parameters, that is lengths variables, which appeared in previous moves.

Since we have a variable $l^o_k$ that does not appear at step $(k+1)$, we extend the phase space at time $(k+1)$ by the pair $(l^o_{k+1},p_o^{k+1})$. The generating function for the 3--2 move reads
\ba\label{c132}
G_{3-2}(l^b_k,p_b^{k+1}; l^e_k,p_e^{k+1};l^o_k,l^o_{k+1})&=&
-\sum_b l^b_{k}p_b^{k+1} -\sum_e l^e_{k}p_e^{k+1}  \, + \, S_\sigma(l^e_{k},l^o_k)\q,
\ea
where the action $S_\sigma$ is given by
\ba
S_\sigma(l^e_{k},l^o_{k})& =&
 \sum_{t^e \subset \sigma} A_{t^e}(l^e_{k},l^o_{k})\left( - \theta^\sigma_{t^e}(l^e_{k},l^o_{k})  \right) + A_{t^n} (l^e_{k},l^o_{k})   \left( \pi - \theta^\sigma_{t^n} (l^e_{k},l^o_{k}) \right) +\nn\\
 &&\q\q\q\q\q\q\q     \sum_{t^o \subset \sigma} A_{t^o} (l^e_{k},l^o_{k})   \left( \pi - \theta^\sigma_{t^o} (l^e_{k},l^o_{k}) \right)  \q .
\ea
The evolution equations amount to
\ba\label{c14}
&&l^b_{k+1}\,=\,l^b_{k}  \q ,\q\q p^{k}_b\,=\,p^{k+1}_b \q , \\
&&l^e_{k+1}\,=\,l^e_{k}\q,\q\q p^{k}_e\,=\,p^{k+1}_e-\frac{\partial S_\sigma}{\partial l^e_{k}}\,=\,
p^{k+1}_e
+\sum_{t^e\subset\sigma} \frac{\p A_{t^e}}{\p l^e_{k}} \theta^\sigma_{t^e} -\frac{\p A_{t^n}}{\p l^e_{k}}(\pi-\theta^\sigma_{t^n} )  \nn\\
&&\q\q\q\q\q\q\q\q\q\q\q\q\q\q\q\q\q\q\q\q\q
-\sum_{t^o\subset \sigma} \frac{\p A_{t^o}}{\p l^e_{k}} (\pi-\theta^\sigma_{t^o})
\q ,\label{c14b}\\
&&p^{k+1}_o\,=\,0 \q,\q\q p^{k}_o\,=\,- \frac{\partial S_\sigma}{\partial l^o_{k}}\q\q\;\,=\,
 -\sum_{t^o\subset \sigma} \frac{\p A_{t^o}}{\p l^o_{k}} (\pi-\theta^\sigma_{t^o})
  \q .\label{c14c}
\ea

This time the second equation in (\ref{c14b}) determines the momenta $p_e^{k+1}$ at step $(k+1)$ as a function of the ones at time $k$ and the old  edge lengths $l^o_{k}$ and $l^e_{k}$.

The second equation in (\ref{c14c}), on the other hand, is a pre--constraint. It requires that the momentum $p^k_o$ be given as a function of the edge lengths at time $k$. This equation is automatically satisfied if the edge $o$ has just been created by a 2--3 move, but in general this condition may not be satisfied. In this case, one either has to tune the new length variables that arose in previous 2--3 or 1--4 moves, or perform a different move.

\subsection{The 4--1 Pachner move}

Finally, let us consider the 4--1 move, which is the inverse of the 1--4 move (see figure \ref{14m}). A complex of four tetrahedra sharing a vertex, four edges $o$ and six triangles $t^o$ is replaced by one tetrahedron, such that the boundary of four triangles $t^e$ and six edges $e$ remains unchanged. The 4--1 gluing move is only possible if these four tetrahedra are part of four different 4--simplices in the underlying triangulation, while the six extrinsic angles at the six $t^o$ are kinematically unrestricted. From the 4D perspective the four `old' edges become bulk edges, consequently, we will have four equations of motion to satisfy. These take the form of pre--constraints and eventually determine the resulting six deficit angles around the $t^o$. The prerequisite for the 4--1 removal move, on the other hand, is clearly that the four relevant tetrahedra reside in the same 4--simplex and the six extrinsic angles at the six $t^o$ are positive.

 We extend the phase space at time $(k+1)$ by the four pairs $(l^o_{k+1},p_o^{k+1})$ and define the generating function
\ba\label{4c41}
G_{4-1}(l^b_{k},p_b^{k+1};\,l^e_{k},p_e^{k+1};\,l^o_k,l^o_{k+1}) &=&- \sum_b l^b_{k}p_b^{k+1} - \sum_e l^e_{k}p_e^{k+1}  \, + \, S_\sigma(l^e_{k},l^o_{k})  \q,
\ea
where we use the action of the single 4--simplex under consideration
\ba
S_\sigma&=& \sum_{t^e \subset \sigma} A_{t^e}(l^e_{k},l^o_{k})\left( - \theta^\sigma_{t^e}(l^e_{k},l^o_{k})  \right) +    \sum_{t^o \subset \sigma} A_{t^o} (l^e_{k},l^o_{k})   \left( \pi - \theta^\sigma_{t^o} (l^e_{k},l^o_{k}) \right)  \; .\q\q
\ea
The equations of motion are then
\ba\label{4c6d}
&&l^b_k\,=\,l^b_{k+1}  \q ,\q\q p^{k}_b\,=\,p^{k+1}_b \q , \\
&&l^e_k\,=\,l^e_{k+1}\q,\q\q p^{k}_e\,=\,p^{k+1}_e-\frac{\partial S_\sigma}{\partial l^e_{k}}\,=\, p^{k+1}_e +   \sum_{t^e\subset \sigma}
 \frac{\p A_{t^e}}{\p l^e_{k}} \theta_{t^e}^\sigma
 -\sum_{t^o\subset \sigma} \frac{\p A_{t^o}}{\p l^e_{k}} \left( \pi-\theta_{t^o}^\sigma\right)
  \,,\q\\
&&p^{k+1}_o\,=\,0 \q,\q\q  p^{k}_o\,=\, -\frac{\partial S_\sigma}{\partial l^o_{k}}\q\q\;\,=\, \q
-\sum_{t^n\subset \sigma} \frac{\p A_{t^o}}{\p l^o_{k}} \left( \pi-\theta_{t^o}^\sigma\right)
. \label{4c6e}
\ea

The $l^o_{k+1}$ remain undetermined, as these do not appear at all. Again, we can choose $l^o_k=l^o_{k+1}$ to carry these data along as boundary data.  The second equation in (\ref{4c6e}) represents four pre--constraints, that have to be satisfied by the canonical data at time $k$, so that the 4--1 move can be performed. This will, in general, restrict the free parameters which appeared in the previous moves.

\subsection{Example: 4D tent moves}\label{tent2}

As a specific example, we can also consider tent moves in 4D which, as already explained in section \ref{sectent1}, can be obtained by a succession of Pachner moves.
In particular, an $N$--valent tent move in 4D can be reproduced through a sequence of one 1--4 move, $(N-3)$ 2--3 moves, $(N-3)$ 3--2 moves and one final 4--1 Pachner move.\footnote{Note that the ordering of the 2--3 and 3--2 Pachner moves is not entirely fixed: one always has to start with a 1--4 move to introduce the new vertex $v_{n+1}$ and a subsequent 2--3 move and finish off with a 3--2 move prior to the final 4--1 move (which removes $v_n$ from $\Sigma_{n+1}$), but the ordering of the 2 - 3 and 3 - 2 moves in between can be chosen freely (according to the given configuration).}

Here, we will discuss two tent move configurations: the first one at a four--valent vertex leads to flat dynamics, while, in contrast to this, the second example involves curvature and the role of the pre-- and post--constraints differs considerably from that in the first example.


To begin with, consider the simplest tent move configuration in 4D, namely the four--valent tent move. Starting from a four-valent vertex $v_0$ in the boundary surface $\Sigma_0$ of a single 4--simplex, perform first one 1--4 and one 2--3 move which introduce five new (freely choosable) edge lengths. Finally, perform one 3--2 move to remove one of the edges $e(v_0i)$, $i\in1,\dots,4$, and one 4--1 move to remove the remaining three edges $e(v_0j)$ and the tent pole $e(v_0v_{1})$ from the new $\Sigma_{1}$ and solve the corresponding five equations of motion or  pre--constraints.

After the 3--2 move the boundary configuration, in fact, corresponds to the configuration of a {\it stacked sphere}, namely to the configuration of one simplex on which one 1--4 move has been performed with the tip of the second simplex pushed ``inwards". A {\it stacked sphere} is a triangulation of the 3--sphere which can be obtained by performing a sequence of 1--4 Pachner moves on the 3D boundary surface of a single 4--simplex and therefore necessarily possesses a 4D flat interior as there are no internal edges. Any triangulation whose boundary corresponds to a stacked sphere configuration possesses flat solutions independent of the existence of internal triangles (up to possible discretization artifacts \cite{bd1}).

That is, the 3--2 Pachner move plays a key role here in that its pre--constraint (or equation of motion) generally imposes flatness of the deficit angle around the internal triangle generated in the course of the previous 2--3 move and thus establishes one non-trivial condition among the five free parameters of the 1--4 and 2--3 moves. Hence, there exists a four parameter family of solutions to the equation of motion of the 3--2 Pachner move, all of which correspond to flat geometries. Therefore, these solutions also automatically solve the equations of motion of the final 4--1 Pachner move (the boundary configuration after the 4--1 move also corresponds to a stacked sphere). Thus, the four pre--constraints of the final 4--1 move are automatically satisfied and no further non--trivial conditions on the remaining four parameters arise. Accordingly, after the 4--1 Pachner move four parameters coordinatizing the position of the vertex remain free and one obtains a four--fold gauge symmetry associated to lapse and shift degrees of freedom.


This situation changes for $N$-valent tent moves with $N\geq5$, since the boundary configuration at none of the individual Pachner move steps corresponds to a  stacked sphere and therefore does not necessarily imply flatness. But non-vanishing curvature in Regge Calculus generically breaks the vertex displacement gauge symmetry \cite{bd1,dithoe1} and all free parameters become fixed. In particular, both the 3--2 and 4--1 move attach internal triangles to the edges in the hypersurface, which generally introduces a dependence on data from other evolution steps, e.g.\ see equation (\ref{y6}). The (post--)constraints after the 1--4 move will, in general, not be preserved by the following Pachner moves. Instead, these transform into pseudo-constraints \cite{dithoe1}. Pseudo--constraints can be understood as equations between the phase space data at time $k$ which further depend (weakly) on phase space data from the neighbouring time steps.

\begin{figure}[hbt!]
\begin{center}
    \psfrag{v}{$v_0$}
    \psfrag{a}{$a_0$}
    \psfrag{b}{$b_0$}
    \psfrag{1}{$1$}
    \psfrag{2}{$2$}
    \psfrag{3}{$3$}
    \psfrag{4}{$4$}
   \psfrag{5}{$5$}
   \includegraphics[scale=0.4]{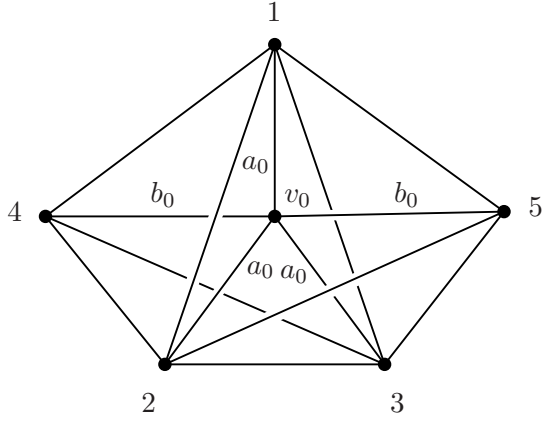}
    \end{center}
    \caption{\label{5vsym}{\small Illustration of the symmetry--reduced 3d star of a five--valent vertex $v_0$, consisting of the six tetrahedra (\ref{tent1}).}}
\end{figure}

As the second example consider, for simplicity, a `symmetry--reduced' tent move at a five--valent vertex $v_0$, also used in \cite{bd1,dithoe1}, where only two dynamical length variables, $a_n$ and $b_n$, arise at each tent move step $n$, apart from the tent pole length. There are six tetrahedra in $\Sigma_0$ with
vertices
\be\label{tent1} v_0124,\q v_0134,\q v_0234, \q  v_0125, \q  v_0135,
\q v_0235  \q .
\ee
Accordingly, we will have nine triangles of the
form $t(v_0ij)$ with $i,j=1,\ldots 5$ in this triangulation, five edges
of the form $e(v_0i)$ and nine edges of the form $e(ij)$  (all possible
ordered combinations of $i,j\in \{1,\ldots 5\}$ with the exception
$45$). This situation is depicted in figure \ref{5vsym}.

\noindent

`Symmetry--reduction' here means that all the lengths of the boundary edges $e(ij)$ are set to $1$ \footnote{This is possible since the vacuum Regge equations are invariant under global rescalings.} and imposing $l^{e(v_ni)}=a_n$, $i=1,2,3$ and $l^{e(v_n4)}=l^{e(v_n5)}=b_n$ at each tent move step $n$.

The 4--simplices associated to the Pachner moves reproducing this tent move are then all of the same type
$\sigma(v_0v_{1}ij\kappa)$, where $i,j$ take values in
$1,2,3$ and $\kappa$ in $4,5$. Hence, the first 1--4 move already introduces all three new (freely choosable) parameters $a_1,b_1$ and the length of the tent pole between $v_0$ and $v_{1}$ which we will call $t_1$. Specifically, let us consider the case where we glue the six 4-simplices onto the tetrahedra (\ref{tent1}) in the following order: 1.\ $\sigma(v_0v_1124)$, 2.\ $\sigma(v_0v_1134)$, 3.\ $\sigma(v_0v_1125)$, 4.\ $\sigma(v_0v_1135)$, 5.\ $\sigma(v_0v_1234)$, 6.\ $\sigma(v_0v_1235)$, which corresponds to one 1--4 move, two subsequent 2--3 moves followed by two 3--2 moves and a final 4--1 move. These Pachner moves are then to be implemented canonically by equations (\ref{4c6},\ref{4c6c},\ref{c13},\ref{c13b},\ref{c13c},\ref{c14},\ref{c14b},\ref{c14c},\ref{4c6d},\ref{4c6e}). The necessary action contribution of each of these six simplices is of the general form
\ba
S_{\sigma(v_0v_{1}ij\kappa)}&=&2A^a_t(c_1\pi-\theta^a_t)+A^b_t(c_2\pi-\theta^b_t)+A^0_a(c_3\pi-\theta^0_a)\nn\\
&&+A^1_a(c_4\pi-\theta^1_a)+2A^0_b(c_5\pi-\theta^0_b)+2A^1_b(c_6\pi-\theta^1_b)-A\theta\q,
\ea
where
\begin{itemize} \parskip -3pt
\item[~] $\theta^0_a,\, A^0_a$ are the dihedral angle and the area of the triangle $\Delta(v_0ij)$,
\item[~] $\theta^0_b,\, A^0_b$ are the dihedral angle and the area of the triangle $\Delta(v_0i\kappa) $,
\item[~] $\theta^a_t,\, A^a_t$ are the dihedral angle and the area of the triangle $\Delta(v_0v_1i )$,
\item[~] $\theta^b_t,\, A^b_t$ are the dihedral angle and the area of the triangle $\Delta(v_0v_1\kappa) $,
\item[~] $\theta^1_a,\, A^1_a$ are the dihedral angle and the area of the triangle $\Delta(v_1ij)$,
\item[~] $\theta^1_b,\, A^1_b$ are the dihedral angle and the area of the triangle $\Delta(v_1i\kappa)$
\item[~] $\theta,\, A$ are the dihedral angle and the area of the triangle $\Delta(ij\kappa)$, 
\end{itemize}
respectively, and $c_m\in\{0,\textstyle{\frac{1}{2}},1\}$, $m=1,\ldots,6$, depending on whether the corresponding triangle has been present prior to the move under consideration, is newly introduced or becomes internal (see the discussion in the paragraph following (\ref{regge2}) in section \ref{secregcal}). 

In the sequel we confine our focus to pre-- rather than post--constraints since the latter are automatically satisfied after the moves. The pre--constraint for $e(v_01)$ of the first 3--2 move coincides with the equation of motion for  the $a_0$--edges, the pre--constraint for $e(v_04)$ of the second 3--2 move with that of the $b_0$--edges and the single pre--constraint of the 4--1 move (recall the `symmetry--reduction') is equivalent to the equation of motion for the length $t_{1}$ of the tent pole $e(v_0v_{1})$. 

In the general (not symmetry--reduced) situation, we would have four parameters introduced by the first 1--4 move and a further 2 parameters introduced by the two following 2--3 moves. The latter two parameters determine the curvature, that is the deficit angles on the triangles which become bulk triangles during the completion of these moves. In the symmetry--reduced situation, on the other hand, the 1--4 move already introduces all three parameters that are allowed by our choice of symmetry--reduction. One of these corresponds to a lapse degree of freedom (determining the height of the tent pole), the other two can be interpreted as determining the value of the curvature. As we shall discuss, in general all parameters introduced by the 1--4 and 2--3 moves will be fixed by the pre--constraints of the subsequent 3--2 and 4--1 moves.

At the level of the momenta, the above symmetry-reduction only holds when a tent move step is complete, however, not for the individual Pachner moves into which the tent move can be decomposed, since the momenta at the different edges (of identical length) get updated in different order depending on the order of the Pachner moves. For this reason, we consider the momenta of each of the edges individually, which we denote by $p^0_i,p^0_\kappa$ for the initial data and $p^1_i,p^1_\kappa$ for those associated to the new edges. (Note that the upper index here counts tent rather than Pachner moves.)

The whole point of elaborating on this example is to demonstrate that, in contrast to the four--valent tent move, it is possible to solve only a subset of the equations of motion (or equivalently, pre--constraints) for $a_0,b_0,t_1$ and violate the remaining ones. This implies that the pre--constraints of the five--valent tent move are, in general, independent, not automatically satisfied and fix parameters which were previously free. 

For instance, consider the following three examples based on the same initial data $a_0=1.1690$, $p^0_1=p^0_2=p^0_3=1.5088$, $b_0=1.1436$, $p^0_4=p^0_5=1.2272$: a) table \ref{tab:1} shows the example of a situation where only the pre--constraint for $a_0$ is solved but not the ones for $b_0$ and $t_1$, b) table \ref{tab:2} provides an example where the pre--constraints for $a_0$ and $b_0$ are solved, but not the one for $t_{1}$, and c) table \ref{tab:3} demonstrates an example where all three pre--constraints are satisfied. Vanishing momenta indicate that a constraint (or equation of motion) is satisfied. In these three examples we choose to use the pre--constraints for $a_0,b_0$ to fix the new lengths $a_1,b_1$, respectively, and the pre--constraint for $t_1$ to fix $t_1$,\footnote{In fact, prior to the 3--2 moves and the implementation of the pre--constraints the new lengths can, in principle, be of any value and only get tuned after imposing the constraints. However, here we set the lengths of the new edges to the value determined by one or more pre--constraints already from the start.} but note that other choices of which length to fix by which constraint are possible.

\begin{table*}[h] \scriptsize{ \caption{\small Momentum updating for the five--valent symmetry--reduced tent move decomposed into Pachner moves. Only the pre--constraint for $a_0$ is eventually satisfied. The rows provide the updated momentum values after the Pachner move given in the left column (following the sequence of the main text). Initial data as given in the text. Two of the three new lengths of the first 1--4 move were chosen as $\ft_1=0.2600$, $b_1=1.3400$. The 3--2 move pre--constraint (\ref{c14c}) for edge $e(v_01)$ (which translates into $p^0_1=0$ after the move) is solved for $a_1=1.4052$. (Due to symmetry, the pre--constraints (\ref{4c6e}) associated to $e(v_02)$ and $e(v_03)$ of the 4--1 move are then automatically satisfied and translate into $p^0_2=0=p^0_3$ after the final move.) The pre--constraint for the edge $e(v_04)$ of the second 3--2 move and the pre--constraints of the 4--1 move for edge $e(v_05)$ and the tent pole are violated, i.e.\ $p^0_4,p^0_5,p_{\ft_1}\neq0$.  }
\vspace{.2cm}
\begin{tabular}{|c|| c|c|c|c|c|c|c|c|c|c|c||} \hline move & $p^0_1$
& $p^0_2$ & $p^0_3$ & $p^0_4$ & $p^0_5$ & $p_{t_1}$ & $p^1_1$
& $p^1_2$ & $p^1_3$ & $p^1_4$ & $p^1_5$\\ \hline\hline
1-4 &2.5119& 2.5119& 1.5088& 0.2082& 1.2272& 5.4351& -1.0316&
  -1.0316& 0& 1.1903& 0\\ \hline
2-3 & -0.1145& 2.5119& 2.5119& -2.2703& 1.2272& 3.1184& 1.5135&
  -1.0316& -1.0316& 3.5106& 0\\ \hline
2-3 & -2.7011& -0.0747& 2.5119& -2.2703& 0.2082&
  -1.0557& 3.9286& 1.3834& -1.0316& 3.5106& 1.1903\\ \hline
3-2 & 0& -0.0747& -0.0747& -2.2703& -2.2703&
  -3.3723& 1.3463& 1.3834& 1.3834& 3.5106& 3.5106\\ \hline
3-2 & 0& -2.7011& -2.7011& -1.4644& -2.2703&
  -7.5464& 1.3463& 3.9286& 3.9286& 2.7552& 3.5106\\ \hline
4-1 & 0& 0& 0& -1.4644& -1.4644&
   -2.1113& 1.3463& 1.3463& 1.3463& 2.7552& 2.7552\\ \hline
\end{tabular}
\label{tab:1}}
\end{table*}

\begin{table*}[h] \scriptsize{ \caption{\small Momentum updating for the five--valent symmetry--reduced tent move decomposed into Pachner moves. Only the pre--constraints for both $a_0, b_0$ are eventually satisfied. Further explanation and initial data are given in the caption of table \ref{tab:1} and the text, respectively. We chose $\ft_1=0.2000$. The pre--constraints for edges $e(v_01)$ and $e(v_04)$ of the two 3--2 moves were solved for $a_1=1.3448$, $b_1=1.2985$. The pre--constraint for the tent pole arising in the final 4--1 move is `almost solved', however, still violated ($p_{\ft_1}=0.0003> 0$, while the other constraints are satisfied to order $10^{-12}$). The latter is a consequence of the fact that the present configuration yields a near--flat geometry in which the vertex displacement symmetry of the flat regime is almost preserved \cite{bd1,dithoe1}. }
\vspace{.2cm}
\begin{tabular}{|c|| c|c|c|c|c|c|c|c|c|c|c||} \hline move & $p^0_1$
& $p^0_2$ & $p^0_3$ & $p^0_4$ & $p^0_5$ & $p_{t_1}$ & $p^1_1$
& $p^1_2$ & $p^1_3$ & $p^1_4$ & $p^1_5$\\ \hline\hline
1-4 &2.1317& 2.1317& 1.5088& 0.7195& 1.2272& 5.1219& -0.6116&
  -0.6116& 0& 0.6253& 0\\ \hline
2-3 &-0.1145& 2.1317& 2.1317& -1.3172& 1.2272& 3.1121& 1.5659&
  -0.6116& -0.6116& 2.5562& 0\\ \hline
2-3 &-2.3208& -0.0747& 2.1317&
  -1.3172& 0.7195& 0.0001& 3.6559& 1.4784&
  -0.6116& 2.5562& 0.6253\\ \hline
3-2 &0& -0.0747& -0.0747& -1.3172& -1.3172&
  -2.0097& 1.4397& 1.4784& 1.4784& 2.5562& 2.5562\\ \hline
3-2 &0& -2.3208& -2.3208& 0& -1.3172&
  -5.1217& 1.4397& 3.6559& 3.6559& 1.3084& 2.5562\\ \hline
4-1 &0& 0&
  0& 0& 0& 0.0003& \
1.4397& 1.4397& 1.4397& 1.3084& 1.3084\\ \hline
\end{tabular}
\label{tab:2}}
\end{table*}

\begin{table*}[h] \scriptsize{\caption{\small Momentum updating for an example of the five-valent symmetry-reduced tent move decomposed into Pachner moves where all three pre--constraints of $a_0, b_0, t_1$ are eventually satisfied. Further explanation and initial data as given in the caption of table \ref{tab:1} and the text, respectively. The three pre--constraints are numerically solved by $t_1=0.3039$, $a_1=1.4387$, $b_1=1.3832$.}
\vspace{.2cm}
\begin{tabular}{|c|| c|c|c|c|c|c|c|c|c|c|c||} \hline move & $p^0_1$
& $p^0_2$ & $p^0_3$ & $p^0_4$ & $p^0_5$ & $p_{t_1}$ & $p^1_1$
& $p^1_2$ & $p^1_3$ & $p^1_4$ & $p^1_5$\\ \hline\hline
1-4 &2.2970& 2.2970& 1.5088& 0.7969& 1.2272& 5.4635& -0.7724&
   -0.7724& 0& 0.6121& 0\\ \hline
2-3 &-0.1145& 2.2970& 2.2970& -1.3947& 1.2272& 3.3397& 1.5390&
  -0.7724& -0.7724& 2.6485& 0\\ \hline
2-3 &-2.4861& -0.0747& 2.2970&
  -1.3947& 0.7969& 0& 3.7361& 1.4247&
  -0.7724& 2.6485& 0.6121\\ \hline
3-2 &0& -0.0747& -0.0747& -1.3947& -1.3947&
  -2.1238& 1.4027& 1.4247& 1.4247& 2.6485& 2.6485\\ \hline
3-2 &0& -2.4861& -2.4861& 0& -1.3947&
  -5.4635& 1.4027& 3.7361& 3.7361& 1.3528& 2.6485\\ \hline
4-1 &0& 0& 0& 0&
  0& 0& 1.4027& 1.4027& 1.4027& \
1.3528& 1.3528\\ \hline
\end{tabular}
\label{tab:3}}
\end{table*}

In conclusion, after the two 3--2 moves two of the three new parameters (i.e.\ lengths) are, in general, fixed and one has a one parameter family of solutions to the two equations of motion of these two moves (with table \ref{tab:2} giving one particular example). However, generally these solutions are {\it not} automatically solutions to the equation of motion of the final 4--1 move. That is, here the three pre--constraints are, in general, independent and completely fix the three free parameters of the initial 1--4 move in which case solutions to all moves exist, but not a parameter family of solutions and therefore no gauge symmetry arises. Nevertheless, the subset of the one parameter family of solutions to the two equations of motion of the 3--2 moves corresponding to flat configurations {\it does} automatically solve the pre--constraint of the 4--1 move and no further non-trivial condition arises on the parameters.\footnote{There are two independent deficit angles in this symmetry--reduced setup, such that flatness imposes {\it two} independent conditions among the three free parameters $a_{1},b_{1},t_{1}$ which allow to write, e.g.\ $a_{1}$ and $b_{1}$ as a function of $t_{1}$. Due to flatness, by (\ref{y2}), the {\it three} equations of motion (or pre--constraints) for $a_0,b_0$ and $t_1$ are automatically satisfied.} That is, in this special case one parameter remains unrestricted after the 4 - 1 move (the length of the tent pole) and one obtains gauge symmetry.

This may be further understood from a covariant point of view, namely an analysis of the Hessian of the action (second partial derivatives of the action with respect to $t_1,a_1,b_1,t_2$) associated to two tent move steps, which was explored in \cite{bd1}. There it was shown numerically that the Hessian possesses a non-vanishing eigenvalue (corresponding to the lapse degree of freedom) in the presence of curvature which approaches zero as the configuration approaches flatness. Furthermore, the momenta associated to the edges are in general given by
\ba
p^n_e=-\frac{\partial\tilde{S}_{n+1}}{\partial l^e_n}\q,
\ea
where $\tilde{S}_{n+1}$ is the effective action of tent move step $n+1$ with the length variable of the tent pole integrated out. In \cite{bd1,dithoe1} it was shown that the matrix of derivatives of these momenta with respect to the $l^e_{n+1}$ is non-invertible too, if the Hessian is degenerate. In this flat case the conditions of the implicit function theorem are violated and we cannot solve for the lengths $l^e_{n+1}$ as a function of $l^e_n$ and $p^n_e$. That is, the data at step $n$ does {\it not} determine the data at step $n+1$ and, in particular, the free parameters of the 1--4 and 2--3 moves are not all fixed, reflecting the gauge symmetry. Conversely, when the Hessian is invertible in the presence of curvature, it {\it is} possible to solve for the $l^e_{n+1}$ as a function of the lengths and momenta at step $n$. For the symmetry--reduced five-valent tent move discussed here the latter means that, given the aim to construct a tent move from the data present at step $n$, the three {\it a priori} free parameters of the 1--4 move are eventually completely fixed by the three pre--constraints of the 3--2 and 4--1 Pachner moves. There is therefore no gauge symmetry left in the presence of curvature.

Consequently, in 4D for non-flat configurations, the pre--constraints provide  conditions among the canonical data (for instance, on the {\it a priori} free parameters introduced in the previous 1--4 and 2--3 moves) and actually implement the equations of motions.  
In 3D the post--constraints arising after the 1--3 moves are preserved by the other moves and exactly match the pre--constraints which come with the 3--1 moves. Hence, in 3D we are in the situation where the pre--constraints are automatically satisfied by canonical data generated by previous Pachner moves, that is the pre--constraints coincide with the post--constraints.

The situation is different in 4D. Here the post--constraints from the 1--4 and 2--3 moves generally do not match up with the pre--constraints of the 4--1 and 3--2 moves. Thus, the parameters introduced by the 1--4 and 2--3 moves will, in general, become fixed by the pre--constraints of the 4--1 and 3--2 moves. (Of course, whether all parameters get fixed, depends, for example, on the number of different moves one is performing.) For initial data leading to flat configurations, however, we have seen that the pre--constraints for the 4--1 moves are automatically satisfied by the canonical data generated by the previous Pachner moves. This converts the four parameters introduced by the 1--4 move to four gauge degrees of freedom (in the sense that the solution for given boundary conditions on an initial and final time slice is not uniquely determined).

\section{Symplectic structures for discrete canonical dynamics} \label{app}

The Pachner move dynamics involves phase spaces of different dimension. Hence, the question arises in which sense these maps can be canonical, i.e.\ preserve a symplectic structure. We will discuss this issue in this section by first considering regular systems, and subsequently general singular systems, where we have to deal with constraints. 
In particular, we will introduce in more detail the concepts of pre-- and post--constraints.
Finally, we will consider the situation for Pachner moves, and, more generally, for dynamics (generated by an action) which involves a change in phase space variables.

\subsection{Regular systems}\label{appreg}

Here we will review the general Lagrangian and Hamiltonian dynamics of discrete systems; for regular systems, which we will define shortly, see the exposition \cite{marsdenwest}. However, we need to specifically discuss singular systems, which appear in the Pachner move dynamics (see also the discussions in \cite{gambini} on constraints in discrete dynamics).

To start with, we will consider regular systems, where by definition constraints will not occur.  Consider a system with configuration spaces ${\cal Q}_k \cong {\cal Q}$ of equal dimension at every time step $k$. We coordinatize ${\cal Q}_k$ by $x_k^i$ where $i$ is in some index set determined by the dimension of $\cal Q$. Time evolution is generated by the discrete action contributions $S_k=S_k(x_{k-1},x_k)$. These discrete actions may be `effective ones', i.e.\ arising from Hamilton's principal function and summarizing several basic time steps into one effective time step (as, e.g., in the tent moves).

Consider three consecutive time steps $k=0,1,2$ and the boundary value problem between times $k=0$ and $k=2$. That is, we have boundary data $x_0^i$ and $x_2^i$ and must extremize
\ba\label{app1}
S:=S_1(x_0,x_1)+S_2(x_1,x_2)
\ea
with respect to $x_1$. This gives the equations of motion
\ba\label{app2}
0&=&\frac{\partial S_1}{\p x_1}+\frac{\p S_2}{\p x_1}\q,
\ea
which we assume to be solvable uniquely for $x_1$ as a function of $x_0,x_2$. In case that
\ba\label{app3}
\text{det}\frac{\p^2 S_2}{\p x_1\,\p x_2} \neq 0
\ea
we can invert these solutions for $(x_1,x_2)$ to obtain the Lagrangian time evolution map
\ba\label{app4}
{\cal L}_1: (x_0,x_1) \mapsto (x_1,x_2)
\ea
from ${\cal Q}_0\times {\cal Q}_1$ to ${\cal Q}_1\times {\cal Q}_2$.
Condition (\ref{app3}) has to hold for regular systems.

From the discrete action we can define the so--called Lagrangian one-- and two--forms. For regular systems these will turn out to be pull--backs via the Legendre transform of the canonical one-- and two--forms. The significance of the Lagrangian two--form is, that it is preserved under the time evolution map (\ref{app4}). This is the reason for the Hamiltonian time evolution map (in phase space) being symplectic with respect to the canonical two--form.

To start with, we define two canonical one--forms on ${\cal Q}_1\times {\cal Q}_2$ from the variation of the action,
\ba\label{app5}
\theta^-_1(x_1,x_2) &=& -\frac{\p S_2}{\p x_1^i} dx_1^i \nn\\
\theta^+_2(x_1,x_2) &=& \frac{\p S_2}{\p x_2^i} dx_2^i\q,
\ea
where we sum over repeated indices $i,j$. Note that $dS_2=\theta_2^+-\theta_1^-$ and, since $d\circ d S_2=0$, we can define the (single) Lagrange two--form
\ba\label{app6}
\Omega_2(x_1,x_2)\,=\,-d\theta^+_2\,=\,-d\theta^-_1&=&- \frac{\p^2 S_2}{\p x_1^i \p x_2^j} \, dx_1^i \, \wedge dx_2^j  \q .
\ea

This time evolution map is preserved under the time evolution map (\ref{app4}), i.e.
\ba\label{app7}
\Omega_1&=&{\cal L}_1^* \Omega_2\q,
\ea
where ${\cal L}_1^*$ denotes the pull--back of ${\cal L}_1$. To see this, consider $S$ from (\ref{app1}) as a function on ${\cal Q}_0\times {\cal Q}_1$ by using for $x_2$ the solutions $x_2(x_0,x_1)$ of the time evolution map (\ref{app4}). Taking the exterior derivative of $S$ on  ${\cal Q}_0\times {\cal Q}_1$, we will obtain only boundary terms because the equations of motion with respect to the inner variable $x_1$ hold. That is,
\ba\label{app8}
dS\,(x_0,x_1)&=& \frac{\p S_1}{\p x^i_0}dx^i_0\,\,+\,\, \frac{\p S_2}{\p x^j_2} \left( \frac{\p x^j_2}{\p x^i_0}dx^i_0 + \frac{\p x^j_2}{\p x^i_1}dx^i_1\right) \nn\\
&=& -\theta_0^- \,+\, {\cal L}_1^* \theta_2^+ \q .
\ea
Again, as $d\circ d=0$ and exterior derivatives commute with pull--backs, we also find
\ba\label{app8a}
\Omega_1&=& {\cal L}_1^* \Omega_2 \q .
\ea
This argument can be easily generalized to any time step difference $(k_1,k_2)$.

In order to discuss the dynamics in phase space, we introduce the Legendre transformations
\ba\label{app9}
&&\mathbb{F}^+S_k:  (x_{k-1},x_k) \mapsto (x_k,{}^{+}p^k)\,=\,\left(x_k,  \frac{\partial S_k}{ \partial x_{k}}\right) \nn\\
&&\mathbb{F}^-S_k:  (x_{k-1},x_k) \mapsto (x_{k-1},{}^{-}p^{k-1})\,=\,\left(x_{k-1},  -\frac{\partial S_k}{ \partial x_{k-1}}\right) \q
\ea
from $\cq_{k-1}\times\cq_k$ to the phase spaces $\cp_{k-1}$ and $\cp_k$ (these are contangent bundles of $\cq_k$ and $\cq_{k-1}$ respectively),  which we call {\it post--} and {\it pre--Legendre transformations}, respectively. The regularity condition (\ref{app3}) ensures that the Legendre transformations are (locally) invertible. The Hamiltonian time evolution
\ba\label{app10}
{\cal H}_1: (x_1,{}^-p_1) \mapsto (x_2, {}^+p_2 )
\ea
can be defined via the Legendre transform from the Lagrangian one (\ref{app4}) and coincides with the time evolution map generated by $S_2$ via the equations (\ref{b3}). Furthermore, it is straightforward to check that the Lagrangian one-- and two--forms, (\ref{app5}) and (\ref{app6}), arise from pulling back the canonical one-- and two--forms, $p_idx^i$ and $ dx^i\wedge dp_i$, respectively,  with the Legendre transformation (\ref{app9}).
This can be used to show that the Hamiltonian time evolution is symplectic, i.e.\ preserves the canonical two--form.

Diagrammetically,
\begin{diagram}
& &\cq_{k-1}\times\cq_k&& \rTo^{\cl_k}&&\cq_k\times\cq_{k+1} &   &\\
&\ldTo^{{\mathbb{F}}^-}& &\rdTo^{{\mathbb{F}}^+}&&\ldTo^{{\mathbb{F}}^-}  && \rdTo^{{\mathbb{F}}^+}&\\
 \cp_{k-1}&& \rTo^{\ch_{k-1}}& &\cp_k&& \rTo^{\ch_{k}}&&  \cp_{k+1}\q.
\end{diagram}

\subsection{Singular systems}\label{sing}

Let us now consider the situation where the regularity condition (\ref{app3}) is violated, yet $\cq_{k-1}\cong\cq_k$, that is, where an equal number of left and right null vectors $L_1^i$ and $R^i_2$ occurs, satisfying
\ba\label{app13}
L_1^i \frac{\p^2 S_2}{\p x_1^i \p x_2^j}\;=0  , \q\q      \frac{\p^2 S_2}{\p x_1^i \p x_2^j} R_2^j\;=0   \q
\ea
in some open neighborhood in ${\cal Q}_1\times {\cal Q}_2$. (To avoid excessively many indices, we will not introduce another index numbering the null vectors in this section.)

Firstly, we will discuss the consequences for the Lagrangian discrete time evolution which is defined as the space of solutions to (\ref{app2}), namely, as the submanifold in ${\cal Q}_0\times {\cal Q}_1\times {\cal Q}_2$ satisfying
\ba\label{app14}
0&=&\frac{\partial S_1}{\p x^i_1}+\frac{\p S_2}{\p x^i_1}.
\ea
Given a particular solution, i.e.\ a configuration $(x_0,x_1,x_2)$ satisfying (\ref{app14}), note that also an infinitesimally displaced configuration $(x_0,x_1,x_2+\varepsilon R_2)$ is a solution. Hence, the solution $x_2$ as a function of $x_0,x_1$ is not uniquely determined and arbitrariness arises. We could call the directions $R_2$ `preliminary gauge directions'. Note, however, that the {\it a priori} free parameters corresponding to these directions may get fixed {\it a posteriori} by entering the equations of motion of later time evolution steps. Therefore, `gauge' can {\it a priori} really only refer to the dynamics of the single time step from $k=1$ to $k=2$. In fact, the null vectors $L_1^i$ and $R^i_2$ of the Lagrangian two--form do not necessarily extend to null vectors of the Hessian of the action which, in turn, define the proper gauge symmetries of the action. 

Notwithstanding the arbitrariness in the solutions, we can define a Lagrangian time evolution map from ${\cal Q}_0\times{\cal Q}_1$ to ${\cal Q}_1\times{\cal Q}_2$. Since $x_2$ is not uniquely determined, however, we either have to fix $N$ {\it a priori} free parameters (if there are $N$ independent null vectors $R_2$) or map to `gauge equivalence classes'. Either way, instead of the time evolution mapping onto a $2Q$ dimensional space, where $Q$ is the dimension of configuration space ${\cal Q}$, it maps at most onto a $(2Q-N)$--dimensional one.
Thus, either the time evolution map is only defined on some (constraint) submanifold of ${\cal Q}_0\times{\cal Q}_1$ or the map is not injective. A combination of both possibilities could also occur.

Assume for the moment that such constraints do not occur, so that we can define a Lagrangian time evolution map $\cl_1$ on the full configuration space ${\cal Q}_0\times {\cal Q}_1$. To this end, just fix some `gauge parameters' $\lambda_2$ in ${\cal Q}_1\times {\cal Q}_2$ to determine $x_2$ uniquely as a function of $x_0$ and $x_1$. Notice that at this step $S(x_0,x_1)$, i.e.\ the action (\ref{app1}) evaluated on the solution as a function of the initial data $x_0,x_1$, will generally depend on this `gauge' choice. In the present case the Lagrangian two--form is, obviously, degenerate: the coordinate expression (\ref{app6}) 
directly shows that it possesses $2N$ null directions $L_1^i$ and $R_2^i$. Nevertheless, the arguments (\ref{app8}) to (\ref{app8a}), showing that the Lagrangian two--form is preserved under the Lagrangian time evolution, still hold true in exactly the same way as for regular systems.

Let us turn to the Hamiltonian dynamics. Due to the $N$ left and $N$ right null vectors (\ref{app13}), the rank of both Legendre transformations
\ba\label{app16}
&&\mathbb{F}^+S_2:  (x_{1},x_2) \mapsto (x_2,{}^{+}p^2)\,=\,\left(x_2,  \frac{\partial S_2}{ \partial x_{2}}\right) \nn\\
&&\mathbb{F}^-S_2:  (x_{1},x_2) \mapsto (x_{1},{}^{-}p^{1})\,=\,\left(x_{1},  -\frac{\partial S_2}{ \partial x_{1}}\right) \q
\ea
is $2Q-N$. Hence, the Legendre transformations are not onto. In general, $\mathbb{F}^\pm S_2$ simultaneously fail to be isomorphisms if and only if condition (\ref{app3}) is violated, i.e.\ if and only if the Lagrangian two--form (\ref{app6}) is degenerate. The image will be given by $(2Q-N)$--dimensional submanifolds in the two phase spaces ${\cal P}_1$ and ${\cal P}_2$ which we will call ${\cal C}_1^-$ and ${\cal C}_2^+$, respectively. We emphasize $\dim\cc^-_1=\dim\cc^+_2$.

\begin{Definition}\label{def_pcons}
The image of the pre--Legendre transform, $\cc_1^-:=\text{\emph{Im}}(\mathbb{F}^-S_2)\subset\cp_1$, is called the \emph{pre--constraint surface}. The image of the post--Legendre transform, $\cc_2^+:=\text{\emph{Im}}(\mathbb{F}^+S_2)\subset\cp_2$, is called the \emph{post--constraint surface}.
\end{Definition}

Note that the {\it pre--constraints} are automatically satisfied by the pre--momenta ${}^-p^1$, yet constitute the conditions that need by satisfied on ${\cal P}_1$ so that a Hamiltonian evolution can be defined. On the other hand, the {\it post--constraints} are automatically satisfied by the post--momenta ${}^+p^2$ after a time evolution from time $k=1$ to time $k=2$ has taken place, but constitute non--trivial conditions for the pre--momenta ${}^-p^2$. In general, the {\it pre--} and {\it post--constraint surfaces} at a given step $k$ do {\it not} coincide, $\cc^+_k\neq\cc^-_k$. Momentum matching requires that one restricts to $\cc^+_k\cap\cc^-_k$.


The $N$ pre--constraints ${ C}_1^-$ are defined through the equation
\ba\label{app16a}
0&=&{ C}_1^-(x_1,{}^-p^1)_{{\big |}  {}^-p^1=  -\frac{\p S_2}{\p x_1}(x_1,x_2)   }
\ea
which has to hold for arbitrary $x_1$ and $x_2$. Differentiating equation (\ref{app16a}) with respect to $x_2$ and with respect to $x_1$, respectively, we find the relations
\ba\label{app16b}
0&=& \frac{\p C_1^-}{\p {}^-p_j^1} \frac{\p^2 S_2}{\p x_1^j \p x_2^i} \\
0&=& \frac{\p C_1^-}{\p x^j_1} \,-\,  \frac{\p C_1^-}{\p {}^-p_j^1} \frac{\p^2 S_2}{\p x_1^j \p x_1^i}  \q .
\ea
Since there are $N$ independent left null vectors $L_1^j$ of $\frac{\p^2 S_2}{\p x_1^j \p x_2^i}$, we can conclude from (\ref{app16b}) that
\ba
 \frac{\p C_1^-}{\p {}^-p_j^1} = \sum_L \gamma^C_L (x_1,p_1)\,  L_1^j\q,
\ea
where we sum over all $N$ left null vectors $L_1$ and $\gamma^C_L(x_1,p_1)$ are appropriately chosen coefficients (which also depend on the constraint under consideration).
Similar relations can be found for the partial derivatives of ${ C}_2^+$. This specifies the gradients of the constraints; the orthogonal subspace is then tangential to the constraint hypersurface ${\cal C}_1^-$, respectively ${\cal C}_2^+$.

The Hamiltonian time evolution map ${\cal H}_1$ will only be defined on the submanifold ${\cal C}_1^-$. As before it can be generated by the discrete action, i.e.
\ba\label{app17}
{}^{-}p^{1}= -\frac{\partial S_{2}}{ \partial x_{1}}  \q  , \q\q  {}^{+}p^{2}  = \frac{\partial S_2}{ \partial x_{2}}   \q .
\ea
This makes the appearance of pre-- and post--constraints explicit. In particular, we can define pre-- and post--constraints independently from the Legendre transformation. Given a Hamiltonian time evolution map, we declare the pre--image of this map as the submanifold of pre--constraints and the image of this time evolution map determines the post--constraints. In the previous sections we have given the Hamiltonian time evolution maps for all the Pachner moves (and, hence, for all evolution maps involving a sequence of Pachner moves) 
based on momentum updating between the respective constraint surfaces which are defined in this manner. For these cases the pre--image and image of the maps could always be specified (but see also section \ref{apppach} below). 

Furthermore, if we find a solution to the first of the equations (\ref{app17}) for $x_2$ as a function of $x_1,{}^-p^1$, then the infinitesimally displaced configuration $x_2+\varepsilon R_2$ is also a solution. That is, as in the Lagrangian picture, $x_2$ is not uniquely defined. (Note that ${}^+p^2$ may also change under the transformation generated by $\varepsilon R_2$.) Rather, we have $N$ `preliminary gauge directions' $R_2$. The same can be said for expressing $x_1$ as a function of $x_2,{}^+p^2$ where now the `gauge' directions are given by $L_1$. Concretely, we have the gauge displacements
\ba\label{app18}
\delta_Lx^i_1=L_1^i\,\, ,\q\q \delta_L p^1_i=-\frac{\p^2S_2}{\p x_1^i \p x_1^j}  L_1^j
\q .
\ea
Note that these vectors are orthogonal to the gradients of the constraints  $C_1^-$, hence, the `gauge' displacements are tangential to the constraint hypersurface ${\cal C}^-_1$.

Given that $\ch_1:\cc^-_1\rightarrow\cc^+_2$, $\ch_1$ cannot be a symplectic mapping. Nevertheless, as regards the preservation of the symplectic structure, consider the canonical two--forms
\ba\label{app19}
\omega_1=dx_1^j\,\wedge\, dp^1_j  , \q\q\omega_2=dx_2^j\,\wedge\,dp^2_j .
\ea
We can pull these two--forms back via the embeddings $\iota_1^-:{\cal C}_1^- \rightarrow {\cal P}_1$ and $\iota_2^+:{\cal C}_2^+ \rightarrow {\cal P}_2$ to two--forms on the constraint surfaces ${\cal C}^-_1$ and ${\cal C}^+_2$, respectively. Notice that the resulting two--forms are pre--symplectic forms. Accordingly, the following theorem proves that discrete Hamiltonian time evolution rather is a pre--symplectic map.

\begin{Theorem}\label{thm_presymp}
The discrete Hamiltonian time evolution map $\ch_1:\cc^-_1\rightarrow\cc^+_2$ satisfies
\ba
(\iota_1^-)^*\omega_1 = {\cal H}_1^*  (\iota_2^+)^*\omega_2  .
\ea
\end{Theorem}

\begin{proof}
It is convenient to introduce coordinates $\{w_k^I, y_k^\alpha, z^k_\alpha\}$ with $k=1,2$ on ${\cal C}_1^-$ and ${\cal C}_2^+$, respectively. The index $I =1,\ldots N$ shall label `preliminary gauge directions' $(L_1)_I^i$ and $(R_2)_I^i$, while $\alpha=1,\ldots, Q-N$ labels coordinates associated to vectors $(N_1)^j_{\alpha},(M_2)^j_{\alpha}$ which are {\it not} null directions of $\Omega_2$. These coordinates can be chosen such that one obtains for the embedding map $\iota_1^-:(w_1,y_1,z^1)\mapsto (x_1,p^1)$ 
\ba\label{app20}
&&\frac{\p x^i_1}{\p w_1^I}=(L_1)_{I}^i  \q, \q\q  \frac{\p p_i^1}{\p  w_1^I}=-\frac{\p^2S_2}{\p x_1^i \p x_1^j}  (L_1)_I^j  \nn\\
&&\frac{\p x^i_1}{\p y_1^\alpha}=(N_1)_\alpha^i       \q, \q \q  \frac{\p p_i^1}{\p  y_1^\alpha}=-\frac{\p^2S_2}{\p x_1^i \p x_1^j}  (N_1)_\alpha^j \nn\\
&&\frac{\p x^i_1}{\p z^1_\alpha}=0                                  \q, \q\q\q \q \,  \frac{\p p_i^1}{\p  z^1_\alpha}=   (T^{-1}_1)^\alpha_j\,,\nn
\ea
and for  $\iota_2^+:(w_2,y_2,z^2)\mapsto (x_2,p^2)$
\ba\label{app20b}
&&\frac{\p x^i_2}{\p w_2^I}=(R_2)_I^i  \q, \q\q  \frac{\p p_i^2}{\p  w_2^I}=\frac{\p^2S_2}{\p x_2^i \p x_2^j}  (R_2)_I^j  \nn\\
&&\frac{\p x^i_2}{\p y_2^\alpha}=(M_2)_\alpha^i       \q, \q \,\,\,\,\,  \frac{\p p_i^2}{\p  y_2^\alpha}=\frac{\p^2S_2}{\p x_2^i \p x_2^j}  (M_2)_\alpha^j \nn\\
&&\frac{\p x^i_2}{\p z^2_\alpha}=0                                  \q, \q\q\q \q  \, \frac{\p p_i^2}{\p  z^2_\alpha}=   (T^{-1}_2)^\alpha_j\, .\nn
\ea
$(T^{-1}_1)_j^{\alpha},(T^{-1}_2)_j^{\alpha}$ are vectors satisfying
\ba\label{invv}
(T^{-1}_1)_j^{\alpha}(L_1)^j_{I}&=&(T^{-1}_2)_j^{\alpha}(R_2)^j_{I}\,\,\,=0,\nn\\
(T^{-1}_1)_j^{\alpha}(N_1)^j_{\alpha'}&=&(T^{-1}_2)_j^{\alpha}(M_2)^j_{\alpha'}=\delta^\alpha_{\alpha'}\q\forall\,I,\alpha.
\ea
The pull--backs of the two forms can then be written as
\ba\label{app21}
(\iota_1^-)^*\omega_1\!\!&=& \sum_j \left((L_1)^j_{I} \,dw_1^I + (N_1)^j_{\alpha}\,dy_1^\alpha\right) \,\wedge \nn\\
&&\left(     -\frac{\p^2S_2}{\p x_1^j \p x_1^i}    \left(( L_1)_{I'}^i   \, dw_1^{I'} +
(N_1)_{\alpha'}^i \, dy_1^{\alpha'} \right) +
( T^{-1}_1)_j^{\alpha'}  \,  dz^1_{\alpha'}   \right)  \nn\\
(\iota_2^+)^*\omega_2\!\!&=& \sum_j \left((R_2)^j_{I} \,dw_2^I + (M_2)^j_{\alpha}\,dy_2^\alpha\right) \,\wedge\nn \\
&&\left(
\frac{\p^2S_2}{\p x_2^j \p x_2^i}
\left(
 (R_2)_{I'}^i  \,dw_2^{I'} +
 (M_2)_{\alpha'}^i \, dy_2^{\alpha'} \right)
+
(T^{-1}_2)_j^{\alpha'}  \,  dz^2_{\alpha'}
\right). \nn
\ea
Firstly, the terms containing the second derivative of the action $S_2$ vanish because they are contracted with an antisymmetric form. Secondly, (\ref{invv}) entails 
\ba
(\iota_1^-)^*\omega_1\!\!&=&dy_1^\alpha    \,\wedge \,
  dz^1_{\alpha},   \label{app22a} \\
(\iota_2^+)^*\omega_2\!\!&=&dy_2^\alpha   \,\wedge \,
dz^2_{\alpha} .\label{app22b}
\ea
These expressions imply that the `preliminary gauge vectors' $\delta_{L_I}$ in (\ref{app18}) are degenerate directions of the two--form $(\iota_1^-)^*\omega_1$, whereas the corresponding $\delta_{R_I}$ are degenerate directions of $(\iota_2^+)^*\omega_2$. Finally, expressing the Hamiltonian time evolution (\ref{app17}) directly in the coordinates $y_k^\alpha$ and $z_k^\alpha$, one can check that the Hamiltonian time evolution preserves the pull--backs of the canonical forms as stated in the theorem.
\end{proof}

Hence, diagrammatically, 
\begin{diagram}
& &\cq_{k-1}\times\cq_k&& \rTo^{\cl_k}&&\cq_k\times\cq_{k+1} &   &\\
&\ldTo^{{\mathbb{F}}^-}& &\rdTo^{{\mathbb{F}}^+}&&\ldTo^{{\mathbb{F}}^-}  && \rdTo^{{\mathbb{F}}^+}&\\
 \cp_{k-1}\supset\cc^-_{k-1}&& \rTo^{\ch_{k-1}}& &\cc^+_k\cap\cc^-_k&& \rTo^{\ch_{k}}&&  \cc^+_{k+1}\subset\cp_{k+1}\q.
\end{diagram}

\subsection{Dynamical systems with changing phase space dimensions}\label{apppach}

Such singular systems as discussed above appear in the dynamics generated by the Pachner moves. There we had to deal with the problem that the number of variables may change from one time step to the next. Let us consider a simple example which will highlight the general principle. Consider again three consecutive time steps with variables $x_0,x_1,x_2$. But now assume that among the variables $x_2$ there is a  `new variable' $x_2^n$, in particular, that the number of variables at time step $k=2$ is $Q+1$, whereas it is $Q$ at times $k=0,1$.  Although this problem might be well posed as a boundary value problem keeping the data at $k=0,2$ fixed, it will, in general, not be possible to turn it into a well posed initial value problem with initial data at times $k=0,1$.

To describe the situation nevertheless by an initial value problem, extend the configuration spaces at times $k=0,1$ by the configuration variables $x_0^n$ and $x_1^n$, respectively. $x_2^n$ can be interpreted as an initial datum which becomes relevant only at step $k=2$, yet which we are allowed to already specify at time $k=0$, e.g.\ as $x_2^n=x_1^n=x_0^n$. The action pieces $S_1(x_0,x_1)$ and $S_2(x_1,x_2)$ neither depend on $x_0^n$ nor on $x_1^n$ because these variables were just introduced for book keeping purposes. Accordingly, the dynamics from $k=0$ via $k=1$ to $k=2$ is singular. Let us assume, for simplicity, that the dynamics is otherwise regular.

The extension does not interfere at all with the dynamics of the other variables for the time step from $k=0$ to $k=1$. Thus, extend also the phase spaces at times $k=0,k=1$ by the pairs $(x_k^n,p^k_n)$; the Legendre transformations will map onto the constraint hypersurfaces $C^-_0={}^-p^0_n=0$ and $C^+_1={}^+p^1_n=0$, respectively. The proof of Theorem \ref{thm_presymp} shows that the canonical two--forms restricted to the constraint hypersurfaces possess degenerate directions, i.e.\ `preliminary gauge directions'. The corresponding variables $x_0^n$ and $x_1^n$ are {\it a priori} free (Lagrange) parameters $\lambda_0,\lambda_1$. Therefore, we can fix $x^n_0,x^n_1$ to some arbitrary value. In particular, we can fix these data to coincide with the value of $x_2^n$. Note that the `reduced phase space', i.e.\ the constraint hypersurface modulo the `preliminary gauge direction' (which is the null direction of the restricted canonical two--form) coincides with the unextended phase space we started with.

Next, let us consider the time step from time $k=1$ to time $k=2$ described on the extended phase space ${\cal P}'_1$ and the phase space ${\cal P}_2$ (which already includes the pair $(x_2^n,p^2_n)$). The matrix
\ba\label{appe1}
\frac{\p^2 S_2}{\p x^i_1 \p x^j_2}
\ea
has (due to our assumption only) one left null vector $L_1^i=\delta^i_n$. Consequently, there is also a right null vector $R_2^i$ and the Lagrangian two--form  (\ref{app6}) possesses these two degenerate directions. The Legendre transformations (\ref{app16}) will again not be surjective; for time $k=1$ it will rather map onto the constraint hypersurface $C_1^-={}^-p^1=0$. This pre--constraint coincides with the post--constraint ${}^+p_1=0$ from the previous time step. Moreover, we will have a post--constraint $C_2^+=0$ on the canonical data at $k=2$.

As we have seen, we will also find `preliminary gauge directions' associated to the constraints. At time $k=1$ the `gauge' direction (\ref{app18}) just corresponds to the coordinate $x^n_2$, i.e.\ $\delta_Lx_1^i=\delta^i_n$ and $\delta_Lp_i^1=0$. At time $k=2$, on the other hand, we have
\ba\label{appe2}
\delta_R x_2^i=R_2^i \q ,\q\q \delta_R p_i^2=\frac{\p^2 S_2}{\p x_2^i \p x_2^j}R^j_2 \q .
\ea
Theorem \ref{thm_presymp} shows that the Hamiltonian time evolution preserves the canonical two--forms restricted to the constraint hypersurfaces. In the present example (\ref{app22a}) at $k=1$ is just the canonical two--form of the original $2Q$--dimensional unextended phase space $\cp_1$ which, however, is used as a pre--symplectic form on a $(2Q+1)$--dimensional constraint surface in the extended phase space $\cp'_1$. On the other hand, (\ref{app22b}) at $k=2$ is the restriction of the canonical two--form of the $(2Q+2)$--dimensional unextended phase space $\cp_2$ to the $(2Q+1)$--dimensional constraint hypersurface $C^+_2=0$. Its single degenerate direction is given in (\ref{appe2}). Consequently, the `reduced phase space' at $k=2$ is $2Q$--dimensional just as the original $\cp_1$. The phase space extension from $\cp_1$ to $\cp'_1$ at $k=1$ can be viewed as embedding the smaller phase space $\cp_1$ into the bigger phase space $\cp_2$; discrete time evolution proceeds such that the (restricted) two--forms are preserved on this bigger phase space.


The case of `old variables'  $x^o_1$ that disappear at time $k=2$ can be discussed in the same way by reversing time direction. 

As an extreme example, consider the cases described in sections \ref{3D} and \ref{4D}, namely the time evolution step from a zero--dimensional phase space to the phase space corresponding to the boundary of a tetrahedron or 4--simplex, respectively.
The initial phase space is zero--dimensional and the phase space at time $k=1$ twelve-- or twenty--dimensional, corresponding to the boundary data of a tetrahedron, respectively 4--simplex. In this case the canonical two--forms at $k=0$ are just the zero forms as the initial phase space is zero--dimensional. Indeed, the time evolution preserves this property, as in both cases we had six, respectively ten constraints, fixing all the momenta as functions of the lengths variables. Moreover, one can check that the canonical two--forms restricted to the constraint hypersurfaces are totally degenerate, that is identically zero.

Finally, let us consider the local Pachner moves explicitly. These are rather directly defined in the Hamiltonian picture. The reason is that the Lagrangian picture requires the knowledge of configurations and velocities at a given time which in the discrete case translates to configuration data of two consecutive time steps. For a Pachner move (or any other evolution move, like the tent move, which does not cover the entire hypersurface), however, the two consecutive time steps overlap, so that we do not have a full set of configuration data. In the Hamiltonian picture the velocities, or in the discrete the configuration data of the second time step, are replaced by the momenta, which are defined at the same time step as the configuration data. That is, we just need one time slice to encode the canonical data. 

For the Pachner moves of type $M$--$N$ with $N>M$ in 3D and 4D we obtain $K$ new edge lengths ($K=1$ for the 2--3 move and $K=N$ for the 1--N move).  We, therefore, extend the phase space at time $k$ by $K$ pairs $(l_k^n,p^k_n)$ which are matched by the $K$ new edge lengths and conjugate momenta $(l_{k+1}^n,p^{k+1}_n)$ at time $k+1$. Additionally, we have pairs $(l^b,p^b)$ which do not change at all during this evolution move and variables $(l^e,p^e)$ for which only the momenta are updated. The Hamiltonian evolution map for this kind of Pachner moves (see, for instance, (\ref{c13}--\ref{c13c})) is given by the following momentum updating
\ba\label{anh1}
l^b_k&=&l^b_{k+1}  \q ,\q\q p^{k+1}_b\,=\,p^k_b \q , \\
l^e_k&=&l^e_{k+1}\q,\q\q p^{k+1}_e\,=\,p^k_e+\frac{\partial S_\sigma(l^e_{k+1}, l^n_{k+1})}{\partial l^e_{k+1}}  
\; ,    \label{anh1b}  \\
p^{k}_n&=&0 \q,\q\q\q p^{k+1}_n\,\,=\, \frac{\partial S_\sigma(l^e_{k+1},l^n_{k+1})  }{\partial l^n_{k+1}}\,\q   
\q .\label{anh1c} 
\ea
Here $S_\sigma$ stands for the action of the glued tetrahedron or 4--simplex and we choose this action to be a function of the length variables at time $k+1$. 

Equations (\ref{anh1c}) contain $K$ pre--constraints, namely $C_n^k=p^k_n$, and $K$ post--constraints
\ba\label{anh2}
 C_n^{k+1}&=&p^{k+1}_n- \frac{\partial S_\sigma(l^e_{k+1},l^n_{k+1})  }{\partial l^n_{k+1}}\q,
\ea
since this equation only involves canonical data from time step $k+1$. Note also that the shape (\ref{anh2}) of the post--constraints ensures that they have vanishing Poisson brackets among each other, i.e.\ they form and Abelian (sub--) algebra.\footnote{Dropping the evolution index $k$, one directly computes
$\{C_n,C_{n'}\}=\frac{\partial^2S_\sigma}{\partial l^n\partial l^{n'}}-\frac{\partial^2S_\sigma}{\partial l^{n'}\partial l^{n}}=0$.} The same holds for the pre--constraints. (This is consistent with the previous discussion showing that to every pre-- or post--constraint we can associate a degenerate direction of the symplectic forms restricted to the pre-- or post--constraint hypersurfaces. This actually means that the constraints are first class.) Notice, however, that the pre--constraints, in general, do not Poisson commute with the post--constraints.

Apart from this, as we discussed for each Pachner move, the Hamiltonian evolution map (\ref{anh1}--\ref{anh1c}) does not specify all variables uniquely and we rather obtain `preliminary gauge directions' both at time $k$ and $k+1$. Note that these are `gauge' only with respect to this evolution move from step $k$ to $k+1$ and, more precisely, with respect to the symplectic forms restricted to the constraint hypersurfaces arising in these moves.
Subsequent moves might result in a different set of (pre--) constraints with respect to which the original `preliminary gauge directions' do not constitute degenerate directions of the restricted symplectic form anymore.

At time $k$ the length variables $l^n_k$ remain undetermined, so that we have $\delta_L x^i_k=\delta^i_n L^n_k$ and $\delta_L p_i^k=0$ (where $i$ stands for any of the indices $b,e,n$ and $L^n_k$ are $K$ linearly independent vector fields). Also at time $k+1$ we have $K$ undetermined variables, namely the edge lengths $l^n_{k+1}$, which we referred to as {\it a priori} free parameters. These correspond to `preliminary gauge directions' 
\ba
\delta_R l^i_{k+1} = \delta_n^i R^n_{k+1}  \q, \q\q  \delta_R p_i^{k+1}=  \frac{\partial^2 S_\sigma  }{\partial l^i_{k+1} \partial l^{n'}_{k+1}}R^{n'}_{k+1}\q,
\ea
where $R^n_{k+1}$ are $K$ linearly independent vector fields. We have seen that the new edge lengths $l^n_{k+1}$ can be interpreted as additional initial data. These can turn either into physical degrees of freedom (as for the 2--3 moves where the new edge length determines the deficit angle at the bulk triangle generated by this move) or into proper gauge degrees of freedom (as for the 1--4 move for initial conditions leading to flat solutions). Additionally, the $l^n_{k}$ remain undetermined, however, in the previous discussions we pointed out that these can be `gauge fixed' to be equal to $l^n_{k+1}$.

As for the case discussed in the previous section \ref{sing}, the `preliminary gauge directions' are tangential to the constraint hypersurfaces defined by the constraints $C_n^k$ and $C_n^{k+1}$, respectively.

We can consider for both time steps the canonical symplectic forms
\be
\omega_k=dl^b_k \wedge dp_b^k + dl^e_k \wedge dp_e^k +dl^n_k \wedge dp_n^k  \q ,  \q\q \omega_{k+1}=dl^b_{k+1} \wedge dp_b^{k+1} + dl^e_{k+1} \wedge dp_e^{k+1} +dl^n_{k+1} \wedge dp_n^{k+1}
\ee
and restrict these to the constraint hypersurfaces in the respective phase spaces. For the (extended) phase space ${\cal P}_k$ at time $k$ we can use the canonical embedding $\iota_k:(l^b_k,p_b^k,l^e_k,p_e^k,l^n_k)\mapsto (l^b_k,p_b^k,l^e_k,p_e^k,l^n_k,p_n^k=0)$ and the pull--back of the symplectic form gives just the symplectic form of the unextended phase space
\ba
(\iota_k)^*\omega_k&=& dl^b_k \wedge dp_b^k + dl^e_k \wedge dp_e^k  \q .
\ea
For time step $k+1$ consider the embedding 
\be
\iota_{k+1}:(l^b_{k+1},p_b^{k+1},l^e_{k+1},p_e^{k+1},l^n_{k+1})\mapsto (l^b_{k+1},p_b^{k+1},l^e_{k+1},p_e^{k+1},l^n_{k+1}, p^n_{k+1}=\frac{\partial S_\sigma}{\partial l^n_{k+1}}) \q .
\ee
In this case, the pull--back of the symplectic form is given by
\ba\label{sy01}
(\iota_{k+1})^*\omega_{k+1} &= & dl^b_{k+1} \wedge dp_b^{k+1} + dl^e_{k+1} \wedge dp_e^{k+1} +dl^n_{k+1} \wedge \frac{\partial^2 S_\sigma}{\partial l^n_{k+1} \partial l^e_{k+1}}dl^e_{k+1}  \q .
\ea
Now the time evolution equations (\ref{anh1}--\ref{anh1c}) express the canonical coordinates at time $(k+1)$ as function of the canonical coordinates at time $k$ and as function of the length variables $l^n_{k+1}$ (as these are not determined by the time evolution map). Using this for the symplectic form (\ref{sy01}) at time $k+1$, we can conclude that the time evolution map ${\cal H}_k$ defined by  (\ref{anh1}--\ref{anh1c}) preserves the restricted symplectic forms:
\ba\label{sy01b}
{\cal H}_k^*(\iota_{k+1})^*\omega_{k+1} &=&dl^b_{k} \wedge dp_b^{k} + dl^e_{k} \wedge dp_e^{k} + dl^e_{k} \wedge  \frac{\partial^2 S_\sigma}{\partial l^e_{k+1} \partial l^n_{k+1}} dl^n_{k+1}+dl^n_{k+1} \wedge \frac{\partial^2 S_\sigma}{\partial l^n_{k+1} \partial l^e_{k+1}}dl^e_{k+1}  \nn\\
&=&dl^b_{k} \wedge dp_b^{k} + dl^e_{k} \wedge dp_e^{k}  \q .
\ea

The Pachner moves of type $N$--$M$ with $N>M$ can be discussed analogously, one only has to reverse the time direction. 
The key difference to those moves with $M>N$ is, however, that pre--constraints now arise before the move. The symplectic form is only preserved when restricted to the post--constraint surface after the move and to the constraint surface defined by both the post--constraints and the new pre--constraints before the move.

 The 2--2 move in 3D is the only Pachner move for which we extended both the initial and final phase space because we had both an `old' and `new' variable pair, see section \ref{22sec}.
The equations of motion were given by
\ba\label{anh21}
l^b_k&=&l^b_{k+1}  \q ,\q\q p^{k+1}_b\,=\,p^k_b \q , \\
l^e_k&=&l^e_{k+1}\q,\q\q p^{k+1}_e\,=\,p^k_e+\frac{\partial S_\tau(  l^e_{k+1},l^o_k,l^n_{k+1})  }{\partial l^e_{k+1}}  
\q ,\\
p^{k}_n&=&0 \q,\q\q\q p^{k+1}_n\,\,\,=\, \frac{\partial S_\tau(  l^e_{k+1},l^o_k,l^n_{k+1}) }{\partial l^n_{k+1}}\,\q 
 , \label{anh21c}\\
p^{k+1}_o &=& 0 \q,\q\q\q p^{k}_o\q\,\,\,=\, -\frac{\partial S_\tau (  l^e_{k+1},l^o_k,l^n_{k+1}) }{\partial l^n_{k+1}}\,\q
 .\q \label{anh21d}
\ea
Here the only pre--constraint is $p^k_n=0$ and the only post--constraint is $p^{k+1}_o=0$. Note that the second equations in both (\ref{anh21c}) and (\ref{anh21d}) do {\it not} constitute constraints as these depend on canonical data of both times $k$ and $k+1$. Correspondingly, the only gauge directions one encounters here arise from the fact that $l_k^n$ and $l_{k+1}^o$ remain undetermined by the equations of motion. Therefore, these can be gauge fixed to arbitrary values, for instance $l^n_k=l^n_{k+1}$ (although the latter length needs to be determined from the equations of motion) and $l^o_{k+1}=l^o_k$. The canonical symplectic form on the extended phase spaces pulled back to the constraint hypersurfaces just gives the canonical forms of the unextended phase spaces. Consequently, we see that in this case the dynamics can be reduced to the unextended phase spaces. The only configuration variables that change are $l^o_k\mapsto l^n_{k+1}$. 

In summary, we have seen that we can define suitably preserved symplectic structures for all Pachner moves. To this end we introduced extended phase spaces to circumvent the problem of changing phase space dimensions. However, we considered the canonical two--forms restricted to the constraint hypersurfaces, and these constraint hypersurfaces, upon factoring out the `preliminary gauge directions', coincided with, or were submanifolds in the original, unextended phase spaces.

An alternative construction is to extend the Hamiltonian time evolution maps, so that these are defined on the full (extended) phase spaces. For instance, considering the $M$--$N$ moves with $N>M$, the relevant changes from (\ref{anh1}--\ref{anh1c}) would be
\ba\label{anh22}
l^n_k=l^n_{k+1} \q ,\q\q p^{k+1}_n&=& p^k_n+ \frac{\partial S_\sigma}{\partial l^n_{k+1}}  \q .
\ea
That is, we obtain for all variable pairs  $i=b,e,n$
\ba\label{anh23}
l^i_{k+1}=l^i_{k} \q ,\q\q p^{k+1}_i&=& p^k_i+ \frac{\partial S_\sigma}{\partial l^i_{k+1}}  \q .
\ea
The time evolution map is defined on the full extended phase space and its image is given by the full phase space at time $k+1$. We can then impose the pre--constraints $p_n^k=0$. The post--constraints are  given as the image of the (pre--) constraint hypersurface under the time evolution map. Maps of the form (\ref{anh23}), which only involve momentum updating via a generating function preserve the canonical two--form (of the extended phase space).\footnote{Using the extended map $\bar{\ch}_{k}$ defined by (\ref{anh23}) and $\omega_{k+1}=dl^i_{k+1}\wedge dp^{k+1}_i$, one immediately verifies by symmetry and antisymmetry $(\bar{\ch}_{k})^*\omega_{k+1}=dl^i_k\wedge d\left(p^k_i+\frac{\partial S_{\sigma}}{\partial l^i_k}\right)=dl^i_k\wedge dp^k_i+\frac{\partial^2S_{\sigma}}{\partial\, l^i_k\partial\,l^{i'}_k}dl^i_k\wedge dl^{i'}_k=dl^i_k\wedge dp^k_i=\omega_k$.}
The previous discussion showed that also the canonical two--form restricted to the constraint hypersurfaces is preserved.

Such a construction is also possible for the 2--2 move, i.e.\ the extended dynamics would also be of the form (\ref{anh23}).  The pre--constraint $p^n_k=0$ and the post--constraint $p^o_{k+1}=0$, however, have to be imposed independently.
The second equation in (\ref{anh22}) then changes into another post--constraint prescribing $p_n^{k+1}$ as a function of the six edge lengths $l_{k+1}^i$ of the tetrahedron, where $i=e,n,o$. Alternatively, we can solve for the length $l^o_{k+1}=l^o_k$ as a function of the five edge lengths $l_{k+1}^n,l_{k+1}^e$ and the momentum $p_n^{k+1}$. Also the second equation of (\ref{anh23}) turns into a (pre--) constraint determining $p^{k}_o$ as a function of the six edge lengths $l_k^i, i=e,n,o$, or, alternatively, $l^n_k=l^n_{k+1}$ as function of the five edge lengths $l^o_k,l^e_k$ and $p_o^k$. That is, we obtain two constraints on each of the extended phase spaces ${\cal P}_k$ and ${\cal P}_{k+1}$. These are second class constraints. Compared to the previous treatment, where we had for, say, ${\cal P}_k$ only $p_n^k=0$ as a constraint and $l_n^k$ as a gauge degree of freedom, we can consider the second constraint between $l^n_k,l^o_k,l^e_k$ and $p^k_o$ as a gauge fixing condition. This shows that also the induced symplectic forms on the (gauge fixed and) constraint hypersurfaces are preserved under time evolution.

\section{Interpretation and role of the constraints}\label{intcon}

In spite of the fact that the set of pre--constraints and the set of post--constraints each form an Abelian (i.e.\ first class) Poisson sub--algebra, they generally do not generate gauge symmetries of the discrete action. Firstly, the pre--constraints do not, in general, Poisson commute with the post--constraints and both sets of constraints have to be imposed at each time step. Secondly, their phase space flows are associated to null vectors of the Lagrangian two--form, but these do not, in general, extend to null vectors of the Hessian of the action. However, a degenerate Hessian on solutions implies the existence of flat directions at the extrema of the action and thereby the presence of gauge symmetries. Rather, these constraints manifest non--uniqueness of solutions and reflect the lack of knowledge at a given time step about the full solution (e.g.\ as a consequence of varying phase space dimensions); in Regge calculus at some step $k$ there may not exist sufficient information about the triangulation at step $k+1$ (or vice versa) and a hypersurface may be `forgetful' about the `past' or `future'. For example, the post--constraints associated to the new edges simply implement the fact that the lengths of these edges can be freely varied in accordance with the canonical data at step $k$, but not that this will eventually be a symmetry of the action.

On the other hand, the gauge symmetries in Regge calculus are given by those variations of the edge lengths which leave the action invariant and correspond to the displacement in flat directions of vertices in the bulk of the triangulation. These symmetries are generically broken in the presence of curvature \cite{bd1,dithoe1,hamberwilliamsgauge,morse}. Such a vertex displacement gauge transformation is specifed by the (four linearly independent) null-eigenvectors of the Hessian of the piece of action corresponding to the star of the vertex in question \cite{bd1,dithoe1,bianca,ditfrespe}. The null-eigenvectors --- if existent --- are, thus, associated to a given vertex and define the flat (gauge) directions in which the vertex may be displaced without changing the action. In contrast to this, {\it a priori} each pre-- or post--constraint of the individual Pachner moves is associated to a given edge in the hypersurface under consideration, rather than any particular vertex. It describes the variation of the corresponding single edge length, however, there is no invariant way of expressing this as the displacement of the one or the other vertex at the ends of the edge in some embedding space. 
Nonetheless, in the canonical formalism the variation of the edges {\it can} be invariantly associated to the displacement of a given vertex $v$, if the constraints of those edges in the hypersurface which are adjacent to $v$ are contracted with the four null-eigenvectors associated to $v$ \cite{dithoe1};\footnote{Internal edges, or edges which do not yet occur at a given step, do not matter because they come with vanishing momenta and solved equations of motion.} this implements a projection in gauge directions and the correct number of gauge generators. At this stage, the contracted constraints generate the gauge transformations defined by the null-eigenvectors. 

This was discussed for tent moves in \cite{dithoe1}, but will be studied in detail and more generally in a forthcoming paper \cite{dhta}, where we will apply the present formalism to the 4D linearized theory, based on an expansion around a flat solution to linear order. This linearized regime, in fact, inherits the gauge freedom of the flat background solution and is relevant for the continuum limit where the diffeomorphism symmetry of classical general relativity ought to be restored and geometries are locally flat. This is where null-eigenvectors of the Hessian necessarily arise which, in turn, can be employed to construct proper gauge transformation generating constraints which are associated to vertices.  

\section{Discussion and outlook}\label{conc}

In the present article we have devised a general canonical formalism for discrete systems involving varying phase space dimensions. To this end we had to extend the general theory of discrete dynamics from the regular to the irregular case in which constraints necessarily arise. The basic ingredient of the new formalism is Hamilton's principal function which may be used as a generating function (of the first kind) for a canonical time evolution map. This time evolution is generally only defined on a constraint surface and maps to another constraint surface; we refer to the pre--image of the time evolution map as {\it pre--constraint} surface, while its image is called {\it post--constraint} surface. The pre--constrainsts, on the one hand, constitute the conditions that need be satisfied in order for the following time evolution step to be allowed. On the other hand, the post--constraints are relations among the phase space data that are automatically realized after an evolution step, however, may provide non-trivial conditions to be satisfied by subsequent time steps. As shown in general, the symplectic forms restricted to the constraint surfaces are preserved under such time evolution maps.

We have applied the new canonical formalism to simplicial gravity, in particular, and thereby constructed a general and fully consistent formulation of (Euclidean) canonical Regge calculus. However, the formalism is general and, therefore, adaptable to other discretization schemes (with additive action). The general idea is to implement a canonical forward time evolution by gluing single simplices step by step to a bulk triangulation, while backward evolution is simply achieved by removing single simplices from the triangulation. The prerequisite for this evolution scheme is additivity of the action. The $(D-1)$-dimensional hypersurfaces then evolve in a discrete `bubble' time through the full $D$-dimensional Regge solution akin to the evolution of hypersurfaces in canonical general relativity. Both the gluing and removal moves allowed in the present scheme can be interpreted entirely within the $(D-1)$-dimensional triangulated hypersurfaces as Pachner moves. Pachner moves are an elementary and ergodic class of piecewise-linear homeomorphisms which can map between any two triangulations of a given topology by finite sequences (in fact, one could also implement `spatial' topology changes --- aka births of baby universes --- by generalizing the class of allowed gluing and removal moves, see section \ref{Pachner}).  

The central issue to be addressed when implementing the Pachner moves in a canonical formalism is the varying phase space dimension since Pachner moves introduce and remove edges. This obstacle can be handled in a straightforward manner by suitable phase space extensions which are controlled by constraints that arise naturally in this framework and constitute the canonical incarnation of the equations of motion of edges to the `future' or `past' of a given hypersurface. In this fashion, phase spaces before and after any Pachner move can be brought to equal dimension, the moves can be implemented via momentum updating as canonical transformations and the symplectic forms restricted to the constraint surfaces are preserved under these moves. 
In fact, in an alternative construction without immediate use of Hamilton's principal function, the elementary time evolution map can also be extended to the full extended phase spaces (i.e.\ disregarding any constraints) via momentum updating and preserves the full symplectic structure. This requires, in addition, to impose the pre-- and post--constraints separately, since momentum updating by itself only specifies the difference of the momenta before and after an elementary move, however, not the specific values.

The implementation of the individual Pachner moves follows one and the same recipe, however, each of the Pachner moves assumes a special role in the evolution of the triangulation. In 3D the 1--3 move introduces one vertex and equips it with three constraints which generate vertex displacements in the flat embedding space; the three new edge lengths thereby assume a lapse and shift type gauge parameter interpretation. The pre-- and post--constraints of the 2--2 move\footnote{Recall that these constraints, in contrast to those from all other moves in 3D and 4D, are only constraints on the extended phase space.} simply preserve and implement flatness, while the pre--constraints of the 3--1 move are automatically satisfied and this move annihilates one vertex and three constraints associated with it. The situation in 3D is special in that all vacuum solutions are necessarily flat and the 3D Regge action is actually a so--called `perfect action' \cite{Bahr:2009qc,Bahr:2011uj}, i.e.\ despite being a discrete action it, nevertheless, reproduces the continuum dynamics and symmetries and is triangulation independent. This has significant repercussions for the initial value problem in the discrete: whereas usually the phase space corresponds to the space of solutions this cannot, in general, be expected when the phase space dimension varies and here sufficient initial data is rather specified in the course of evolution. Notwithstanding this issue, in 3D the space of initial data {\it does} correspond to the space of solutions (modulo gauge) because there are only global physical degrees of freedom (and the Pachner moves do not change the `spatial' topology), while the bulk is necessarily flat and there exists a three-fold gauge symmetry at each vertex in the triangulation. The latter symmetry corresponds to lattice diffeomorphisms and is a consequence of the action being perfect. 

In 4D, however, the situation changes because of local degrees of freedom and generically broken (diffeomorphism) symmetries; the 4D Regge action is {\it not} perfect and we face the issue of non--uniqueness of solutions in the sense that different solutions arising from the same initial data cannot be mapped into each other by gauge transformations because of the broken symmetries and are thus inequivalent. 
As an extreme example of this non--uniqueness for geometries with spherical `spatial' topology take the boundary of a 4--simplex as initial slice which fits into any such triangulation and whose data thus by no means specifies the ensuing geometry (or, even more extremely, consider the `no boundary proposal' for triangulations). This is also manifested in the particular roles of the individual Pachner moves in 4D:  the 1--4 move\footnote{Here we focus solely on gluing moves.} introduces a vertex along with four associated post--constraints which are automatically satisfied after the move and reflect the {\it a priori} free choice of the four new edge lengths. The 2--3 move generates a bulk triangle while only adding a new boundary edge which comes with a new post--constraint. Again, the length of this new edge, and thus the value of the curvature angle around the internal triangle, is {\it a priori} free; the move thereby introduces a (free) curvature degree of freedom. These first two moves are the only Pachner moves for 4D simplicial gravity which introduce new edges, however, at the same time do not invoke any equations of motion and, consequently, the new edge lengths are {\it a priori} free parameters. In addition, the 3--2 move is only allowed if the pre--constraint corresponding to the equation of motion of the edge that is rendered internal in the course of the move is satisfied. Likewise, prior to the 4--1 move which annihilates one vertex, the four pre--constraints associated to the four edges adjacent to the vertex which become internal must be fulfilled. The pre--constraints of these last two moves impose non-trivial conditions which, in general, require a tuning of the free parameters of previous moves (see also the example of the symmetry--reduced five--valent tent move in 4D in section \ref{tent2}). In short, in 4D all new edge lengths are {\it a priori} free data which, however, may become fixed {\it a posteriori} by the pre--constraints of subsequent 3--2 and 4--1 moves. 

As regards the occurrence of gauge symmetries, we have argued in section \ref{intcon} that the pre-- and post--constraints of the Pachner moves generally do not generate gauge transformations. However, the vertex displacement gauge symmetry of Regge calculus exists on flat configurations and for this regime, the post--constraints of the 1--4 moves --- in analogy to the 1--3 moves in 3D --- {\it do} constitute the gauge generators. In general, it depends on the particular configuration and, in particular, on choices of data during subsequent moves, whether the 1--4 move introduces gauge generators. A specific example of this was treated in section \ref{tent2}.

The formalism introduced here is very general and allows to also consider evolution maps in which arbitrary pieces of $D$--dimensional triangulations are glued onto $(D-1)$--dimensional hypersurfaces (under certain regularity conditions). These can be built up by Pachner moves, but may also be considered in an `effective' picture, where one time step would involve a more complicated piece of triangulation than just a simplex. For the example of the so--called tent moves we have discussed the `effective evolution' in 3D in section \ref{sectent1} and in 4D in section \ref{tent2}.

The new general canonical formalism for discrete systems involving varying phase space dimensions, clearly, warrants further research. In particular, it remains to be investigated under which precise conditions {\it a priori} free parameters in 4D canonical Regge calculus remain free throughout evolution, i.e.\ are true gauge parameters, or, equivalently, under which conditions post--constraints of previous steps match pre--constraints of later steps (but involving the same hypersurface). This is a prerequisite for providing a constraint classification with clear distinction between proper equations of motion and (pseudo-) constraints generating (broken) gauge symmetries. The latter question stresses the close connection between constraints and equations of motion, which often can be transformed into each other by relabeling time steps (see the discussion in section \ref{candisc}) or by changing the slicing into equal time hypersurfaces. Therefore, the question arises whether different choices of `effective' time steps (or slicings of the bulk space-time) may influence the constraints associated to a given hypersurface. This question will be important for quantization as one usually attempts to find a quantization of the constraint hypersurfaces. These issues will be the topic of a separate paper \cite{dhta2}. 

In another forthcoming paper \cite{dhta}, we will address some of these issues explicitly in the linearized theory which inherits the gauge symmetry of the flat background. The specific role of each of the Pachner moves becomes clear in this regime, namely: the 1--4 move generates four gauge degrees of freedom and the same number of gauge generating constraints, the 2--3 move generates one `graviton', the 3--2 move imposes the only proper, non-trivial equation of motion and annihilates one `graviton' and, finally, the 4--1 move annihilates four gauge degrees of freedom and the same number of gauge generating constraints which are automatically satisfied prior to the move. This allows for an explicit counting of both gauge and observable degrees of freedom at each evolution step in a scenario of an evolving `spatial' hypersurface. For reasons of continuity, the role of the individual Pachner moves must also be the same in a neighbourhood of the linearized regime.

Furthermore, the new formalism can also be viewed as a specific algorithm to generate solutions and might thereby be interesting for numerical investigations.
During the evolution it may happen that some variables can be solved only in the complex domain. This problem also appears in the consistent discretization approach \cite{bahrgambinipullin}, where lapse and shift, which are determined by the equations of motion
obtain complex values. 

The interpretation of this phenomenon is that the initial data chosen are not suitable for the discretization. It is related to the fact, that gauge symmetries are broken and therefore the corresponding constraints on initial data do not have to be implemented
strictly. However, there is still a remnant of the constraints: namely that a large set of initial data might be unsuitable as it leads to complex (or unphysical large) lapse and shift parameters.

Although in this work we do not fully address the main underlying problem, namely the fact that diffeomorphism symmetry is broken in 4D Regge gravity \cite{bd1}, this problem might appear to be less severe than in an approach where the discretization
is fixed from the outset. The advantage of the present approach is that the choice of triangulation and therefore discretization can be adjusted during evolution, i.e.\ a given sequence of Pachner moves might lead to complex solutions, whereas another sequence does
not. For instance one would expect that to avoid complex solutions, one has to choose a finer triangulation for regions with high curvature. The advantage of the approach proposed here is that such an adjustment can be implemented during evolution.

In view of Loop Quantum Gravity (LQG) which involves changing spatial graphs and Spin Foam models which employ triangulations, it seems promising to attempt a quantization of the Pachner moves and implement them as (possibly unitary) transformations on some Hilbert space of `spatial' geometries. This may help in investigating the connection between LQG and Spin Foams (see also \cite{Bonzom:2011tf} for recent advances from the Spin Foam side); due to the fact that the classical canonical formalism is based on using the action as a generating function, a canonical quantization thereof should coincide with a covariant path-integral quantization of the theory. Usually, a canonical program proceeds by quantizing the phase space (or rather the constraint surface); here the additional challenge emerges of how to incorporate the fact that initial data continues to be specified during evolution and certain {\it a priori} free parameters become fixed later on.
These questions are related to the `general boundary proposal' \cite{oeckl}, which attempts to generalize the notion of Cauchy hypersurfaces to arbitrary surfaces for quantum field theories. A quantization of the Pachner moves may also shed some light on the `$e^{\pm iS}$ vs.\ $\cos S$' debate in the Spin Foam community: because gluing and removal moves add and subtract action contributions of simplices, respectively, an amplitude with a single exponential of the action should correspond either to pure forward or backward evolution, while an amplitude with the cosine of the action incorporates a superposition of both directions of evolution. 

A final interesting question arises: can one construct from the ideas presented here a canonical formulation of Causal Dynamical Triangulations (CDT) (see also \cite{Arnsdorf:2001wh})? Given that {\it a priori} all new lengths in 4D canonical Regge calculus can be freely chosen and, in particular, all be fixed equal to one while the births of baby universes are disallowed, can one construct `classical CDT solutions' from this, such that the pre-- and post--constraints determine the connectivity by rejecting or accepting certain Pachner moves? Presumably, this will generically not work and one might have to allow, instead, for the lengths to take value in a small interval $(1-\epsilon,1+\epsilon)$. It would be interesting to investigate whether a quantization with such restrictions can be connected to CDT. This also requires an extension of the present scheme to Lorentzian signature and an implementation of the condition that spacelike hypersurfaces remain spacelike throughout evolution.

\section*{Acknowledgements}

B.D.\ thanks Laurent Freidel for collaboration on \cite{draft}. Both authors thank Benjamin Bahr and Renate Loll for interesting discussions. Furthermore, P.A.H.\ is grateful for Exchange Grant 3068 within the ESF activity entitled `Quantum Geometry and Quantum Gravity' which made this collaboration feasible and thanks the Albert Einstein Institute in Potsdam for hospitality in the early and final stages of this work.

\end{document}